\documentclass[11pt,a4paper]{article} 
\usepackage{jheppub,multirow}
\usepackage{amsthm}
\usepackage{xcolor}
\usepackage{float}
\usepackage{url}
\usepackage{microtype}
\usepackage{subcaption}
\usepackage{empheq}
\usepackage{booktabs}  
\usepackage[makeroom]{cancel}
\usepackage{dsfont} 
\usepackage{ytableau} 
\usepackage{setspace}
\usepackage[skins]{tcolorbox}
\usepackage[utf8]{inputenc}
\DeclareUnicodeCharacter{2212}{\ensuremath{{}{-}}}

\newcommand{\beq}{\begin{equation}}
\newcommand{\eeq}{\end{equation}}

\newcommand{\fb}{\mathfrak{b}}
\newcommand{\fg}{\mathfrak{g}}
\newcommand{\fa}{\mathfrak{a}}
\newcommand{\fh}{\mathfrak{h}}
\newcommand{\fl}{\mathfrak{l}}
\newcommand{\fm}{\mathfrak{m}}

\newcommand{\fn}{\mathfrak{n}}
\newcommand{\fp}{\mathfrak{p}}

\newcommand{\fu}{\mathfrak{u}}

\newcommand{\cC}{\mathcal{C}}
\newcommand{\cN}{\mathcal{N}}

\newcommand{\cW}{\mathcal{W}}
\newcommand{\cO}{\mathcal{O}}

\newcommand{\cS}{\mathcal{S}}

\def\ie{\begin{equation}\begin{aligned}}
\def\fe{\end{aligned}\end{equation}}

\newcommand*\xbar[1]{%
	\hbox{%
		\vbox{%
			\hrule height 0.5pt 
			\kern0.3ex
			\hbox{%
				\kern-0.1em
				\ensuremath{#1}%
				\kern-0.05em
			}%
		}
	}
}

\newtheoremstyle{fullit}
{\topsep}      
{\topsep}      
{\normalfont}  
{0pt}          
{\itshape}     
{.\ }          
{0pt}          
{\thmname{#1} \thmnumber{#2}}             

\newtheorem{thm}{Theorem}[section]

\theoremstyle{definition}

\theoremstyle{fullit}
\newtheorem{example}[thm]{Example}

\DeclareCaptionSubType*[alph]{figure}
\title{The ABCDEFG of Little Strings}
\author{Nathan Haouzi$^a$,}
\author{Can Koz\c{c}az$^{b,c,d,}$\footnote{Current affiliation Bo\u{g}azi\c{c}i University}}
\affiliation{$^a$Center for Theoretical Physics,
University of California, Berkeley, USA}
\affiliation{$^b$Department of Physics, Bo\u{g}azi\c{c}i University, Istanbul, Turkey}
\affiliation{$^c$Center of Mathematical Sciences and Applications,
Harvard University, USA}
\affiliation{$^d$Jefferson Physical Laboratory,
Harvard University, USA}
\emailAdd{nathanhaouzi@berkeley.edu,can.kozcaz@boun.edu.tr}

\abstract{Starting from type IIB string theory on an $ADE$ singularity, the $(2,0)$ little string  arises when one takes the string coupling $g_s$ to 0. In this setup, we give a unified description of the codimension-two defects of the little string, labeled by a simple Lie algebra $\fg$. Geometrically, these are D5 branes wrapping 2-cycles of the singularity, subject to a certain folding operation when the algebra is non simply-laced. Equivalently, the defects are specified by a certain set of weights of $^L \fg$, the Langlands dual of $\fg$. As a first application, we show that the instanton partition function of the $\fg$-type quiver gauge theory on the defect is equal to a 3-point conformal block of the $\fg$-type deformed Toda theory in the Coulomb gas formalism. As a second application, we argue that in the $(2,0)$ CFT limit, the Coulomb branch of the defects flows to a nilpotent orbit of $\fg$.}
\begin{document}
\maketitle
\setlength{\parindent}{0pt}
\clearpage


\newpage

\section{Introduction}

In the landscape of quantum field theories, six-dimensional superconformal field theories (SCFTs) hold a privileged place: six  is the highest number of dimensions where a conformal field theory with supersymmetry can exist. 
Those SCFTs are truly exotic in many regards: in Physics, the theories with $(1,0)$ or $(2,0)$ supersymmetry, for instance, have no description in terms of an action functional. Furthermore, the degrees of freedom of those theories are described by tensionless strings. On the Mathematics side, there is no precise definition of the theory, which makes it problematic to study. 

It proves useful instead to analyze a deformation of the 6d SCFT: the six-dimensional little string is such a theory. The $(2,0)$ little string is labeled by a Lie algebra $\fg$, of $ADE$ type. One can obtain it from type IIB string theory compactified on a surface $X$ with an $ADE$ type singularity, and by sending the string coupling $g_s$ to zero. The $(2,0)$ CFT is recovered by sending to infinity the only scale left in the theory, the string mass $m_s$ (while keeping the moduli of the $(2,0)$ theory fixed in the process.)\\

To illustrate this principle, consider for example the so-called Alday--Gaiotto--Tachikawa (AGT) conjecture \cite{Alday:2009aq}. Namely, the statement that certain 4d ${\cal N}=2$ theories whose origin is the 6d $(2,0)$ SCFT compactified on a punctured Riemann surface $\cC$, are related to a 2d conformal field theory on the surface, of Toda type, with ${\cal W}(\fg)$-algebra symmetry.

The little string setup enables one to state a version of this correspondence precisely, generalize it, and even prove it: the partition function (supersymmetric index) of the $(2,0)$ little string on a Riemann surface ${\cal C}$, with certain brane defects at points on ${\cal C}$, is in fact equal to a ``quantum" deformation of the Toda CFT conformal block on the surface ${\cal C}$, in the Coulomb gas formalism \cite{Aganagic:2015cta}. The deformed Toda theory no longer has conformal symmetry, but displays instead a ${\cal W}_{q,t}(\fg)$-algebra symmetry, depending on two deformation parameters $q$ and $t$, and was first analyzed in the 90's  \cite{Shiraishi:1995rp,Awata:1995zk,Frenkel:1998}. The vertex operators in the deformed Toda theory are determined by the positions and the types of defects.  Specifically, one introduces D5 branes as codimension two defects that are points on the Riemann surface $\cC$ and wrap non-compact 2-cycles of the resolved singularity $X$.\\

A first result of our paper is the extension of the above analysis to surface defects labeled by a non simply-laced Lie algebra. These non simply-laced theories arise in type IIB string theory from a non-trivial fibration of the $ADE$ surface $X$ over $\mathbb{C}^2\times \cC$. We define a non simply-laced defect as a ``folding" operation on D5 branes wrapping the 2-cycles of $X$, after acting on them with the outer automorphism group action of the $ADE$ Lie algebra.  

For definiteness, let us fix the Riemann surface $\cC$ to be an infinite cylinder, and let $\fg$ be a simple Lie algebra.  When $\fg$ is simply-laced, the little string compactified on  $\cC$, with D5 branes at points on it, is described at low energies by a 5d  $\cN=1$  quiver gauge theory of shape the Dynkin diagram of $A$, $D$, or $E$. When $\fg$ is non simply-laced, making sense of the low energy theory as a gauge theory is a delicate affair. We will show that it is rather natural to label it as a non simply-laced quiver, of shape the Dynkin diagram of $B$, $C$, $F$ or $G$, obtained by folding an $A$, $D$, or $E$ quiver. It turns out that one can still formally define the Coulomb branch of the non simply-laced theory, and even count instantons in this setup; see also the recent work \cite{Kimura:2017hez}. Then, we are able to show that the partition function of a $\fg$-type quiver is a conformal block of  $\fg$-type deformed Toda, with insertion of certain vertex operators. The vertex operators are labeled by a collection of coweights of $\fg$, or equivalently, of weights taken in the Langlands dual algebra $^L\fg$, obeying certain constraints that we specify explicitly.\\

The above correspondence is in fact a triality: to see this, one needs to turn on vortex flux in the 5d theory, and realize that there exists an effective description of the theory on the vortices. In the little string picture, the vortices are realized as additional D3 branes that are points on the cylinder $\cC$, and which wrap compact 2-cycles in $X$. At low energies, the theory on these branes is a 3d  $\cN=2$ quiver theory, again of shape the Dynkin diagram of $\fg$.

We show that the partition function of the 3d theory is equal to the partition function of the 5d theory, when the 5d Coulomb moduli are frozen to masses, with some amount of vortex flux turned on. In particular, this  gives a version of gauge/vortex duality when $\fg$ is non simply-laced. The triality follows because the 3d theory's partition function, written as an integral over its Coulomb moduli, is manifestly a conformal block in deformed Toda, where the number of D3 branes becomes the number of screening charges.\\

Our second result is the classification of these D5 brane defects, in terms of what we define as polarized and unpolarized sets of coweights of $\fg$. The polarized defects have a direct interpretation in terms of parabolic subalgebras of $\fg$, while the unpolarized ones do not.\\

Finally, we characterize the defects in the CFT limit, after taking the string mass $m_s$ to infinity. It is well known today that codimension 2 defects of the 6d $(2,0)$ CFT are classified by nilpotent orbits \cite{Chacaltana:2012zy,Chacaltana:2010ks,Chacaltana:2011ze,Chacaltana:2013oka,Chacaltana:2014jba,Chacaltana:2015bna,Chacaltana:2016shw,Chacaltana:2012ch}. Here, we conjecture that in the $m_s\rightarrow\infty$  limit, the Coulomb branch of a 5d $\fg$-type quiver gauge theory living on D5 branes flows to a nilpotent orbit of $\fg$. Furthermore, the coweight data of the D5 branes defining those quivers encodes the Bala--Carter labeling of a nilpotent orbit in the CFT limit.\\

Let us mention that the results of this paper are relevant to a correspondence known as geometric Langlands, which aims to prove an equivalence between specific categories associated to a connected complex Lie group and its Langlands dual. This duality can be phrased in the context of two-dimensional conformal field theories on a Riemann surface: on one side, one considers the center of the affine Kac-Moody algebra $\widehat{{^L \fg}}$ at level $^L k = - ^L h^{\vee}$; on the other side, one considers the classical $W$-algebra ${\cal W}_{\infty}(\fg)$.
Recently, a two-parameter deformation of the geometric Langlands correspondence has been  proposed \cite{Aganagic:2017smx}: the first side of the duality becomes the quantum affine algebra $U_{\hbar}(\widehat{^L \fg})$, a quantum deformation by the parameter $\hbar$ of the universal enveloping algebra of $\widehat{^L \fg}$. The other side becomes the $W$-algebra ${\cal W}_{q,t}(\fg)$ mentioned above:
\[
{\cal U}_\hbar(\widehat{^L \fg}) \longleftrightarrow {\cal W}_{q,t}(\fg)
\]
In particular, evidence was found that the conformal blocks of the two theories should be the same. A natural and important generalization is to introduce ramifications at points on the Riemann surface in this picture. These ramifications are nothing but the D5 branes (surface defects) we study in this paper, so our results provide an explicit realization of the objects on the right-hand side of the duality. We leave it to future work to analyze the left-hand side and prove the correspondence with ramifications.\\

The paper is organized as follows: in Section 2, we  review the $(2,0)$ little string theory and the defects for the simply-laced Lie algebras. Then we extend the setup to include the non simply-laced algebras as well. In Section 3, we provide a unified treatment of triality for an arbitrary simple Lie algebra, and prove it. In Section 4, we take the string mass $m_s$ to infinity in the little string and make contact with the CFT classification of defects as nilpotent orbits. Section 5 is dedicated to various examples illustrating  the statements of the paper.

\newpage

\section{$(2,0)$ Little String and Codimension 2 Defects}
\label{sec:review}

The $(2,0)$  little string is a six dimensional theory, labeled by a $\fg_0=ADE$ Lie algebra. It  therefore has 16 supercharges. The little string theory is not a local QFT, and the strings have a finite tension $m_s^2$. The theory was originally discovered and analyzed in \cite{Seiberg:1997zk,Losev:1997hx}\footnote{For a pedagogical review of the theory's main features, see  \cite{Aharony:1999ks}.}. Recently, there has been a renewed interest in its study, in a variety of contexts \cite{Kim:2015gha,Bhardwaj:2015oru,DelZotto:2015rca,Lin:2015zea,Hohenegger:2015btj,Hohenegger:2016eqy,Hohenegger:2016yuv,Aganagic:2016jmx,Bastian:2017ing,Bastian:2017ary}.\\ 

One construction of the theory starts in ten-dimensional type IIB compactified on a  two complex-dimensional surface $X$ labeled by the simply-laced Lie algebra $\fg_0$. This surface $X$ is a hyperk\"ahler manifold, which one constructs by resolving a ${\mathbb C}^2/\Gamma$ singularity;  $\Gamma$ is a discrete subgroup of $SU(2)$, and the fact that such discrete subgroup is labeled by the simply-laced Lie algebra $\fg_0$ is known as the McKay correspondence.  One then sends the string coupling $g_s$ to zero, which decouples the gravitational interactions in the type IIB string theory; that does not make the theory  trivial: the degrees of freedom supported near the surface $X$ remain. Indeed, the moduli space of the $(2,0)$ little string is $(\mathbb{R}^4 \times S^1)^{\mbox{rk}(\fg_0)}/W$, with $W$ the Weyl group of $\fg_0$, and the different scalars parameterizing this moduli space come from the moduli of the metric on $X$, as well as the NSNS and RR B-fields of the type IIB theory. 
At energies well below the string scale $m_s$, the little string reduces to a $(2,0)$ 6d CFT.

\subsection{$ADE$-type Defects}
\label{ssec:ade}

Our focus will be on studying a class of surface defects in the $(2,0)$ little string. Let us briefly review how these arise in our context \cite{Aganagic:2015cta}. We compactify the  little string theory on a Riemann surface $\cC$; in this paper, the Riemann surface will be the cylinder   $\cC= \mathbb{R} \times S^1(\hat{R})$. Note that $X\times \cC$ is a solution of the type IIB string theory, since the cylinder has a flat metric. Codimension 2 defects are realized as D5 branes in the original type IIB. Namely, we take D5 branes to wrap 2-cycles in $X$ and  $\mathbb{C}^2$, and to be points  on the cylinder $\cC$. A single D5 brane will brake half of the supersymmetry, leaving us with  8 supercharges; since we will ultimately be considering a collection of many D5 branes, it is important that they preserve the same supersymmetry. This can be done by setting to zero the periods of certain self-dual 2-forms defined on $X$.  The tension of the D5 branes remains finite in the little string limit $g_s \rightarrow 0$, so one can study the theory on the branes explicitly. At low energies, below the string scale, it takes the form of a five-dimensional quiver gauge theory, with $\cN=1$ supersymmetry, on $\cC= \mathbb{R} \times S^1(R)$, where $R=1/(m_s^2\,\hat{R})$\footnote{The 5d nature of the theory, as opposed, say, to a 4d $\cN=2$ theory, is due to the presence here of the circle $S^1(\hat{R})$ making up the cylinder $\cC$. Even though the D5 branes are points on this $S^1(\hat{R})$, they still feel its presence as KK modes on the T-dual circle of radius $R=1/(m_s^2\,\hat{R})$. The low energy physics is therefore an honest 5d theory on a circle. Had we chosen the Riemann surface $\cC=\mathbb{C}$, the low energy physics would have been a 4d theory; had we chosen $\cC=T^2$, the low energy physics would have been a 6d theory on the T-dual torus.}. The quiver has the shape of the  Dynkin diagram of $\fg_0$ \cite{Douglas:1996sw}\footnote{In that work, the quiver gauge theory on the branes was an affine $ADE$ quiver. Here, taking the limit $g_s\rightarrow 0$ decouples the affine node.}, with unitary gauge groups. In the rest of this paper, we will call this gauge theory $T^{5d}$.\\

The dictionary from geometry to gauge theory data is as follows:
the moduli of the 6d $(2,0)$ theory become gauge couplings of $T^{5d}$. The D5 branes  which wrap compact 2-cycles  of $X$ are dynamical, and as such, their positions on $\cC$ are the Coulomb moduli of the gauge theory. The D5 branes  which wrap  non-compact 2-cycles of $X$ are frozen, by virtue of the fact that these 2-cycles extend to infinity; as such, the position of these D5 branes on $\cC$ realizes mass parameters for fundamental hypermultiplets.

\subsection{$BCFG$-type Defects}
\label{ssec:bcfg}

Let $\fg_0$ be a simply-laced Lie algebra. We define $\fg$ to be a subalgebra of $\fg_0$ invariant under the outer automorphism group action of $\fg_0$. It is well known that such outer automorphisms of $\fg_0$ are in one-to-one correspondence with the automorphisms of the Dynkin diagram of $\fg_0$. The resulting subalgebras  $\fg$ are called non simply-laced. Let $A$ be a nontrivial outer automorphism group of $\fg_0$. Then either $A=\mathbb{Z}_2$ or $A=\mathbb{Z}_3$, where the precise group action is shown in Figure \ref{folding} below:

\begin{figure}[h!]
	\begin{center}
		\includegraphics[width=0.95\textwidth]{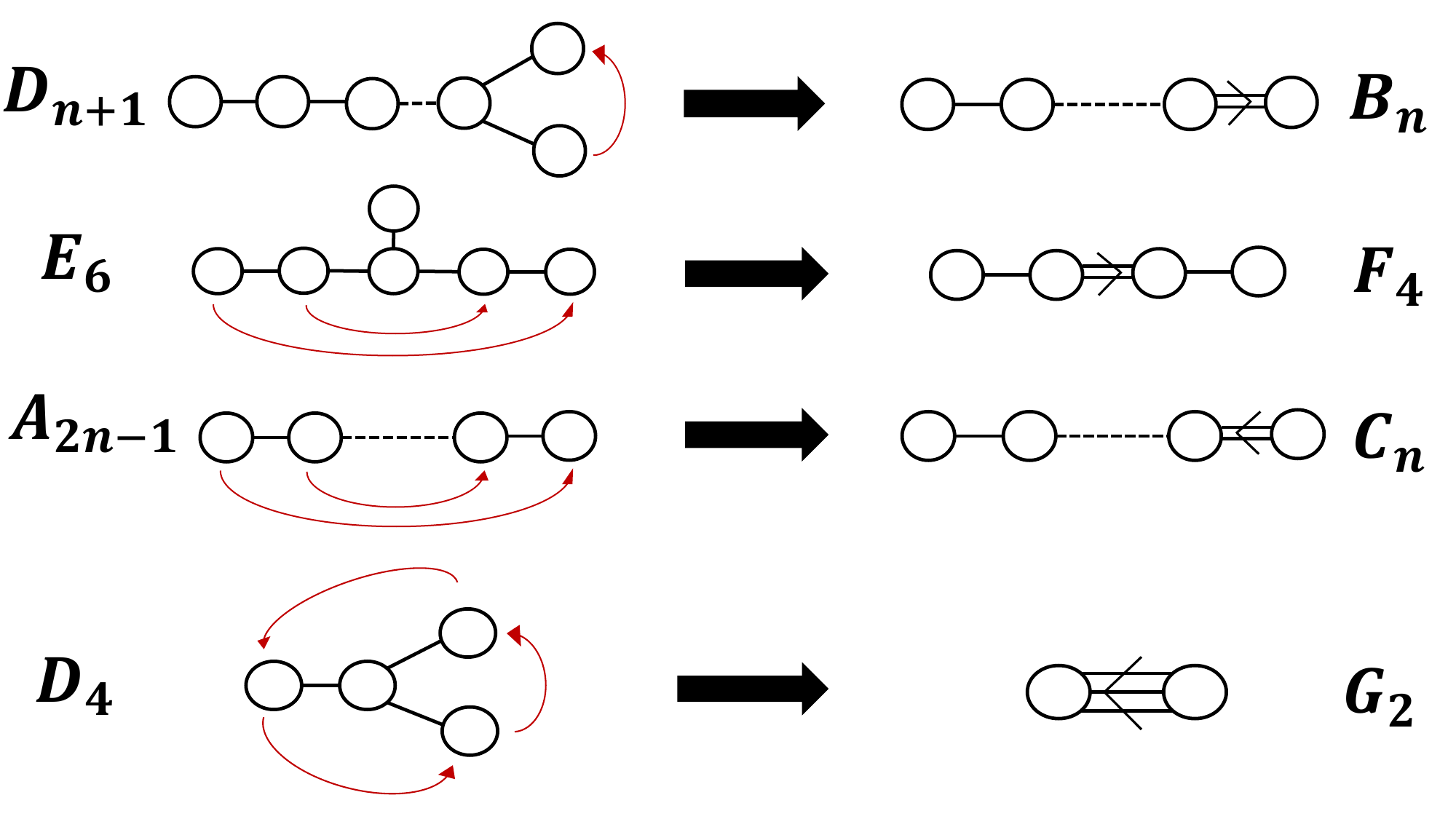}
        	\end{center}
        \caption{The action of the outer automorphism group $A$ on the simply-laced Lie algberas. In the case of $D_4$, the outer automorphism  can be either $\mathbb{Z}_2$ (resulting in $B_3$) or  $\mathbb{Z}_3$ (resulting in $G_2$).}
\label{folding}
\end{figure}

From our discussion of the McKay correspondence, it is clear how one can engineer non simply-laced theories in the little string context \cite{Aspinwall:1996nk}. Namely, consider the following nontrivial fibration of $X$ over $\mathbb{C}^2 \times \cC$: as one goes around the origin of one of the complex planes $\mathbb{C}$ wrapped by the D5 branes, we require that $X$ goes back to itself, up to the action of the group $A$. This action will permute some of the compact 2-cycles, according to Figure \ref{folding}, and there is a corresponding action on the root lattice of $\fg_0$. Let $a\in A$. If the set of simple roots of $\fg_0$ is denoted $\Delta$, then the simple roots of $\fg$ are grouped into two sets: 
\begin{align}
\Delta_l = \{\alpha \;\; / \alpha\in\Delta,\; \alpha=a(\alpha) \}
\end{align}
is the set of roots of $\fg_0$ invariant under the action of $A$. They are called the long roots of $\fg$, and we set them to have length squared $2$. The remaining simple roots of $\fg$ are constructed as follows:

\begin{align}
\mbox{If}\; A=\mathbb{Z}_2,\;\; \Delta_s &= \{\frac{1}{2}\left(\alpha+a(\alpha)\right)\qquad / \alpha\in\Delta,\; \alpha\neq a(\alpha) \}\\
\mbox{If}\; A=\mathbb{Z}_3,\;\; \Delta_s &= \{\frac{1}{3}\left(\alpha+a(\alpha)+a^2(\alpha)\right) \;\;\; / \alpha\in\Delta,\; \alpha\neq a(\alpha) \}
\end{align}
They are called the short roots of $\fg$, and have length squared $2/r$, with $r$ the lacing number of $\fg$ ($r=2$ if $A=\mathbb{Z}_2$ and $r=3$ if $A=\mathbb{Z}_3$).

Denoting the Cartan-Killing form by $\langle\cdot, \cdot\rangle$, note we have assumed that the length squared $\langle\alpha_a,\alpha_a\rangle$ of the simple root $\alpha_a$ in $\fg_0$ is equal to 2. The simple coroots of $\fg$ are defined by $\alpha_a^{\vee}=2\alpha_a/\langle\alpha_a,\alpha_a\rangle$, and the Cartan matrix of $\fg$ is $C_{ab}=\langle\alpha_a,\alpha_b^{\vee}\rangle$.

Not all D5 brane configurations as described in Section \ref{ssec:ade} will represent defects in the nontrivial fibration of $X$ over $\mathbb{C}^2 \times \cC$; only the D5 branes that wrap 2-cycles left invariant under $A$-action are allowed. This implies the following for the quiver theory $T^{5d}$ describing the D5 branes: starting with a simply-laced quiver theory, the ranks of the flavor and gauge groups which lie in a given orbit of $A$ must be equal. A non simply-laced defect is then well-defined.\\

A fundamental coweight  $w_a^{\vee}$ of $\fg$ is in fact a sum of fundamental weights of $\fg_0$, all belonging in the same $A$ orbit. So fundamental coweights are appropriate to label the D5 branes wrapping non-compact 2-cycles of the fibered geometry. They are defined by $\langle w_a^{\vee} , \alpha_b \rangle=\delta_{ab}$, with $\alpha_b$ a simple root of $\fg$, and $a, b=1, \ldots , \mbox{rank}(\fg)$. Furthermore, the simple coroots are the adequate objects to label the D5 branes wrapping compact 2-cycles of the geometry.\\

Note the fundamental coweights of $\fg$ are the fundamental weights of $^L\fg$, and the simple coroots of $\fg$ are the simple roots of $^L\fg$. We can therefore equally well label the D5 brane defects using the fundamental weights and simple roots of $^L\fg$ if we wish to do so \footnote{In particular, when $\fg$ is simply-laced, the coweight lattice (respectively coroot lattice) is the same as the weight lattice (respectively  root lattice).}.

\subsection{Coweights Description of Defects}
\label{ssec:d5defects}

Though in principle a very generic assortment of D5 branes can be studied in this setup (with the only requirement that the branes preserve the same supersymmetry), a beautiful structure emerges when one imposes a vanishing flux constraint on the set of branes. In turn, this imposes constraints on which specific non-compact 2-cycles of $X$ should be wrapped  by the D5 branes.

First, we choose a class $[S^*]$ in the coweight lattice $\Lambda_*^{\vee}$ of $\fg$. Each coweight thus specifies the charge of D5 branes wrapping non-compact 2-cycles of $X$. We can expand a given set of coweights, identified here with a non-compact homology class $[S^*]$, in terms of fundamental coweights:
\beq\label{ncomp}
[S^*] = -\sum_{a=1}^{n} \, m_a \, w_a^{\vee} \;\;  \in\, \Lambda_*^{\vee}\; ,
\eeq
with  $m_a$ non-negative integers and $n=\mbox{rank}(\fg)$. The $w_a^{\vee}$ are the $n$ fundamental coweights of $\fg$. Each fundamental coweight  is conveniently written with Dynkin labels as a vector of size $n$, with a 1 in the $a$-th entry and 0 everywhere else. For instance, $w_2^\vee=[\phantom{-}0, \phantom{-}1, \phantom{-}0, \phantom{-}0, \ldots, \phantom{-}0]$. In what follows, all coweights will be written in this fundamental coweight basis.

To get the brane flux to vanish, we then need to add some D5 branes wrapping a compact homology class $[S]$ in the coroot lattice of $\fg$;  we have the following expansion in terms of simple positive coroots:
\beq\label{comp}
[S] = \sum_{a=1}^n  \,d_a\,\alpha_a^{\vee}\;\;  \in  \,\Lambda^{\vee} \; ,
\eeq
with $d_a$ non-negative integers. The vanishing flux constraint takes the form:
\beq\label{conf}
[S+S^*] =0 \, .
\eeq
Now, if $S+S^*$ vanishes in homology, then $\# (S_a \cap (S+S_*))$ also vanishes, for all $a=1, \, \ldots \, , n$.  After a little algebra, we can rewrite \eqref{conf} as the constraint
\beq\label{conformal}
\sum_{b=1}^n C_{ab} \;d_b = m_a\; ,
\eeq
with $C_{ab}$  the Cartan matrix of $\fg$.  Note that in a 4d $\cN=2$ context, this constraint is familiar as a conformality condition.\\

Alternatively, the D5 branes can be made to wrap non-compact 2-cycles exclusively. Their charge is then encoded in various coweights, all taken in fundamental representations of $^L\fg$. To achieve this, we reshuffle the branes and arrange them in a configuration wrapping a set of non-compact cycles $S_i^*$; their homology classes $\omega_i$ now live in the coweight lattice $\Lambda_*^\vee$:
\beq\label{weightsfr}
\omega_i  = [S_i^*] \qquad \in\, \Lambda_*^\vee.
\eeq 
We assemble the coweights in a set,
\begin{equation}
\label{WS}
{\cW}_{\cS} = \{ \omega_i\}\;,
\end{equation}
whose size is the rank of the flavor group of $T^{5d}$:  $\left|{\cW}_{\cS}\right|=\sum_{a=1}^n m_a$. 
Explicitly, each coweight can be decomposed as
\begin{align}
\omega_i=-w_a^\vee+\sum_{b=1}^n h_{ib}\, \alpha_b^\vee,
\end{align}
where $-w_a^\vee$ is the negative of the $a$-th fundamental coweight, $h_{ib}$ are non-negative integers, and $\alpha_b^\vee$ is a positive simple coroot. 

The constraint \eqref{conf} (or \eqref{conformal}) is then equvalent to imposing
\begin{equation}
\label{sumweights}
\sum_{\omega_i\in{\cW}_{\cS}}\omega_i=0\; .
\end{equation}

In this paper, we will limit our analysis to sets of size 
\[
1\leq |{\cW}_{\cS}|\leq n+1 \; ,
\]
since as we will later see, the most generic defect of the $\fg$-type little string can always be described by at most $n+1$ coweights satisfying equation \eqref{sumweights}. A given coweight is allowed to appear with multiplicity higher than one in the set, and null coweights are allowed, as long as they are present in a fundamental representation of $^L\fg$. In particular, note that if $|{\cW}_{\cS}|=1$, then the set is necessarily made up of a single null coweight, by our constraint.\\

In conclusion, by choosing distinct sets of coweights ${\cW}_{\cS}$, we get an explicit realization of all the defects of the little string satisfying \eqref{conformal}. However, it would be nice to have a finer classification of the defects. It turns out that there is an elegant answer to this problem, which was analyzed when $\fg=ADE$ in  \cite{Haouzi:2016ohr,Haouzi:2016yyg}; we now extend the analysis to the case where $\fg$ is an arbitrary simple Lie algebra.

\subsection{Polarized and Unpolarized Defects}
\label{ssec:polar}

D5 brane defects are divided into two groups,  as follows:
Consider the set ${\cW}_{\cS}$ of coweights in a representation of $^L\fg$ generated by (minus) some fundamental coweight $-w^\vee_a$ for some $a$. If there exists a coweight $\omega\in {\cW}_{\cS}$ which is \emph{not} in the Weyl group orbit of $-w^\vee_a$, then we will say that ${\cW}_{\cS}$ makes up an \emph{unpolarized} defect of the little string. Otherwise, we call the defect polarized\footnote{The terminology will be explained in Section \ref{sec:CFTLIMIT}, and is directly related to the definition of the parabolic subalgebras of $\fg$.}. To fully characterize an unpolarized defect, it is then necessary (and sufficient) to further specify the representation $\omega$  belongs in.\\ 

\begin{example}[Polarized and Unpolarized Defects] 
-- Consider the following set of coweights of $G_2$:
$$\cW_\cS=\{\omega_1=[\phantom{-}0, \phantom{-}1], \;\;\; \omega_2=[\phantom{-}0, -1]\}\; ,$$
written here in the fundamental coweights basis. One can check at once that both coweights satisfy the condition to make $\cW_{\cS}$ a polarized defect.\\

-- Consider now the following set with a single coweight of $F_4$:
$$\cW_\cS=\{\omega=[\phantom{-}0, \phantom{-}0, \phantom{-}0, \phantom{-}0]_1\} \; .$$
This is an unploarized defect of the $F_4$ little string theory, since it cannot possibly be in the Weyl group orbit or any nonzero coweight.
Note that the null coweight is present in all four of the fundamental representations of $F_4$, and each one of these designates a distinct defect, so we added an extra label to specify which null coweight we are considering. In the present case, $[\phantom{-}0, \phantom{-}0, \phantom{-}0, \phantom{-}0]_1$ means that $\omega=-w_1^\vee+\ldots$ , where the dots ``$\ldots$" stand for a sum of simple positive coroots.\\

-- As a final example, consider the following set of (co)weights of $D_5$:
$$\cW_\cS=\{\omega_1=[\phantom{-}1, \phantom{-}0, \phantom{-}0,\phantom{-}0, \phantom{-}0]_1, \;\;\; \omega_2=[-1, \phantom{-}0, \phantom{-}0,\phantom{-}0, \phantom{-}0]_3\}\; ,$$
The weight $\omega_2$ is taken in (minus) the third fundamental representation:  $\omega_2=-w_3+\ldots$, hence the extra label ``3" (the dots once again stand for a sum of simple positive (co)roots). However, note that $\omega_2$ is in the Weyl group orbit of $-w_1$. The set $\cW_\cS$  therefore contains at least one coweight which satisfies the unpolarized condition, and we call the resulting defect as a whole unpolarized.
\end{example}

When $\fg$ is non simply-laced, we name the effective description of the defect a non simply-laced quiver gauge theory, in the sense that it is obtained from the folding of a $\fg_0$-type simply-laced quiver gauge theory, following the discussion of section \ref{ssec:bcfg}. In particular, the $\fg$-type Coulomb branch can always be understood from folding a $\fg_0$-type one. Its dimension is defined as $\sum_{a=1}^n d_a$, where the integers $d_a$ were defined in \eqref{comp} and determined from \eqref{conformal}. They are the ranks of the unitary gauge groups\footnote{Technically, a $U(1)$ in each node of the quiver is nondynamical in 5d, so one should really subtract $n$ to this sum to get the number of normalizable Coulomb moduli. We will keep this subtlety in mind but it will not affect our results.} in the quiver $T^{5d}$.

When a given set of coweights $\cW_{\cS}$ defines a polarized defect, there is yet another way to identify the dimension of $T^{5d}$'s Coulomb branch, since then a direct computation shows that:
\begin{equation}
\label{count}
\sum_{a=1}^n d_a=\sum\limits_{\langle e_\gamma,\omega_i\rangle<0} \left\vert\langle e_\gamma,\omega_i\rangle\right\vert \; .
\end{equation}
The sum on the right-hand side runs over all positive roots $e_\gamma$ of $\fg$, and the coweights $\omega_i$ to be included in the sum must satisfy $\langle e_\gamma,\omega_i\rangle<0$.

If $\cW_{\cS}$ defines an unpolarized defect, the left-hand side of \eqref{count} is still a valid way to evaluate its Coulomb branch dimension, but the right-hand side is no longer applicable. We will have more to say about these defects after explaining the physics of triality.\\

We want to stress that many distinct sets $\cW_{\cS}$ often result in one and the same quiver theory $T^{5d}$ at low energies; in other words, the map from $\cW_{\cS}$ to $T^{5d}$ is not one-to-one, and therefore quivers do not provide an adequate definition of little string defects. Crucially, the coweights of $\cW_{\cS}$ tell us which precise 2-cycles are wrapped by which D5 branes, and this data is lost in the quiver description of $T^{5d}$.

\section{$\fg$-type Triality}
\label{sec:triality}

A triality between $T^{5d}$ and Toda theory was first discussed in in the case $\fg=A_{1}$ \cite{Aganagic:2013tta}, before being extended to $\fg=A_{n}$ \cite{Aganagic:2014oia} and later  $\fg=D_n, E_n$  \cite{Aganagic:2015cta}. It states the equivalence of the instanton partition function of the 5d $ADE$ quiver gauge theory $T^{5d}$ on ${\mathbb R}^{4}\times S^{1}(R)$ with the partition function of its vortices in 3d at a point of its moduli space where the Coulomb branch is integer points away from the root of the Higgs branch (the locus where the Coulomb and Higgs branches meet). Furthermore, the 3d vortex partition function is nothing but the Coulomb gas representation of the deformed $ADE$ Toda conformal blocks, with ${\cal W}_{q,t}(\fg)$ algebra symmetry.

The triality can in fact be derived from the study of the $(2,0)$ $\fg=ADE$ little string on $\cC$ with polarized brane defects. This is because the dynamics of the little string in the presence of D5 brane defects localize to the defects themselves. In particular, the little string partition function with D5 brane defects is the partition function of the 5d $ADE$ quiver $T^{5d}$.   Our goal is now to extend the triality to include non simply-laced Lie algebras, in a general framework that applies to defects labeled by an arbitrary simple Lie algebra $\fg$. To this end, we need to compute the instanton partition function of folded quivers.

\subsection{5d Gauge Theory Partition Function}
We compute the partition function of the $(2,0)$ little string theory on ${\cal C} \times {\mathbb C}^2$ with a collection of defects at points of ${\cal C}$.  As we argued above, this is the partition function of a 5d quiver theory on a circle with twisted boundary conditions, $Z_{5d}\,({\mathbb C}^2\times S^1(R))$ \cite{Moore:1997dj,Nekrasov:2002qd}.  As we go around the circle $S^{1}(R)$ of the (T-dual) cylinder we rotate different ${\mathbb C}$'s by different angles, $\epsilon_{1}$ and $\epsilon_{2}$:
\begin{align}
z_{1}\mapsto e^{i\, R\, \epsilon_{1}}z_{1}\equiv q\,z_{1}, \qquad \,z_{2}\mapsto e^{i\, R\, \epsilon_{2}}z_{2}\equiv t^{-1}\,z_{2}.
\end{align}
The partition function for 5d ADE type quiver gauge theories compactified on a circle was computed in \cite{Nekrasov:2012xe}. For the simply-laced quivers all nodes in the quiver designate simple roots that are on an equal footing. However, if the quiver is given by a non simply-laced Lie algebra $\fg$, the nodes label either  short or long roots of $\fg$. In \cite{Kimura:2017hez}, the partition function for quivers that have an arbitrary lacing number is computed  using  equivariant localization. Such  quivers are called fractional, and non simply-laced Dynkin diagrams  fall into this category. An integer $r_{a}$ is assigned to each node $a$ to distinguish its relative length squared from the other nodes'. In particular, the partition function will reduce to the simply-laced case when all $r_{a}$'s are equal to one. There, it was argued that the action of only one of the rotation generators is modified to account for the contribution of a given node $a$: 
\begin{align}
z_{1}\mapsto e^{i\, R\, r_{a}\,\epsilon_{1}}z_{1}\equiv q^{r_{a}}\,z_{1}, \qquad z_{2}\mapsto e^{i\, R\, \epsilon_{2}}z_{2}\equiv t^{-1}\,z_{2}.
\end{align}
Note this prescription is in full agreement with the string theory picture of section \ref{ssec:bcfg}.\\

The partition function is an index
$$
Z_{5d}\,({\mathbb C}^2\times S^1(R))  = {\rm tr}\, (-1)^F \; g \;\; ,
$$
where $g=q^{r_{a}(S_1 - S_R)} t^{-S_2 + S_R}$; $S_1$ and $S_2$ are the generators of the two rotations around the complex planes in ${\mathbb C}^2$ defined above.  $F$ is the fermion number. Finally, $S_R$ is the generator of the $U(1)_R\subset  SU(2)_R$ charge of the R-symmetry\footnote{When considering type IIB string theory on the surface $X$, an $SO(5)_R$ R-symmetry is preserved. The D5 branes will only preserve an $SU(2)_R$ subgroup of this R-symmetry, and only a $U(1)_R$ subset is relevant here.}. We twist by this R-symmetry to preserve supersymmetry.

This index can be computed using equivariant integration, and written as a sum over fixed point contributions on the instanton moduli space labeled by Young diagrams:
\beq\label{bN}
{Z}_{5d} = r_{5d} \;\sum_{\{\mu\}}I_{5d,\{\mu\}}(q,t; a, m, \tau).
\eeq
The normalization factor $r_{5d}$ contains the tree level and the one loop contributions to the partition function. We have used the following shorthand notation for Young diagrams to express the equivariant fixed points of the relevant torus action:
\begin{align}
\{\mu\} = \{\mu^{a}_{I,i}\}_{ a=1, \ldots n;\;\; I=1, \ldots d_{a};\;\; i=1,\ldots,\infty},
\end{align}
where the number of nodes in the quiver is given by $n$, and rank of the $a$-th node is given by $d_{a}$. The variable $i$ denotes the $i$-th row of the Young diagram. Note that only a finite number of rows is nonzero in a given diagram; we let $i$ tun ro infinity, with the understanding that $\mu^{a}_{I,i}$ vanishes when $i$ becomes greater than the number of rows. Sometimes we prefer to suppress one or both subscrpits to avoid cumbersome notation and hope that our notation will be clear from the context. The gauge theory partition function will depend on more parameters than just $q$ and $t$; as we reviewed in Section \ref{ssec:ade}, there are gauge couplings $\tau_a$, which come from certain moduli of the $(2,0)$ theory in six dimensions; there are also fundamental hypermultiplets  masses, which originate from the positions of non-compact D5 branes on ${\cal C}$, and Coulomb moduli, which are the positions of the compact D5 branes on ${\cal C}$.

The contributions of the different ``multiplets" at a node $a$ will depend on the integer $r_a=1,2, \,\mbox{or} \; 3$. The quotation marks here stand for the abuse of notation we use throughout our discussion when $\fg$ is non simply-laced: when we speak of a $U(d_a)$ gauge node in a $BCFG$ quiver, we really mean the folding of $r_a$ gauge nodes in the associated $ADE$ quiver gauge theory, as defined in the previous sections. Moreover, $a$ labels a short root $\fg$, the various fields in the node $a$ multiplets should be understood as the sum of $r_a$ fields in the corresponding multiplets of the simply-laced theory; meanwhile, is $a$ labels a long root of $\fg$, the fields in the node $a$ multiplets are the same as in the simply-laced theory, with $U(d_a)$ an honest gauge node, since a long root is by definition invariant under the outer automorphism action.

With this caveat understood, the fixed point contributions $I_{5d,\{\mu\}}$ have the following generic form:

\beq\label{5dp}
I_{5d, \{ \mu\}}= \; e^{\tau \cdot \mu} \;\; \cdot \prod_{a=1}^n z^{5d}_{V_a, {\vec \mu}^{a}} \; z^{5d}_{H_a, {\vec \mu}^{a}}\; z^{5d}_{CS, {\vec \mu}^{a}} \; \cdot \prod^n_{a,b=1}
z^{5d}_{H_{ab}, {\vec \mu}^{a},  {\vec \mu}^{b}},
\eeq
where $z^{5d}_{V_a, {\vec \mu}^{a}}$ and $z^{5d}_{H_a, {\vec \mu}^{a}}$ are the contributions of the vector and hyper multiplets at node $a$, respectively. $z_{CS, {\vec \mu}^{a}}$ stands for the topological Chern-Simons factors. We also have bifundamental matter multiplets charged under two distinct nodes, say $a$ and $b$, and we label them with $z^{5d}_{H_{ab}, {\vec \mu}^{a},  {\vec \mu}^{b}}$. We assume that $z^{5d}_{H_{ab}, {\vec \mu}^{a},  {\vec \mu}^{b}}$ is $1$ if there is no bifundamental hypermultiplet between nodes $a$ and $b$.

The various equivariant fixed point contributions for all  multiplets can be written in terms of a single function, which we dub the Nekrasov function\footnote{We suppress the explicit dependence on $r_{a}$ and $r_{b}$ to avoid clutter in our notation, and refer only to $r_{ab}$ since it determines the type of the $q$-Pochhammer symbol. Moreover, $r_{ab}$ is generic in the definition of the Nekrasov function, but will be specialized when we introduce  the bifundamental contributions.}:
\beq\label{nekrasovN}
N_{\mu^{a}\mu^{b}}(Q;q^{r_{ab}}) = \prod\limits_{i,j = 1}^{\infty} 
\dfrac{\big( Q q^{r_{a}\mu^{a}_i-r_{b}\mu^{b}_j} t^{j - i + 1};q^{r_{ab}} \big)_{\infty}}{\big( Q q^{r_{a}\mu^{a}_i-r_{b}\mu^{b}_j} t^{j - i} ;q^{r_{ab}}\big)_{\infty}} \
\dfrac{\big( Q t^{j - i};q^{r_{ab}} \big)_{\infty}}{\big( Q t^{j - i + 1} ;q^{r_{ab}}\big)_{\infty}}.
\eeq
where $r_{ab}$ is a positive integer divisor of $r_{a}$ and $r_{b}$ for now, and $(x;q)_{\infty}=\prod_{i=0}^{\infty}(1-x\,q^{i})$ is the $q$-Pochhammer symbol. Let us summarize the contributions from the different multiplets. At each node $a$, we have a node of rank $d_a$ with  vector multiplet contribution
\begin{align}
z^{5d}_{V_a, {\vec \mu}^{a}}= \prod_{1\leq I,J\leq d_{a}}[N_{\mu^{a}_{I}\mu^{a}_{J}}(e_{a,I}/e_{a,J};q^{r_{a}})]^{-1}.
\end{align}
Here, the $e_{a,I} = \exp(R \,\fa_{a,I})$ encode the $d_a$ exponentiated Coulomb branch parameters at node $a$. We can also couple $m_{a}$ fundamental hypermultiplets with masses $\beta_a$'s. These contribute
\begin{align}
z^{5d}_{H_a, {\vec \mu}^{a}} = \prod_{1\leq \alpha \leq m_a} \prod_{1\leq I\leq d_a} N_{\emptyset \mu^{a}_{I}}( v_a^2\, f_{a, \alpha}/e_{a, I};q^{r_{a}}).
\end{align}

The corresponding exponentiated masses are denoted as $f_{a, \alpha}= \exp(R\, \beta_{a,\alpha})$, where $\alpha$ takes $m_a$ values, and $v_a\equiv\sqrt{q^{r_a}/t}$. Note that $\sum_{a=1}^n m_a=|\cW_{\cS}|$. For every pair of nodes $a,b$ connected by an edge in the Dynkin diagram, we get a bifundamental hypermultiplet. Its contribution to the partition function is:
\begin{align}
z^{5d}_{H_{ab}, {\vec \mu}^{a},  {\vec \mu}^{b}}= \prod_{1\leq I\leq d_{a}}\prod_{1\leq J\leq d_{b}}[N_{\mu^{a}_{I} \mu^{b}_{J}}(e_{a,I}/e_{b,J};q^{r_{ab}})]^{\Delta_{ab}}.
\end{align}
where $\Delta$ is a matrix whose entries $\Delta_{ab}$ are equal to either $1$ or $0$, depending on whether the $a$'th and the $b$'th nodes are connected or not. There is an important subtlety arising for fractional quivers: the bifundamental matter can be coupled to gauge nodes corresponding to different length roots. For those multiplets, we have $r_{ab}=\mbox{gcd}(r_{a},r_{b})$, the greatest common divisor of $r_{a}$ and $r_{b}$.  

In a 5d theory we can turn on Chern-Simons term of $k^{CS}_{a}$ units, and their contribution to node $a$ reads
\begin{align}
z^{5d}_{CS, {\vec \mu}^{a}} = \prod\limits_{1\leq I\leq d_a} \big(T_{\mu^{a}_{I}}\big)^{k^{CS}_a}
\end{align}
Here, $T_{\mu}$ is defined as $ T_\mu =(-1)^{|\mu|} q^{\Arrowvert \mu\Arrowvert^{2}/2}t^{-\Arrowvert \mu^{t}\Arrowvert^{2}/2}$. The bare Chern-Simons term in this paper is set to 0, and the above contribution is uniquely fixed by our constraint \eqref{conformal}. In particular, with the partition function as written above, $k^{CS}_{a}$ on the $a$-the node is the difference of ranks between successive nodes $d_a - d_{a+1}$.
The gauge couplings keep track of the total instanton charge, via the combination

\begin{align}
\tau \cdot \mu = \sum_{a=1}^n  \sum_{I=1}^{d_a}\;\tau_a\; |\mu^{a}_{I}|.
\end{align}

\subsection{3d Gauge Theory Partition Function}
\label{ssec:3dgaugepart}

Just like the D5 branes we studied, D3 branes of type IIB retain a finite tension in the little string limit $g_s\rightarrow 0$. Here, we wrap the D3 branes on one of the two complex planes $\mathbb{C}$, and compact 2-cycles of the resolved $ADE$ singularity $X$. As such, the D3 branes are dynamical. First, let us focus on the case $\fg_0=ADE$; the low energy theory on the D3 branes is once again an $ADE$ quiver gauge theory \cite{Douglas:1996sw}. It is a 3d theory with ${\cal N}=4$ supersymmetry on $\mathbb{C}\times S^1(R)$\footnote{Similar to the D5 branes, the D3 branes also feel the stringy effects due to the presence of the transverse circle in the cylinder ${\cal C}$ and the tower of states resulting from it. Therefore, the theory is really three-dimensional on a circle at low energies.}. The cubic superpotential is entirely fixed by the supersymmetry. Let $N_{a}$ be the number of D3 branes wrapping the $a$-th compact 2-cycle, belonging to the second homology group of $X$, isomorphic to the root lattice of $\fg_0$. Then, the 3d quiver theory has gauge content $\prod_{a=1}^n U(N_{a})$. We further obtain bifundamental matter hypermultiplets by quantizing strings streched between adjacent nodes in the associated Dynkin diagram, described by the previously defined matrix $\Delta_{ab}$.

We now consider the effects of the previously introduced D5 branes, which break supersymmetry to $\cN=2$ from the point of view of the D3 brane theory. Recall that in our presentation, the D5 branes can be made to exclusively wrap non-compact 2-cycles, and their charge is encoded in a set ${\cal W}_{S}$ of (co)weights $\omega_i$ satisfying $\sum_{\omega_i\in{\cW}_{\cS}}\omega_i=0$. This is the presentation which is useful here.
Indeed, from the 5d gauge theory standpoint, the Coulomb moduli labeled by positive simple roots $\alpha_a$ are now frozen to some mass labeled by (minus) a fundamental weight  $-w_a$, which algebraically translates to
\begin{align}
\omega_i=-w_a+\sum_{b=1}^n h_{ib}\, \alpha_b \; .
\end{align}
with $h_{ib}$ non-negative integers. Physically, this describes the root of the Higgs branch. Geometrically, the above decomposition of $\omega_i$ has the interpretation of having different compact and non-compact D5 branes bind together. In particular, the position of all the compact branes now coincide with the position of at least one of the non-compact D5 branes on $\cC$.

We need to quantize the strings strechted between D3 and D5 branes too, and those give rise to chiral and anti-chiral multiplets of ${\cal N}=2$ supersymmetry at the intersection points of compact 2-cycles wrapped by D3 branes and non-compact 2-cycles wrapped by the D5 branes. This matter content makes the 3d theory a handsaw quiver variety \cite{Aganagic:2015cta}, where the details of the ``teeth" are uniquely determined by the set ${\cal W}_{S}$.

From the D5 brane point of view, the D3 branes are codimension 2 and realize vortices. Their charge gives the magnetic flux in the remaining directions transverse to D3 branes. For an arbitrary collection of vortices to be BPS, the FI parameters which are the moduli of little strings need to be aligned at each node of the 5d quiver theory, which is the case here. The chiral multiplets coming from D3-D5 strings get expectation values due to non-zero 3d FI parameters in the supersymmetric vacua. Turning on $N_a$ units of lifts us off the root of the Higgs branch, back on the Coulomb branch, but on an integer lattice of spacing proportional to $N_a$.\\

If we perform the folding operation of section \ref{ssec:bcfg}, we can also make sense of a D3 brane theory which at low energies is described by a non simply-laced quiver gauge theory. Let us write the corresponding $BCFG$ algebra as $\fg$. The above discussion still applies, with the caveat that roots should now be understood as coroots and weights as coweights. More importantly, there is an important subtlety: there is no notion of Higgs branch for the 5d $\fg$-type quiver gauge theory, so it is unclear what it means to study its vortices. In particular, freezing the 5d Coulomb moduli to some masses should no longer describe the root of a Higgs branch. Nevertheless, the procedure is at least formally algebraically sound, and we will see a fortiori it is the correct picture to make contact with $BCFG$-type Toda. We then \emph{define} vortices in this case as the vortices of the original simply-laced theory $\fg_0$, followed by the folding of the 3d $\fg_0$-quiver theory by the relevant outer automorphism group action. By abuse of notation, we call the resulting 3d theory a vortex theory of non simply-laced type, though we stress once again that the 5d $\fg$-type theory has no well-defined Higgs branch to start with.\\

Then, for a general simple Lie algebra $\fg$, let us call the low energy effective 3d $\cN=2$ $\fg$-type quiver gauge theory $G^{3d}$.

Once subjected to the $\Omega$-background, we can compute the partition function of $G^{3d}$ using localization \cite{Shadchin:2006yz,Hama:2011ea,Kapustin:2011jm,Beem:2012mb}. The equivariant action that we used to compute the 5d partition function can be used for the 3d one too. We choose the D3 brane to extend on the plane rotated by the parameter $q$, and to be transverse to the plane rotated by $t$.

The partition function is again an index:
\begin{align}
Z_{3d}(\mathbb{C}_q\times S^{1}(R))={\rm tr}\, (-1)^{F}\,g \; ,
\end{align}
where $g=q^{r_{a}(S_{1}-S_{R})}t^{-S_{2}+S_{R}}$ consists of rotations $S_{1,2}$ acting on the two different planes, and $S_{R}$ is the R-symmetry rotation.  We placed the D3 branes such that $S_{2}$ acts on the transverse plane to the branes, and is therefore an R-symmetry generator from the 3d theory perspective. The theory can have at most $U(1)_R$ symmetry, so $S_{R}-S_{2}$ is a global symmetry. Localization allows us to write the 3d partition function as a sum over Young diagram just as in the case of the 5d theory. This form will be crucial to establish the connection between $T^{5d}$ and $G^{3d}$. However, there also exists an integral representation of the 3d partition function, where the integration is performed over the 3d Coulomb moduli.  Ultimately, the two representations of the partition function are related by performing the latter integration via residues. In the integral representation, the partition function reads 
\begin{align}
Z_{3d}=\int dx\,I_{3d}(x) \; ,
\end{align}
where the integrand $I_{3d}(x)$ can  easily be read off from the quiver description of the theory. It is given by the product of individual contributions coming from vector multiplets and different types of matter multiplets coupled to the gauge groups on the nodes. Generically, it has the following form, 
\begin{align}
I_{3d}(x)=r_{3d}\prod_{a=1}^{n}z^{3d}_{V_{a}}(x_{a})\,z^{3d}_{H_{a}}(x_{a},f)\prod_{a<b}z^{3d}_{H_{ab}}(x_{a},x_{b}).
\end{align}
$r_{3d}$ is a normalization factor whose precise form is not important for our purposes. The contributions of each type of multiplet is known, and we collect them here for completeness. The ${\cal N}=4$ vector multiplet for a unitary gauge group $U(N_{a})$ is given by 
\begin{align}
\label{wowvec}
z^{3d}_{V_{a}}(x_{a})=e^{\sum_{I=1}^{N_{a}}\tau_{a}x_{a,I}}\prod_{1\leq I\neq J\leq N_{a}}\frac{(e^{x_{a,I}-x_{a,J}};q^{r_{a}})_{\infty}}{(t\,e^{x_{a,I}-x_{a,J}};q^{r_{a}})_{\infty}},
\end{align}
where as before, $r_a\equiv r\,\langle\alpha_a, \alpha^\vee_a\rangle/2$ for each node $a$, with $r$ be the highest number of arrows linking two adjacent nodes in the Dynkin diagram of $\fg$  (and $\langle\alpha_a, \alpha^\vee_a\rangle=2$ for long roots, in our normalization). The numerator consist of contribution coming from the gauge bosons, and the denominator takes into account the adjoint chiral multiplets within the vector multiplet. The bifundamental hypermultiplets give a similar contribution to the 5d case,
\begin{align}
\label{wowbif}
z^{3d}_{H_{ab}}(x_{a},x_{b})=\prod_{1\leq I \leq N_{a}}\prod_{1\leq J \leq N_{b}}\left [ \frac{(v_{ab} t\, e^{x_{a,I}-x_{b,J}};q^{r_{ab}})_{\infty}}{(v_{ab} \, e^{x_{a,I}-x_{b,J}};q^{r_{ab}})_{\infty}}\right]^{\Delta_{ab}},
\end{align}
where again $\Delta$ describes how the nodes are connected to each other.  $r_{ab}$ is the greatest common divisor of $r_a$ and $r_b$ for neighboring nodes $a$ and $b$. The factor $v_{ab}$ is a modified refined factor for non simply-laced Lie algebras: $v_{ab}=\sqrt{q_{ab}/t}$ with $q_{ab}=q^{r}$ if both nodes $a$ and $b$ correspond to long roots; otherwise, $v_{ab}=v=\sqrt{q/t}$. Chiral multiplets in the fundamental representation of the $a$-th gauge group, with $S_{R}$ R-charge $-r/2$ (not to be confused with the lacing number of $\fg$), contribute $\prod_{1\leq I\leq N_{a}}(v_{a}^r\,f_{a,i}e^{-x_{a,I}};q^{r_{a}})^{-1}_{\infty}$ to the partition function, while anti-chiral multiplets contribute $\prod_{1\leq I\leq N_{a}}(v_{a}^r\,f_{a,i}e^{-x_{a,I}};q^{r_{a}})_{\infty}$, with $f_{a,i}$ the associated flavor. 

The integral runs over all the Coulomb branch moduli of the $n$ gauge groups in the quiver. To perform this integral, one needs to select a vacuum and pick a contour. We will not attempt to give a precise contour prescription from first principles in this paper, and will simply conjecture what they should be based  on the input from the 5d theory.

\subsection{$\fg$-type Toda and its deformation}
\label{ssec:Toda}
We now review a last important piece of physics that is needed to establish a triality, the Toda conformal field theory on the Riemann surface $\cC$. The partition function of the gauge theory on D3 branes presented above is in fact equal to a certain canonical ``deformation" of the  Toda CFT conformal block on ${\cal C}$. Let us first briefly review the Toda CFT, and then its deformation.

\subsubsection{Free Field Toda CFT} 

Let $\fg$ be a simple Lie algebra.
$\fg$-type Toda field theory can be written in terms of $n={\rm rk}({\fg})$ free bosons in two dimensions; there is a background charge contribution, and an exponential potential that couples the bosons to that charge:
\beq\label{Todaaction}S_{Toda} =  \int dz d{\bar z} \;\sqrt g \; g^{z{\bar z}}[\langle\partial_z \varphi,  \partial_{\bar z} \varphi\rangle+  \langle Q, \varphi\rangle\, R + \sum_{a=1}^n e^{\langle\alpha_a^{\vee},\varphi\rangle/b} ].
\eeq
The bosonic field $\varphi$ is a vector in the $n$-dimensional coweight space, whose modes obey a Heisenberg algebra. The bracket $\langle\cdot,\cdot\rangle$ is the Cartan-Killing form on the Cartan subalgebra of $\fg$, and $Q=\rho\, b+ \rho^\vee/b$ is the background charge. $\rho$ is the Weyl vector of $\fg$, and $\rho^\vee$ is the Weyl vector of $^L\fg$. As before, $\alpha_a^{\vee}$ label the simple positive coroots of $\fg$.
 
The Toda CFT has a ${\cal W}({{\fg}})$ algebra symmetry (see \cite{Bouwknegt:1992wg} for a review). When $\fg= \mbox{su(2)}$, the CFT is called Liouville theory, with Virasoro symmetry. The ${\cal W}({{\fg}})$ symmetry of Toda is generated by the spin 2 Virasoro stress energy tensor, and additional higher spin currents.

The free field formalism of the Toda CFT was first introduced in \cite{Dotsenko:1984nm}. It was then studied in our context in \cite{Dijkgraaf:2009pc,Itoyama:2009sc,Mironov:2010zs,Morozov:2010cq, Maruyoshi:2014eja}. We label the primary vertex operators of the ${\cal W}({{\fg}})$ algebra by an $n$-dimensional vector of momenta $\beta$, and given by:
\beq\label{primary}
V^{\vee}_{\beta}(z) = e^{\langle\beta, \varphi(z)\rangle}.
\eeq
The conformal blocks of the Toda CFT in free field formalism take the following form:
\beq\label{expect}
\langle V^{\vee}_{\beta_1}(z_1) \ldots V^{\vee}_{\beta_k}(z_k) \;\;\prod_{a=1}^{n} (Q^{\vee}_{a})^{N_a} \rangle_{free} \, .
\eeq
In the above, we have defined the screening charges
$$
Q^{\vee}_{a} \equiv \oint dx \,S^{\vee}_{a}(x) \, .
$$
These $n$ charges are integrals over the $n$ screening current operators $S^{\vee}_a(x)$:
\beq\label{scc}
S^{\vee}_{a}(z) = e^{ \langle\alpha^{\vee}_{a}, \phi(z)\rangle/b} \, .
\eeq
The ${\cal W}({\bf g})$ algebra can then be defined as a complete set of currents that commute with the screening charges.
For a derivation of the conformal block expression \eqref{expect}, we refer the reader to \cite{Fateev:2007ab}.

Momentum conservation imposes the following constraint:
\beq\label{constraint}
\sum_{i=1}^k \beta_i+ \sum_{a=1}^n N_a \alpha^{\vee}_{a}/b = 2Q.
\eeq 
The right-hand side comes from the background charge on a sphere, induced by the curvature term in \eqref{Todaaction}. Thus, the above constraint tells us that one of the momenta, say $\beta_{\infty}$, corresponding to a vertex operator insertion at $z={\infty}$, is fixed in terms of the momenta $\beta_i$ of  the other vertex operators, and the number of screening charges $N_a$.\\

The correlators of the theory can be computed by Wick contractions, and the conformal block \eqref{expect} takes the form of an integral over the positions $x$ of the $N_a$ screening currents:
\beq\label{conf1}
Z_{Toda} = \int dx \;I_{Toda}(x) \; .
\eeq
The integrand $I_{Toda}(x,z)$ is a product over various two-point functions:
\beq\label{cbasic}
I_{Toda}(x,z) = \prod_{a=1}^n I^{Toda}_a(x_a)\cdot I_{a, V}(x_a, z) \cdot \prod_{a<b } I^{Toda}_{ab}(x_a, x_b)
\eeq
The two-point functions of screening currents with themselves at a given node of the Dynkin diagram of $\fg$ give: 
\beq\label{cbasic1}
I^{Toda}_{a} = \prod_{1\leq I \neq J \leq N_{a}}  \langle S^\vee_a(x_{a,I}) S^\vee_a(x_{a,J}) \rangle_{free}.
\eeq
These are the vector multiplet contributions at node $a$.
The two-point functions of screening currents between two distinct nodes $a$ and $b$ is in turn given by:
\beq\label{cbasic2}
I^{Toda}_{ab} = \prod_{1\leq I\leq  N_{a}}\prod_{1\leq J\leq N_{b}}  \langle S^\vee_a(x_{a,I}) S^\vee_b(x_{b,J}) \rangle_{free}.
\eeq
These are the bifundamental hypermultiplet contributions. Finally, the two-point functions of screening currents at a given node with all the vertex operators.
\beq\label{cbasic3}
 I^{Toda}_{a, V} = \prod_{i=1}^k \prod_{1\leq I \leq N_{a}}\   \langle S^\vee_a(x_{a,I})  V^\vee_{\beta_i}(z_i) \rangle_{free},
\eeq
will correspond to chiral matter contributions. 
The two-point functions  are readily evaluated to be:
\beq\label{sct}
 \langle S_a^\vee(x) S_b^\vee(x') \rangle_{free} = (x-x')^{b^2 \langle\alpha^\vee_{a},\alpha^\vee_{b}\rangle}
\eeq
\beq\label{vt}
\langle S_a^\vee(x) V^\vee_{\beta} (z) \rangle_{free} = (x-z)^{-\langle\alpha^\vee_a,\beta\rangle}
\eeq

\subsubsection{Deformed Toda Theory}

In \cite{Frenkel:1998}, a deformation of ${\cal W}({\fg})$ algebras was given by defining screening currents depending on two ``quantum" parameters $q$ and $t$.
Starting with the definition of the quantum number
\begin{align}
[n]_q = \frac{q^{n}-q^{-n}}{q-q^{-1}} \; ,
\end{align}
and the incidence matrix $I_{ab}=2 \, \delta_{ab} - C_{ab}$, one defines the $(q,t)$-deformed Cartan matrix, $C_{ab}(q,t)= \left(q^{r_a}t^{-1} +q^{-r_a}t\right) \, \delta_{ab}- [I_{ab}]_q$. 
The number $r_a$ is defined as before: $r_a\equiv r\,\langle\alpha_a, \alpha^\vee_a\rangle/2$, with $r$ the lacing number of $\fg$.

If the Lie algebra $\fg$ is non simply-laced, its Cartan matrix $C_{ab}$ is not symmetric. Then, we first need to introduce the matrix $B_{ab}(q,t)$, which is the symmetrization of $C_{ab}(q,t)$. It is obtained as follows; the symmetrized Cartan matrix is then given by:
$$
B_{ab}=r_a \, C_{ab} \; .
$$
Its $(q,t)$-deformation is simply:
$$
B_{ab}(q,t)=[r_a]_q \, C_{ab}(q,t) \; .
$$

We are now able to construct a deformed Heisenberg algebra, generated by $n$ ``simple coroot generators" $\alpha_a$, satisfying
\begin{align}
[\alpha_a[k], \alpha_b[m]] = {1\over k} (q^{k\over 2} - q^{-{k\over 2}})(t^{{k\over 2} }-t^{-{k\over 2} })B_{ab}(q^{k\over 2} , t^{k\over 2} ) \delta_{k, -m} \; .
\end{align}

The Fock space representation of the Heisenberg algebra is given by acting on a vacuum state $|\lambda\rangle$ with the simple coroot generators:
\begin{align}\nonumber
\alpha_a[0] |\lambda\rangle &= \langle\lambda, \alpha_a\rangle |\lambda\rangle\\
\alpha_a[k] |\lambda\rangle &= 0\, , \qquad\qquad\;\; \mbox{for} \; k>0.
\end{align}

From these generators, one can define the (magnetic) screening charge operators: 
\beq\label{screendef}
S^\vee_a(x) = x^{-\alpha_a[0]/r_a}\,: \exp\Bigl(\sum_{k\neq 0}{ \alpha_a[k] \over q^{k\, r_a\over 2} - q^{-\,{k \, r_a \over 2}}} \, e^{kx}\Bigr): \; .
\eeq
The ${\cal W}_{q,t}({\bf g})$ algebra is then defined as a complete set of operators which commute with the screening charges\footnote{One can also define a set of ``electric" screenings  \cite{Frenkel:1998}, in the parameter $t$ instead of $q$, but they will not be needed here.}.
Next, one introduces ``fundamental coweight generators" $w_a[m]$, through the commutation relation:
\begin{align}
[\alpha_a[k], w_b[m]] ={1\over k} (q^{k \, r_a\over 2}  - q^{-{k \, r_a\over 2} })(t^{{k \over 2}}-t^{-{k \over 2} })\,\delta_{ab}\,\delta_{k, -m} \, ,
\end{align}
such that
\beq\label{etow}
\alpha_a[k] = \sum_{b=1}^n C_{ab}(q^{k},t^k) w_b[k]\; .
\eeq
Correspondingly, we define (magnetic) degenerate vertex operators:
\beq\label{vertexdef}
V^\vee_a(x) = x^{w_a[0]/r_a}\,: \exp\Bigl(-\, \sum_{k\neq 0}{ w_a[k] \over q^{k\, r_a\over 2} - q^{-\,{k \, r_a \over 2}}} \, e^{kx}\Bigr): \; .
\eeq

Using the notation  $\langle \mathellipsis\rangle$ for a vacuum expectation value, and making use of the theta function definition $\theta_{q^{r_{a}}}(x)= (x \,;\, q^{r_{a}})_\infty\,(q^{r_{a}}/x \,;\, q^{r_{a}})_\infty$, we obtain the following two-point functions:

For a given node $a$,
\beq\label{sctda}
 \langle S^\vee_a(x)\, S^\vee_a(x') \rangle_{free} = {( e^{x-x'}; q^{r_{a}})_\infty \over (   t \,e^{x -x'}; q^{r_{a}})_\infty}  \; {( e^{x'-x}; q^{r_{a}})_\infty \over (   t \,e^{x'-x}; q^{r_{a}})_\infty}\; { \theta_{q^{r_{a}}}(t \, e^{x-x'})\over \theta_{q^{r_{a}}}(e^{x-x'})}\; .
\eeq
When $a$ and $b$ are distinct nodes connected by a link,
\beq\label{sctdab}
 \langle S^\vee_a(x)\, S^\vee_b(x') \rangle_{free} = {( t\, v_{ab} \, e^{x-x'}; q^{r_{ab}})_\infty\over (   v_{ab} \,e^{x -x'}; q^{r_{ab}})_\infty} \; .
 \eeq

The  two-point of a screening with a ``fundamental" vector operator is given by:
\beq\label{sctdb}
 \langle S^\vee_a(x)\, V^\vee_{b}(x') \rangle_{free} = {( t\, v_a \, e^{x'-x}; q^{r_{a}})_\infty\over (   v_a \,e^{x' -x}; q^{r_{a}})_\infty} \; .
 \eeq 
 
In the above, we have $v_{a}\equiv\sqrt{q^{r_{a}}/t}$ and $v_{ab}\equiv\sqrt{q^{r_{ab}}/t}$. Recall that if either node $a$ or node $b$ denotes a short root, then $r_{ab}=1$, while both  nodes denote long roots, then $r_{ab}=r$.

The vertex operators that are relevant to us are not exactly the operators $V_{a}(x')$ introduced above in \eqref{vertexdef}. Rather, each vertex operator,  labeled as $V^\vee_{\omega_i}(x_i)$,  is a normal ordered product of rescaled  ``fundamental coweight operators",  
\beq\label{vvertexdef}
W_a(x) = : \exp\Bigl(\, \sum_{k\neq 0}{ w_a[k] \over (q^{k\, r_a\over 2} - q^{-\,{k \, r_a \over 2}})(t^{k\over 2} - t^{-\,{k \over 2}})} \, e^{kx}\Bigr): \; ,
\eeq
and rescaled ``simple coroot operators", 
\beq\label{sscreendef}
E_a(x) = : \exp\Bigl(\sum_{k\neq 0}{ \alpha_a[k] \over (q^{k\, r_a\over 2} - q^{-\,{k \, r_a \over 2}})(t^{k\over 2} - t^{-\,{k \over 2}})} \, e^{kx}\Bigr): \; ,
\eeq
where we dropped the zero mode contributions in the above definitions, since we will not need them in what follows\footnote{More precisely, they correspond to a redefinition of the vacuum in the 3d gauge theory. We will not keep track of such a shift of the vacuum definition in our discussion.}.
The fundamental vertex operators ${W_{a}}^{\pm 1}(f\, v_a^r \, x)$ have two point functions with the screening currents $S^\vee_a(x')$ that are equal to the contributions of either chiral or anti-chiral multiplets of R-charge $-r/2$ as described in Section \ref{ssec:3dgaugepart}.\\

We now consider a set $\cW_S$ of coweights $\omega_i$ in the coweight space of $\fg$, taken in fundamental representations of $^L\fg$ and satisfying $\sum_{i=1}^{|\cW_S|} \omega_i=0$; to this set $\cW_S$,  we associate a primary vertex operator:
\beq\label{cv}
:\prod_{i=1}^{|\cW_S|} V^\vee_{\omega_i}(x_i): \; ,
\eeq
where each $V^\vee_{\omega_i}(x_i)$ is constructed out of the ``fundamental weight" and ``simple root" vertex operators.

To fully specify the conformal block, we also need to make a choice of contour in \eqref{conf1}. In particular, it is worth noting that for a given theory, the number of contours generically increases after $q$-deformation, when ${\bf g}\neq A_n$. This is because the number of contours in the undeformed case is equal to the number of solutions to certain hypergeometric equations satisfied by the conformal blocks, while the number of countours in the $q$-deformed theory is instead the number of solutions to $q$-hypergeometric equations, which is generically bigger.
Giving a prescription for the integration contours when ${\bf g}\neq A_n$   is an open problem in the matrix model community, and we will not address this question here.\\

Recovering the undeformed theory is straightforward: we  let
$q= \exp(R \epsilon_1),\, t=\exp(-R \epsilon_2)$, and take the $R$ to zero limit.  In this limit, $q$ and $t$ tend to $1$. The individual $V^\vee_{\omega_i}(x_i)$ do not have a good conformal limit, but the products do:
$$
:\prod_{i=1}^{|\cW_S|} V^\vee_{\omega_i}(x_i): \qquad \rightarrow \qquad V^\vee_{\beta}(z).
$$
The momentum $\beta$ carried by $V^\vee_{\beta}(z)$ is:
\begin{equation}
\label{momentumpunc}
\beta = \sum_{i=1}^{|\cW_S|} \beta_i \,\omega_i\; .
\end{equation}
Then, we set the argument of the vertex operators to be:
\beq\label{cftposm}
e^{x_{i}} = z \, q^{-\beta_{i}} \; .
\eeq
Then the correlator
\beq\label{this}
\langle S^\vee_a(x)\, :\prod_{i=1}^{|\cW_S|} V^\vee_{\omega_i}(x_i): \rangle_{free}
\eeq
becomes the undeformed two-point  \eqref{vt} of the vertex operator with the $a$-th screening current: $(1- e^x/z)^{-\langle\alpha_a^\vee, \beta\rangle}$, with$\beta$ defined above. In this way, one is able to realize the insertion of any number of primary vertex operators, and have complete control over how the insertion scales in the undeformed limit. Any collection of primary vertex operators with either arbitrary or (partially) degenerate momenta can be analyzed in this way.

\subsection{Proof of Triality}
\label{ssec:Proof}

In this section, we give the proof of triality for $\fg$ a simple Lie algebra. 
The proof can be divided into two parts. First, we show that the 5d theory partition function reduces to the 3d vortex partition function at a special place in the moduli space. Second, we show that the integral representation of the 3d partition function is nothing but the Coulomb gas representation of the conformal blocks in deformed Toda theory.

\subsubsection{3d-5d Partition functions}
\par{For the first part of the proof, the integral representation of the 3d partition function is not very useful. Instead, we would like to explicitly perform the integrals. Once the appropriate contour is chosen, the contributing poles turn out to be labeled by Young diagrams. Therefore, the 3d partition function can also be expressed as a sum over Young diagrams:}

\begin{align}
Z_{3d}=\int dx\, I_{3d}(x)=\sum_{\{ \mu \}}\mbox{res}_{\{ \mu \}}\, I_{3d}(x).
\end{align}
The summand can be easiest computed after normalizing it by the residue of the pole at $\{\varnothing\}$:

\begin{align}
\label{3dres}
\mbox{res}_{\{ \mu \}}I_{3d}(x)/\mbox{res}_{\{ \varnothing\}}I_{3d}(x)=I_{3d}(x_{\{\mu \}})/I_{3d}(x_{\{\varnothing \}}),
\end{align}
where $x_{\{\mu \}}$ denote $\mu$ dependent substitution for the Coulomb branch parameters:
\begin{align}
\{e^{x_{\mu}} \}=\{e_{a,I}\,q^{\mu^{a}_{I,i}}t^{\rho_{i}}\,v^{\#_{a,I,i}}q^{\#'_{a,I,i}} \},
\end{align}
where $\#_{a,I,i}$ and $\#'_{a,I,i}$ are integers, which we will determine explicitly. The integers $\#'_{a,I,i}$ are only nonzero if $\fg$ is non simply-laced.

To be explicit, let us start in 5d. We derive the partition function of $G^{3d}$ from the the one of $T^{5d}$ by moving to a point in the moduli space of the 5d theory where the Coulomb moduli are frozen. To this end, we tune the Coulomb branch parameters to equate some of the masses of the hypermultiplets:
\begin{align}\label{specialize}
e_{a,I}=f_{i}\,t^{N_{a,I}}v^{\#_{a,I,i}}q^{\#'_{a,I,i}}.
\end{align}
Here, we have introduced  $N_{a,I}$ positive integers, which are  units of vortex flux. Effectively, then, one can get back on the Coulomb branch of $T^{5d}$, but only on an integer-valued lattice. The above effect of turning on vortex flux as a shift of the Coulomb moduli by units of $t$ is as expected in the Omega-background \cite{Nekrasov:2010ka}.

Now, recall that the 5d partition function is a sum over Young diagrams. Let us assume that one of the diagrams, say $\mu$, labeling the generalized Nekrasov factor $N_{\mu\nu}(Q;q^{r_{\mu\nu}})$, has at most $N$ rows. If we set $Q=q^{r_{\nu}}t^{-(M+1)}$,  one can then show that the Nekrasov function vanishes unless the length of $\nu$ is bounded by $N+M$, i.e. $\ell(\nu)\leq N+M$. Furthermore, we make use of the following identity, which results from the properties of the $q$-Pochhammer symbol:
\begin{align}
N_{\mu\nu}(Q;q)=\prod_{a=0}^{r_{\mu\nu}-1}N_{\mu\nu}(q^{a}Q;q^{r_{\mu\nu}})
\end{align}

We previously mentioned that the fixed points of the equivariant action used in computing 5d instantons are labeled by Young diagrams, and these Young diagrams are allowed to be of any size. At this point \eqref{specialize} of the moduli space, it is not hard to show that the only non-zero contributions to the partition function come from Young diagrams which have less than or equal to $N_{a}$ rows. We can find a truncation pattern, and easily deduce that each Young diagram is limited in length by an integer. This truncation behavior can be checked directly by studying the generalized Nekrasov functions $N_{\mu\nu}(Q;q^{r_{\mu\nu}})$. 

Once we know that the Young diagrams $\mu$ and $\nu$ are truncated such that $\ell(\mu)\leq N_{\mu}$ and $\ell(\nu)\leq N_{\nu}$, it is easy to show the Nekrasov function can be rewritten as

\begin{align}
\nonumber
N_{\mu\nu}(Q;q^{r_{\mu\nu}})&=\prod_{i=1}^{N_{\mu}}\prod_{j=1}^{N_{\nu}}\frac{(Q\,q^{r_{\mu}\mu_{i}-r_{\nu}\nu_{j}}t^{j-i+1};q^{r_{\mu\nu}})_{\infty}(Q\,t^{j-i};q^{r_{\mu\nu}})_{\infty}}{(Q\,q^{r_{\mu}\mu_{i}-r_{\nu}\nu_{j}}t^{j-i};q^{r})_{\infty}(Q\,t^{j-i+1};q^{r})_{\infty}}\\
&\times N_{\mu\varnothing}(Q\, t^{N_{\nu}};q^{r_{\mu\nu}})\,N_{\varnothing\nu}(Q\,t^{-N_{\mu}},q^{r_{\mu\nu}}).
\end{align}

This identity is crucial in establishing the equivalence between the 3d and 5d partition functions. For definiteness, suppose that $T^{5d}$ is engineered from a given \textit{polarized} set  $\cW_{\cS}$ of coweights of $\fg$, in the sense of Section \ref{ssec:polar}. We then  look at the decomposition of the various  multiplet  contributions after imposing \eqref{specialize}. The 5d vector multiplets become
\begin{align}
\prod_{a=1}^{n}z^{5d}_{V_{a},\mu^{a}}=\prod_{a=1}^{n}\frac{z^{3d}_{V_{a}}(x_{\mu^{a}})}{z^{3d}_{V_{a}}(x_{\varnothing})}\cdot V_{vect},
\end{align}
where the first factor is nothing but the vector multiplet contribution for 3d theory. $V_{vect}$ is all the remaining factors from 5d vector multiplet. Similarly, we can also reduce and isolate factors from bifundamental multiplets that make up 3d bifundamental contribution and a leftover factor $V_{bifund}$:
\begin{align}
\prod_{a<b}z^{5d}_{H_{ab},\mu^{a},\mu^{b}}=\prod_{a<b}\left [\frac{z^{3d}_{H_{ab}}(x_{\mu^{a}},x_{\mu^{b}})}{z^{3d}_{H_{ab}}(x_{\varnothing},x_{\varnothing}) }\right]^{\Delta_{ab}}\cdot V_{bifund} \; .
\end{align}
We write the following for the contributions of the fundamental hypermultiplets and Chern-Simons term:
\begin{align}
\prod_{a=1}^{n}z^{5d}_{H_{a},\mu^{a}}&=V_{fund},\\
\prod_{a=1}^{n}z^{5d}_{CS,\mu^{a}}&=V_{CS}.
\end{align}

We now collect all the leftover factors from the above reduction. After many cancellations, one can show that these factors make up a 3d hypermultiplet contribution,
\begin{align}
V_{vect}V_{bifund}V_{fund}V_{CS}=\prod_{a=1}^{n} \frac{z^{3d}_{H_{a}}(x_{\mu^{a}})}{z^{3d}_{H_{a}}(x_{\varnothing})} \; ,
\end{align}
where $z^{3d}_{H_{a}}(x_{\mu^{a}})$ can be written compactly as:
\beq\label{hyper}
z^{3d}_{H_{a}}(x_{a},\{f_{i}\}) =  \prod_{1\leq I \leq N_a}\,\prod_{j=1}^{|\cW_{\cS}|} \left[( v_a^{\widetilde{\#}_{a,I}} \, e^{x^{(a)}_I}/f_{j};q^{r_{a}})_{\infty}\right]^{\omega_{j,a}}
\eeq
Here, $\omega_{j,a}$ is the $a$'th Dynkin label of the $j$'th coweight in $\cW_{\cS}$, with the coweights  expanded in terms of fundamental coweights. For example, the $\fg=B_3$ coweight $\omega_1=[-1, \phantom{-}1, \phantom{-}0]$ has $\omega_{1,1}=-1$,\; $\omega_{1,2}=1$,\; $\omega_{1,3}=0$, and is to be understood as minus the first fundamental coweight plus the second fundamental coweight of $B_3$. The requirement that the sum of the coweights in $\cW_{\cS}$ add up to zero implies that the matter contribution \eqref{hyper} is in fact a ratio of q-Pochhammer's. From the point of view of $T^{5d}$, the various integers $\widetilde{\#}_{a,I}$ are uniquely determined from the requirement that the partition function truncates. From the point of view of $G^{3d}$, those integers are fixed by R-charge conservation.
In the end, the summand of the 5d gauge theory partition function becomes the summand of the 3d partition function \eqref{3dres}, establishing the first half of the proof for the case of polarized defects.\\

\subsubsection{Deformed Toda Conformal Block and 3d Partition Function}

The second part of the proof is more straightforward, and consists in simply comparing the deformed Toda conformal block with the partition function of $G^{3d}$: the two-point functions of screenings  in \eqref{sctda}, \eqref{sctdab},  are the contributions of the ${\cal N}=4$ vector and bifundamental multiplets in \eqref{wowvec} and \eqref{wowbif} to the D3 brane partition function, respectively. The number $N_a$ of D3 branes on the $a$-th node maps to the number of screening charge insertions. The evaluation of the two-point of a screening and a vertex operator \eqref{this} becomes the 3d $\cN=2$  hypermultiplet matter contribution \eqref{hyper}.

\subsection{The case of Unpolarized Defects}
\label{sec:zeroweight}

The above proof of triality was technically only written for the case of polarized defects of the little string. Is triality still true when $\cW_{\cS}$ defines an unpolarized defect? The answer is affirmative, and the proof is exactly as we presented it above, with one caveat; the matter content on node $a$ of the 3d theory, which was previously given by  \eqref{hyper} for a polarized defect, no longer has a closed form expression.

More precisely, the matter factor $z^{3d}_{H_{a}}(x_{a},\{f_{i}\})$ is no longer expressible in terms of the coweights of $\cW_{\cS}$. To illustrate this statement, consider the unpolarized defect $\cW_{\cS}=[0,0,0,0]$ in the second fundamental representation of $D_4$. We find in that case that the 3d matter factor after truncation of the 5d partition function is nontrivial. Moreover, multiple truncation schemes exist in that case, resulting in distinct 3d theories differing by their matter content. We suspect that each resulting matter factor corresponds to a distinct weight in a finite-dimensional irreducible representation of the quantum affine algebra $U_\hbar(\widehat{^L \fg})$. In our example, the weight $[0,0,0,0]$, initially with multiplicity 4 in the second fundamental representation of $D_4$, appears ``quantized" in 5 different ways (the original 4 plus a trivial representation) in $U_\hbar(\widehat{D_4})$.

A fascinating question is whether there exists a one-to-one map between the number of truncations in 5d and the various weights appearing in finite-dimensional representations of $U_\hbar(\widehat{^L \fg})$. We leave this question to future work; see also the study of $q$ and $qq$ characters of quantum affine algebras \cite{Frenkel:qch,Frenkel:1998,Nekrasov:2015wsu,Kimura:2015rgi}.

\section{$(2,0)$ CFT Limit and Nilpotent Orbits Classification of the Coulomb Branch}
\label{sec:Bala}

Because it has a scale $m_s$, the  $(2,0)$ little string theory on ${\cal C}$ is not conformal. To recover the $(2,0)$ 6d CFT theory on ${\cal C}$, we take this string scale $m_s$ to infinity, while keeping all  moduli of the $(2,0)$ theory  fixed in the process. Furthermore, if we denote by $\Delta x$ the relative position of the $|\cW_{\cS}|$ D5 branes on ${\cal C}$,  we then take the product $\Delta x \; m_s$ to zero.

\subsection{Description of the Defects}
\label{sec:CFTLIMIT}

The quiver gauge theory description of the defects is only valid at finite  $m_s$.  
Taking the $(2,0)$ CFT limit $m_s \rightarrow \infty$ has drastic effects on the physics; most notably, the radius of the 5d circle $R=1/(m_s^2 \,\hat{R})$ vanishes in the limit (the original cylinder radius $\hat{R}$ is fixed), so the theory becomes four-dimensional. We call the resulting theory $T^{5d}_{m_s\rightarrow \infty}\equiv T^{4d}.$ The 4d inverse gauge couplings $\tau_a$ vanishes as well, because the combinations $\tau_a \, m_s^2$ turn out to be moduli of the $(2,0)$ CFT, which are fixed in the limit. In other words, in the $ADE$ case, there is no longer a Lagrangian\footnote{Note there was no proper $BCFG$ 5d Lagrangian to begin with.} describing the theory on the D5 branes\footnote{Our 4d limit does not describe the theories of \cite{Nekrasov:2012xe}; there, one has a 4d quiver gauge theory, with the same quiver as for $T^{5d}$, by keeping the inverse gauge couplings $\tau_a$ finite.  This is \textit{not}  the $(2,0)$ theory on ${\cal C}$, since the moduli $\tau_a m_s^2$ then become infinite.}. Though an effective description as a quiver gauge theory is no longer available, a lot can be deduced about the resulting 4d theory in that limit, as we now explain.\\

Most notably, an essential feature is that when $m_s \rightarrow \infty$, the Coulomb branch of an $ADE$ defect theory $T^{4d}$ flows to a nilpotent orbit of $\fg$ \cite{Haouzi:2016ohr, Haouzi:2016yyg}. This can be for instance argued  based on the analysis of the Seiberg-Witten curve of the theories. We conjecture the same to be true for non simply-laced defects. In this paper, we will  perform  basic checks of this conjecture, such as dimension counting of the Coulomb branch, and matching of the Bala-Carter labeling of nilpotent orbits.\\

To arrive at nilpotent orbits, however, we must first remind the reader of the  elegant connection that exists between the coweights defining a little string defects and the so-called parabolic subalgebras of $\fg$.
We will need two facts from representation theory: First, a Borel subalgebra of a Lie algebra $\fg$ is a maximal solvable subalgebra. We note that the Borel subalgebra can always be written as the direct sum $\fb=\fh\oplus\fm$; here, $\fh$ is a Cartan subalgebra of $\fg$, and $\fm=\sum_{\alpha\in\Phi^+}\fg_\alpha$, with $\fg_\alpha$ the root spaces associated to a given set of positive roots $\Phi^+$. We fix the set  $\Phi^+$, which in turn fixes the Borel subalgebra $\fb$, for a given Lie algebra $\fg$.

Second, a parabolic subalgebra $\fp_\Theta$ is defined as a subalgebra of $\fg$ which contains the Borel subalgebra $\fb$. It also obeys a direct sum decomposition:  
\begin{equation}
\label{decomp}
\fp_\Theta=\fl_\Theta\oplus\fn_\Theta\, .
\end{equation}
In our notation, $\Theta$ is a subset of the set of positive simple roots of $\fg$. We introduced $\fn_\Theta=\sum_{\alpha\in\Phi^+ \backslash \langle\Theta\rangle^+}\fg_\alpha$ is the nilradical of $\fp_\Theta$, while $\fl_\Theta=\fh\ \oplus\sum_{\alpha\in\langle\Theta\rangle}\fg_\alpha$ is called a Levi subalgebra; the subroot system  $\langle\Theta\rangle$ is generated by the simple roots in $\Theta$, while $\langle\Theta\rangle^+$ is built out of the positive roots of $\langle\Theta\rangle$. Then, it follows that $\fn_{\Theta}\cong \fg/\fp_{\Theta}$.\\

We can now explain how parabolic subalgebras of $\fg$ arise from noncompact D5 branes:
Consider a set of coweights defining a puncture,
\begin{equation*}
{\cW}_{\cS} = \{ \omega_i\} \; .
\end{equation*}
As we explained in Section \ref{ssec:d5defects}, each coweight $\omega_i$ represents a distinct D5 brane.
A set of simple roots $\Theta$, as defined in the previous paragraph, is constructed as the subset of all simple roots of $\fg$ that have a zero inner product with every coweight of ${\cW}_{\cS}$. 

Among the many possible sets of coweights, we look in particular for a set in the Weyl group orbit of $\fg$ for which $|\Theta|$ is the largest. We call such a set of coweights \textit{distinguished}. In the rest of this paper, the sets of coweights ${\cW}_{\cS}$ we  consider are all taken to be distinguished. If a given set  is not distinguished, acting simultaneously on all its coweights with the Weyl group of $\fg$ will always turn it into a distinguished set.\\

\begin{example}[$F_4$ example 1]
	Let us consider the following set of $F_4$ coweights, expanded in terms of fundamental coweights as: 
$$\cW_\cS=\{\omega_1=[\phantom{-}0, \phantom{-}0, \phantom{-}1, -1],\;\; \omega_2=[\phantom{-}0, \phantom{-}0,  -1, \phantom{-}1]\}.$$ 
Both coweights have a zero inner product with $\alpha_1, \alpha_2$, so $|\Theta|=2$. A Weyl relfection about the simple root $\alpha_4$ turns the set into:
$$\cW'_\cS=\{\omega_1=[\phantom{-}0, \phantom{-}0, \phantom{-}0, \phantom{-}1], \;\; \omega_2=[\phantom{-}0, \phantom{-}0, \phantom{-}0, -1]\}$$ 
Note that this time, $|\Theta|=3$, and that is the maximal size of $\Theta$ for these two coweights. Therefore, we call the set $\cW'_\cS$ distinguished.
\end{example}

A nilradical  $\fn_{\Theta}$ of $\fg$ occuring in the direct sum decomposition \eqref{decomp} always specifies the Coulomb branch of some defect $T^{4d}$ \footnote{For a related discussion in the context of the codimension 2 defects of 4d $N=4$ SYM, see the work of Gukov and Witten \cite{Gukov:2008sn}; there, defects are described as the sigma model $T^*(G/P)$, with $P$ a parabolic subgroup of the Lie group $G$. Our discussion is related to that setup by further compactifying the little string on a torus \cite{Haouzi:2016ohr}, and then taking the limit $m_s\rightarrow\infty$.}. 
Starting from the weight data of the defect, the nilradical is extracted as follows: it is the direct sum of the root spaces associated to a set of positive roots $\{e_{\gamma}\}$ in $\fg$, such that
\beq
\langle e_{\gamma}, \omega_i\rangle <0
\eeq
for at least one coweight $\omega_i$ of $\cW_{\cS}$. The bracket $\langle \cdot, \cdot \rangle$ is the Cartan-Killing form of $\fg$. In particular, the size of this set is the complex dimension of the Coulomb branch of $T^{4d}$. It is important to note that the Coulomb branch is generically smaller than  at finite $m_s$, for  $T^{5d}$, where we had  \eqref{count}:
$$\sum\limits_{\langle e_\gamma,\omega_i\rangle<0} \left\vert\langle e_\gamma,\omega_i\rangle\right\vert \; .
$$
Indeed, in the little string formula above,  positive roots are counted with multiplicity, while this is not the case in the CFT limit. As a consequence, the Coulomb branch dimension of  $T^{4d}$ is at most the number of all positive roots of $\fg$.  This decrease of the Coulomb branch is  directly related to an effect we pointed out in  Toda theory: there, the number of contours in the evaluation of conformal blocks was conjectured to be bigger in deformed Toda, as opposed to the undeformed case.

Though we do not have a direct proof of the above prescription for computing the Coulomb branch dimension of $T^{4d}$, we checked it explicitly for the defects of all exceptional algebras, and up to a high rank for the classical algebras.\\

\begin{example}[$F_4$ example 2]
	Let us illustrate the above statements for the $F_4$ defect 
$$\cW_\cS=\{\omega_1=[\phantom{-}0, \phantom{-}0, \phantom{-}0, \phantom{-}1], \;\;\omega_2=[\phantom{-}0, \phantom{-}0, \phantom{-}0, -1]\}$$ 
First, let us compute the Coulomb branch dimension of $T^{5d}$  in the little string and of $T^{4d}$ in the CFT limit.
 $\omega_1$ has no negative inner product with any of the positive roots, so it does not contribute to the Coulomb branch counting.

$\omega_2$ has an inner product equal to -2 with 7 of the positive roots, and an inner product equal to -1 with 8 of the positive roots. Summing up the absolute value of these inner products, we deduce that the complex Coulomb branch dimension of $T^{5d}$ is 22. Writing down the quiver engineered by $\cW_\cS$, the Coulomb content from the gauge nodes is indeed $4+8+6+4=22$.
Furthermore, we can conclude that 15 of the positive roots have a negative inner product with at least one of the weights. Thus, the Coulomb branch of $T^{4d}$ has complex dimension 15.

The set $\cW_\cS$ is distinguished, and $\omega_1$ and $\omega_2$  both clearly have a zero inner product with the three simple roots $\alpha_1, \alpha_2, \alpha_3$ (they have common zeros for their first three Dynkin labels). We conclude at once that $\Theta=\{\alpha_1, \alpha_2, \alpha_3\}$. Therefore, the parabolic subalgebra associated to this defect is  $\fp_{\{\alpha_1, \alpha_2, \alpha_3\}}$.
\end{example}

The above discussion leads us straight to the consideration of nilpotent orbits.
The fact that surface defects of 6d $(2,0)$ CFTs are described by these orbits has been studied in detail in \cite{Chacaltana:2012zy}. A useful reference for the rest of this section is the book \cite{Collingwood:1993}.\\

Having chosen a fixed faithful representation for $\fg$, we say that $X\in \fg$ is  nilpotent if the matrix that represents $X$ is nilpotent. Then, all the  elements in the $G$-adjoint \footnote{Note it is the adjoint Lie \textit{group} action that is used here, not the Lie algebra.} orbit $\cO_X$ of $X$ are nilpotent, and we call $\cO_X$ a nilpotent orbit.  Consider a Levi subalgebra arising from the direct sum decomposition of the parabolic subalgebra $\fp=\fl\oplus\fn$, the nilpotent orbit $\mathcal{O}_\fl$ associated to $\fp$ is the maximal orbit containing a representative $X\in \mathcal{O}_\fl$ for which $X\in \fn$.

The Coulomb branch of a physical defect theory $T^{4d}$ is intimately related to the existence of a \textit{duality map} which acts on the nilpotent orbits. The Spaltenstein function \cite{Spaltenstein:1982} is such a map; it reorganizes nilpotent orbits of $\fg$ by mapping an orbit $\cO$ of $\fg$ to another orbit of $\fg$ (which is sometimes the same as $\cO$).   The map is many-to-one for all algebras except $\fg=A_n$ \footnote{Since the defects of the little string live in the weight lattice of $^L\fg$, one can also choose to work with a slightly different map, the Spaltenstein-Barbasch-Vogan map \cite{10.2307/1971193}, which sends nilpotent orbits of $\fg$ to orbits of $^L\fg$. Ultimately, there is no difference in the resulting physics, so we choose to work with the Spaltenstein map instead, denoting defects as living in the coweight lattice of $\fg$, as we have done in the rest of this paper.}. Before explaining the relevance of the Spaltenstein map in our context, let us address the question of classification of nilpotent orbits.

\subsection{Bala--Carter Labeling of Nilpotent Orbits}
\label{ssec:balalabel}

There exist different ways to label nilpotent orbits  in the Mathematics literature; the characterization that turns out to arise naturally in the $m_s\rightarrow\infty$ limit of the little string was developed by  Bala and Carter, and is applicable to any semi-simple Lie algebra \cite{bala:1976msaa,bala:1976msab}. We only need to borrow the end result of their analysis, so we will be brief in describing their construction. It relies on the use of the Levi subalgebras of $\fg$. 


The Bala--Carter prescription is to label a nilpotent orbit $\mathcal{O}$ by the smallest Levi subalgebra $\fl\subset\fg$ that contains some representative of that orbit. When  $\fg \neq A_n$, it can happen that this Levi subalgebra does not specify uniquely  $\cO$, so extra data is needed. The prescription is as follows: suppose a parabolic subalgebra $\fp$ has the usual direct sum decompostion into Levi and nilradical parts, $\fp=\fl^\prime\oplus\fu$. We say $\fp$ is distinguished if $\dim \fl^\prime=\dim \left(\fu/[\fu,\fu]\right)$ (an example of such a $\fp$ is the Borel subalgebra of $\fl$.) Then, one can show that a nilpotent orbit $\cO$ is uniquely determined by the Levi subalgebra $\fl$ \textit{and} by a distinguished parabolic subalgebra of $[\fl,\fl]$.

If $\fl$ is sufficient to uniquely specify a nilpotent orbit $\cO$, meaning $\fl$ contains a unique distinguished parabolic subalgebra, then  $\cO$ is said to have Bala--Carter label $\fl$.  The  orbit  $\cO$  is called the \emph{principal nilpotent orbit} of $\fl$. 
 If the orbit $\cO$ is not uniquely determined by $\fl$, an additional label specifying a distinguished parabolic subalgebra of $[\fl,\fl]$ is needed (it is usually given as the number of simple roots in a Levi subalgebra of $\fp$).\\

It is remarkable that  one can read off the Bala--Carter label of a nilpotent orbit just from the Dynkin labels of the coweights specifying a D5 brane defect in little string theory. To be precise, we find the following general result, for  ${\cW}_{\cS}$  a distinguished set of coweights of $\fg$, and $\Theta$ its associated set of simple roots, as  defined in the previous section:
\begin{itemize}
\item If ${\cW}_{\cS}$ denotes a polarized defect of the little string, then one can identify the set $\Theta$ with the Bala--Carter label of the defect in the CFT limit. Specifically, the union of all  elements of the set $\Theta$ is a subquiver of $\fg$, called the Bala--Carter label of this defect, written as $\fl_{\Theta}$. The Coulomb branch of $T^{4d}$ is then a resolution of the Spaltenstein dual of $\cO$, where $\cO$ is the nilpotent orbit labeled by the Bala--Carter label $\fl_{\Theta}$. The orbit $\cO$ is the principal nilpotent orbit of the Levi subalgebra $\fl_{\Theta}$.

\item If ${\cW}_{\cS}$ denotes an unpolarized defect of the little string \textit{and} $\fg$ is simply-laced,  then one can identify the set $\Theta$ with part of the Bala--Carter label of the defect in the CFT limit. To fully characterize the defect, one must also indicate which fundamental representation the coweights of ${\cW}_{\cS}$ belong in. This additional prescription is in one-to-one correspondence with specifying the extra data needed to denote the Bala--Carter label of a non-principal nilpotent orbit. Furthermore, the Coulomb branch of $T^{4d}$ is not in general in the image of the Spaltenstein map. 

When $\fg$ is non simply-laced, it can happen that an unpolarized defect ${\cW}_{\cS}$ in the CFT limit has no relation to the labeling of nilpotent orbits predicted by Bala and Carter (the nilpotent orbit is still realized physically as a Coulomb branch of some theory $T^{4d}$, but the Bala--Carter label for it is not readable from the simple roots set $\Theta$ of ${\cW}_{\cS}$).
\end{itemize}

\begin{example}[$F_4$ example 3]
	Let us consider again our $F_4$ defect,
$$\cW_\cS=\{\omega_1=[\phantom{-}0, \phantom{-}0, \phantom{-}0, \phantom{-}1], \;\;\omega_2=[\phantom{-}0, \phantom{-}0, \phantom{-}0, -1]\}.$$ 
One can easily check that the defect is polarized. Furthermore,
we identified in the previous example that $\Theta=\{\alpha_1, \alpha_2, \alpha_3\}$. Therefore, the Bala--Carter label for the defect is $B_3$, and the Coulomb branch of the defect in the CFT limit is the Spaltenstein dual of the nilpotent orbit $B_3$, which is the orbit $A_{2_s}$. The orbit $A_{2_s}$ has complex dimension 15, which confirms our previous computation of the dimension from a different method.
\end{example}

Some comments are in order: First, the above points imply that the $m_s\rightarrow\infty$ limit of a nilpotent orbit  realized as the  of the Coulomb branch of some Dynkin-shaped quiver gauge theory, with unitary gauge groups. Second, the coweight data of the D5 branes defining those quivers almost always provides a physical realization of the Bala--Carter classification of nilpotent orbits, with a few exceptions: for some non simply-laced unpolarized defects, the \textit{labeling} predicted by Bala and Carter is not readily readable from the little string Physics perspective.  We will illustrate this feature in detail for $\fg=G_2$ in section \ref{ssec:g2list}.

\subsection{Some Comments on Classification}
\label{ssec:more}

It turns out that the Bala--Carter label of a polarized defect can also be obtained as the union of some simple roots $\alpha_i$ in the undeformed $\fg$-type Toda theory, as a null state condition at level 1 \footnote{It was first pointed out in \cite{Kanno:2009ga} that when $\fg=A_n$, surface defects of $\cN=4$ SYM are characterized in Toda theory by level 1 null states.}:
\begin{equation}
\label{levelone}
\langle\beta,\alpha_i\rangle=0 \quad \forall \alpha_i\in\Theta \; ,
\end{equation}
with  $\Theta$  a subset of simple roots of $\fg$, and $|\vec\beta\rangle$ a highest coweight state of the  $\cW(\fg)$-algebra. By the state-operator correspondence, the momentum $\beta$ carried by the vertex operator $V^\vee_{\beta}(z)$ is simply $\beta=\sum_{i=1}^{|\cW_{\cS}|}\beta_i\; \omega_i$, as we wrote  previously in equation \eqref{momentumpunc}.

In the notation of Section \ref{ssec:balalabel}, a subalgebra $\fp_\Theta$ can then be associated to $\beta$, with $\Theta$ characterizing the level 1 null state above. An induced nilpotent orbit can then be extracted from it, following the procedure described in the previous section.\\

Finally, let us make contact with the so-called weighted Dynkin diagrams that appear in the literature as yet another way to classify nilpotent orbits.

Weighted Dynkin diagrams are vectors of integers $r_i\in\{0,1,2\}$, where $i=1,\ldots , n$; thus, we get one number for each node in the Dynkin diagram of $\fg$. We can associate such a vector to each nilpotent orbit of $\fg$, and each nilpotent orbit has a unique weighted Dynkin diagram. Note, however, that not all such labellings of the Dynkin diagram  have a nilpotent orbit corresponding to it.\\

A notable curiosity is that all weighted Dynkin diagrams can strikingly be interpreted as physical quiver defect theories of the little string, with flavor symmetry $\prod_{i=1}^{n}U(r_i)$. Namely, the labels on the nodes of a weighted Dynkin diagram can be understood as the rank of a flavor symmetry group in a quiver gauge theory. The quivers one reads off in this way all turn out to satisfy the constraint \eqref{conformal}. For instance, the full puncture, or maximal nilpotent orbit, which is always denoted by the weighted Dynkin diagram $(2,2,\ldots,2,2)$, can be understood as a little string quiver gauge theory with a $U(2)$ flavor attached to each node, for all simple Lie algebras. The surprise is that even though the quiver theories are defined in the little string, at finite $m_s$, in the CFT limit their Coulomb branch flows to the nilpotent orbit precisely labeled by that weighted Dynkin diagram.

Furthermore, this analogy gives another way to compute the dimension of a nilpotent orbit. We interpret the weighted Dynkin diagram of a nilpotent orbit $\cO$ as a coweight $\omega$, written down in fundamental coweight basis. We then compute the sum of the inner products of all the positive roots of $\fg$ with this coweight. This gives a vector of non-negative integers. Truncating the entries of this vector at 2 and taking the sum of the entries gives the (real) dimension of $\cO$. The proof in the simply-laced case is given in \cite{Haouzi:2016yyg}, and generalizes straightforwardly to all simple Lie algebras.\\

\begin{example}[$F_4$ example 4] 
Let us look at the weighted Dynkin diagram (0,0,0,2) in the algebra $\fg= F_4$. We therefore consider the coweight $\omega=[0,0,0,2]$, which happens to be twice the fourth fundamental coweight of $F_4$. Let  ${\Phi^+}^\vee$ be the set of the 24 positive roots of $F_4$.
Calculating the inner product of all of these positive roots with $\omega$ gives:
\begin{align*}
\langle {\Phi^+}^\vee,\omega\rangle=(4,4,4,4,2,4,2,4,4,0,2,2,0,2,0,2,0,0,2,0,0,0,2,0).
\end{align*}
Truncating at multiplicity 2, the sum of the inner products is $2\times 7+2 \times 8 = 30$, which is the correct real dimension of the nilpotent orbit denoted by the diagram $(0,0,0,2)$.  It is quite amazing that at finite $m_s$, in the little string, the gauge theory whose Coulomb branch  flows to this orbit in the CFT limit is precisely the quiver with mass content $(0,0,0,2)$. This is just the quiver engineered in the previous examples, from the set:
$$\cW_\cS=\{\omega_1=[\phantom{-}0, \phantom{-}0, \phantom{-}0, \phantom{-}1], \;\;\omega_2=[\phantom{-}0, \phantom{-}0, \phantom{-}0, -1]\}.$$
\end{example}


\section{Examples}
\label{sec:examples}

We now illustrate the various results of the paper, for $\fg$ a simple Lie algebra. 

\subsection{Sphere with 3 full punctures}
\label{ssec:3full}

We start by showcasing the Triality of Section \ref{sec:triality}, for a Riemann surface being a sphere with three full punctures, in the terminology of \cite{Gaiotto:2009we}; in this paper, we compactified the little string on a cylinder $\cC$. This is equivalent to choosing $\cC$ to be the sphere with two full punctures that come for free. A given set of D5 branes at points on $\cC$  will characterize additional arbitrary punctures. In what follows, $n\equiv \mbox{rank}\,(\fg)$.\\
 
In particular, in order to construct a single full puncture defect out of D5 branes,  we pick a set $\cW_{\cS}$ of $n+1$ coweight vectors adding up to 0, such that in the CFT limit $m_s\rightarrow\infty$, the Bala-Carter label for this set is $\varnothing$. Such a defect is always polarized, in the terminology of Section \ref{ssec:polar}. Out of the many sets of coweights that a priori satisfy this condition,  we will present a ``canonical" set  $\cW_{\cS}$ with a generic matter content.\\

The 5d gauge theory $T^{5d}$ on the D5 branes, the 3d  gauge theory $G^{3d}$ on the D3 branes, and the collection of vertex operators in deformed Toda, are related by triality. We will see explicitly that the partition function of $T^{5d}$ truncates to a  3-point function in the deformed Toda theory,  with 3 primary operator insertions of generic momenta. 

For each $\omega_i$, we pick a point on the Riemann surface ${\cal C}$, with coordinate $x_i=R \, \beta_i.$  This $x_i$ specifies the position of the D5 brane wrapping $\omega_i = [S_i^*]$ on ${\cal C}$, and the masses $\beta_i$ of the various  matter fields in the 5d and 3d gauge theories. In the Toda theory, these parameters specify the $n$ momenta and the position of the puncture on ${\cal C}$.   The $n$ 5d gauge couplings $\tau_a$ become the 3d FI parameters, or equivalently the momentum  of the puncture at $z=0$ in the Toda picture\footnote{The $n$ non-normalizable Coulomb moduli coming from the $U(1)$ centers in the gauge groups of $T^{5d}$ become the ranks $N_a$ of the 3d gauge groups of $G^{3d}$, which is also the number of screening charges in Toda theory; this specifies the momentum of the puncture located at $z=\infty$.}.\\

As we explained, the vertex operator $:\prod V^\vee_{\omega_i}(x_i):$  is the $(q,t)$-deformation of the primary operator $V_{\beta}(z)$, through the relations 
\begin{equation}
\beta = \sum_{i=1}^{|\cW_S|} \beta_i \,\omega_i\; 
\end{equation}
and
\beq
e^{x_{i}} = z \, q^{-\beta_{i}} \; .
\eeq
The R-charges of the 3d chiral multiplets determine all $v_a=\sqrt{q^{r_a}/t}$ factors which will appear in the argument of the vertex operators. These multiplets are generated from strings stretching between a D3 branewrapping $S_a$ and a D5 brane wrapping $S_i^*$.

In the undeformed Toda CFT, the three-point of ${\cal W}$-algebra primaries is labeled by three momenta $\beta_{0}, \,\beta, \,\beta_{\infty}$:
\begin{align}\label{3pt}
\langle V_{\beta_{0}}(0)\, V_{\beta}(z) \, V_{\beta_{\infty}}(\infty)\rangle \; .
\end{align}
If $\beta_{\infty} = -\beta_0-\beta -\sum_{a=1}^n N_a \, \alpha^\vee_a/b$ for positive integers $N_a$ (which are the ranks of the gauge groups of) we can compute  the three-point function \eqref{3pt} in free field formalism: we simply insert $N_a$ screening charge operators $Q^\vee_a = \int dx \,S^\vee_a(x)$:
\beq
 \label{3point}
\langle V^\vee_{\beta_0}(0) V^\vee_{\beta}(z) V^\vee_{\beta_{\infty}}(\infty) \prod_{a=1}^n (Q^\vee_a)^{N_a}\rangle_{free}.
\eeq
Once we replace the screening charges and the vertex operators by their $(q,t)$-deformed counterparts, we obtain a deformed conformal block of the ${\cal W}_{q,t}(\bf g)$ algebra, as described in Section \ref{ssec:Toda}.\\

When $\fg=A, D, E$, the sphere with three full punctures was analyzed  in \cite{Aganagic:2015cta}, so we will be brief in describing those cases; we focus instead on the defects of the little string when  $\fg$ is non simply-laced. For definiteness, we will use the following definitions of the Cartan matrices in the examples:

\begin{align*}
& C_{ab}^{B_n} = \left( \begin{array}{cccccc}
2 & -1 & 0 & 0  & \ldots & 0\\ 
-1 & 2 & -1 & 0 & \ldots & 0\\
\vdots & \vdots & \vdots & \vdots & \vdots & \ddots\\
0 & \ldots & -1 & 2 & -1 & 0\\
0 & \ldots & 0 & -1 & 2 & -2\\
0 & \ldots  & 0 & 0 & -1 & 2
\end{array} \right) \qquad\qquad  C_{ab}^{C_n} = \left( \begin{array}{cccccc}
2 & -1 & 0 & 0  & \ldots & 0\\ 
-1 & 2 & -1 & 0 & \ldots & 0\\
\vdots & \vdots & \vdots & \vdots & \vdots & \ddots\\
0 & \ldots & -1 & 2 & -1 & 0\\
0 & \ldots & 0 & -1 & 2 & -1\\
0 & \ldots  & 0 & 0 & -2 & 2
\end{array} \right)\\
\\
& \qquad\qquad C_{ab}^{G_2} = \left( \begin{array}{cc}
2 & -1\\ 
-3 & 2
\end{array} \right) \qquad\qquad\qquad\qquad\;\;\;  C_{ab}^{F_4} = \left( \begin{array}{cccc}
2 & -1 & 0 & 0\\ 
-1 & 2 & -2 & 0\\
0 & -1 & 2 & -1\\
0 & 0 & -1 & 2
\end{array} \right)
\end{align*}

\subsubsection{$A_n$ Full Puncture}

\begin{figure}[h!]
	\begin{center}
		\includegraphics[width=0.95\textwidth]{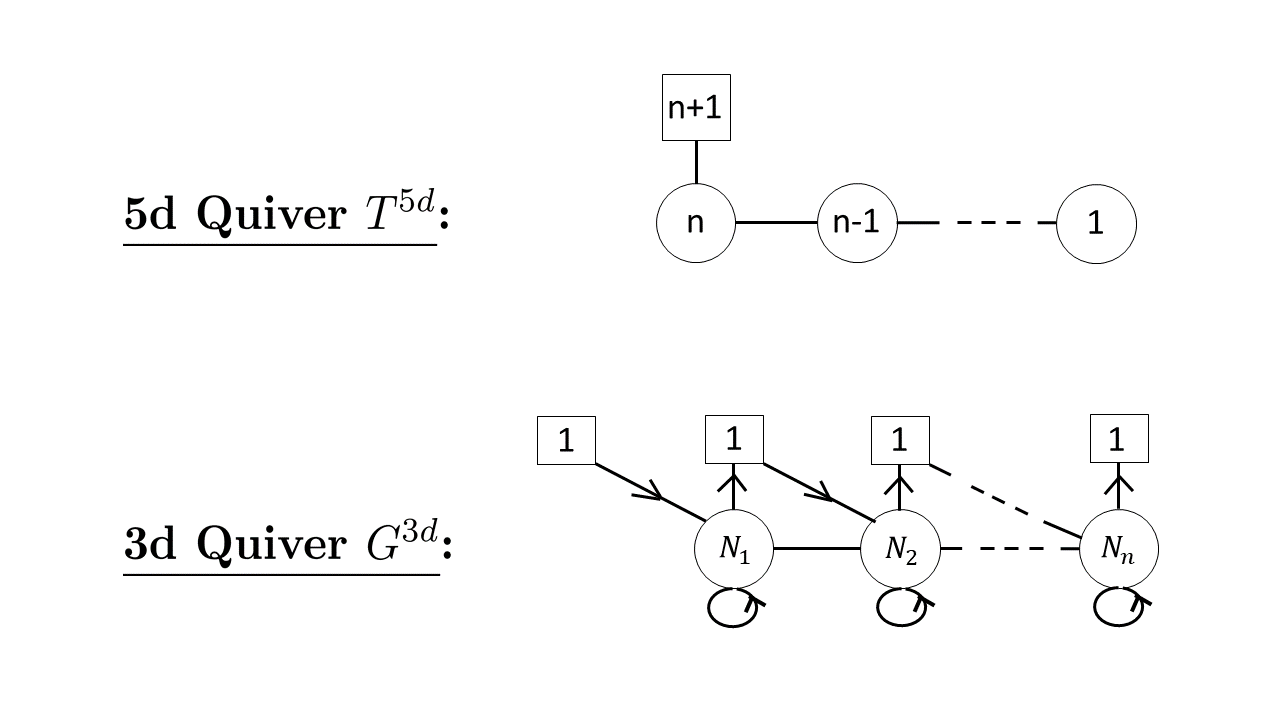}
        	\end{center}
        \caption{Sphere with 3 full $A_n$  punctures: 5d theory $T^{5d}$ and 3d theory $G^{3d}$ resulting from $\cW_{\cS}$.}
\label{AnFull}
\end{figure}

For the simply-laced case, we use the words weights (respectively roots) and coweights (respectively coroots) interchangeably.
For $\fg=A_n$, a full puncture is realized by the following  set $\cW_{\cS}$ of $n+1$ coweights: 
\beq
 \begin{aligned}\label{fund}
&\omega_1= -w_1^\vee\\
&\omega_2 = -w_1^\vee+\alpha_1^\vee\\
&  \vdots\\ 
&\omega_{n} = -w_1^\vee+\alpha_1^\vee+\ldots + \alpha_{n-1}^\vee\\
&\omega_{n+1} = -w_1^\vee+\alpha_1^\vee+\ldots+ \alpha_{n-1}^\vee+\alpha_n^\vee
\end{aligned}
\eeq
$w_a^\vee$ is the $a$-th fundamental coweight, and $\alpha_a^\vee$ is the $a$-th simple coroot. Note that the set $\cW_{\cS}$ spans the coweight lattice. Each one of the coweights $\omega_a$ represents a distinct D5 brane wrapping a non-compact 2-cycle and some compact 2-cycles. At low energies, one can directly read off the 5d $\cN=1$ $A_n$ quiver gauge theory living on the branes: the coefficients of the $\alpha_a^\vee$ give the rank of the gauge group, while the rank of the flavor symmetry in the fundamental representation of the $a$-th gauge group is given by the number of occurrences of $-w_a$. The resulting 5d quiver gauge theory is shown in figure \ref{AnFull}.\\

We superimpose the compact D5 branes on top of the non-compact ones and add D3 branes wrapping the compact 2-cycles. The strings that stretch between D3 branes realize a 3d $\cN=4$ $A_n$ quiver gauge theory, with gauge content $\prod_{a=1}^n U(N_a)$. Supersymmetry is broken to $\cN=2$ due to the strings stretching between the D3 and D5 branes, resulting in chiral and anti-chiral multiplets in fundamental representation of the various gauge groups. The Dynkin labels of the coweights \eqref{fund} written in the fundamental coweight basis encode the precise matter content of the 3d theory:
\beq
 \begin{aligned}
\omega_1 &= [-1,\phantom{-}0, \phantom{-}0, \ldots,   \phantom{-}0,  \phantom{-}0]\\
\omega_2 &= [\phantom{-}1, -1, \phantom{-}0, \ldots,   \phantom{-}0,  \phantom{-}0]\\
&  \vdots\\ 
\omega_{n} &= [\phantom{-}0,  \phantom{-}0, \phantom{-}0, \ldots,   \phantom{-}1, -1]\\
\omega_{n+1} &= [\phantom{-}0,  \phantom{-}0, \phantom{-}0, \ldots,   \phantom{-}0, \phantom{-}1]
\end{aligned}
\eeq
We obtain the 3d $\cN=2$ quiver gauge theory $G^{3d}$ shown in figure \ref{AnFull}. Note in passing that the set $\cW_{\cS}$  has no common zeros in the above notation. Acting on this set with the Weyl group will not change that, so the set is distinguished and the defect is indeed a full puncture.\\

The deformed vertex operator that realizes the full puncture is the product $:\prod_{i=1}^{n+1} V^\vee_{\omega_i}(x_i):$, where
\beq
 \begin{aligned}\label{vertan}
&V^\vee_{\omega_1} (x)= W_1^{-1}(x), \\
&V^\vee_{\omega_2} (x)= : W_1^{-1}(x) E_1(xv^{-1}):\\
&  \vdots\\ 
&V^\vee_{\omega_n} (x)= : W_1^{-1}(x) E_1(xv^{-1}) E_2(x v^{-2}) \ldots  E_{n-1}(xv^{1-n}):\\
&V^\vee_{\omega_{n+1}} (x)= : W_1^{-1}(x) E_1(xv^{-1}) E_2(x v^{-2}) \ldots  E_{n-1}(xv^{1-n}) E_n(x v^{-n}):
\end{aligned}
\eeq
The above ``fundamental coweight" and ``simple coroot" vertex operators were defined in section \ref{ssec:Toda}; the expression is a $(q, t)$ refinement of the relation \eqref{fund}. The dependence on the $v$-factors above encodes the value of the Coulomb moduli at the triality point. Namely, let $v^{\#_{a,i}}$ be the various $v$-factors in the arguments of the $E_a$ operators \eqref{vertan}. Then, the Coulomb moduli of the 5d gauge theory that truncate the partition function to the $A_n$ deformed conformal block are given by 
$$e_{a,i}=f_i\, t^{N_{a,i}} \,v^{\#_{a,i}} \, v^{-a}\, , \qquad a=1, \ldots, n.$$

The Coulomb branch of the 5d theory has complex dimension $\sum_{a=1}^n d_a=n(n+1)/2$, with $d_a$ the ranks of the $n$ gauge groups. This can also be obtained from \eqref{count}:
$$\sum\limits_{\langle e_\gamma,\omega_i\rangle<0} \left\vert\langle e_\gamma,\omega_i\rangle\right\vert=\frac{n(n+1)}{2}
$$
In the above sum, one counts all positive roots that have a negative inner product with at least one of the coweights, including multiplicity. Here, all positive roots of $A_n$ satisfy this condition, with multiplicity 1, so the right-hand side is simply the number of positive roots of $A_n$. This is also the number of supersymmetric vacua (or equivalently, integration contours) of the 3d theory, and the number of parameters needed to  specify the 3-point of the deformed $\cW_{q,t}(A_n)$ algebra.\\

In the CFT limit, when $m_s\rightarrow\infty$, the counting is done without multiplicity, but since each positive roots is counted once in the little string, the Coulomb branch dimension does not change. The Coulomb branch of the resulting theory $T^{4d}$ is the maximal nilpotent orbit of $A_n$, with Bala-Carter label $A_n$. This orbit is in the image  by the Spaltenstein map of the orbit denoted by $\varnothing$. This pre-image Bala--Carter label $\varnothing$ is identified at once since the set  $\cW_{\cS}$ has no common zeros in the Dynkin labels of the different coweights, as we pointed out.

\subsubsection{$D_n$ Full Puncture}

\begin{figure}[h!]
	\begin{center}
		\includegraphics[width=0.95\textwidth]{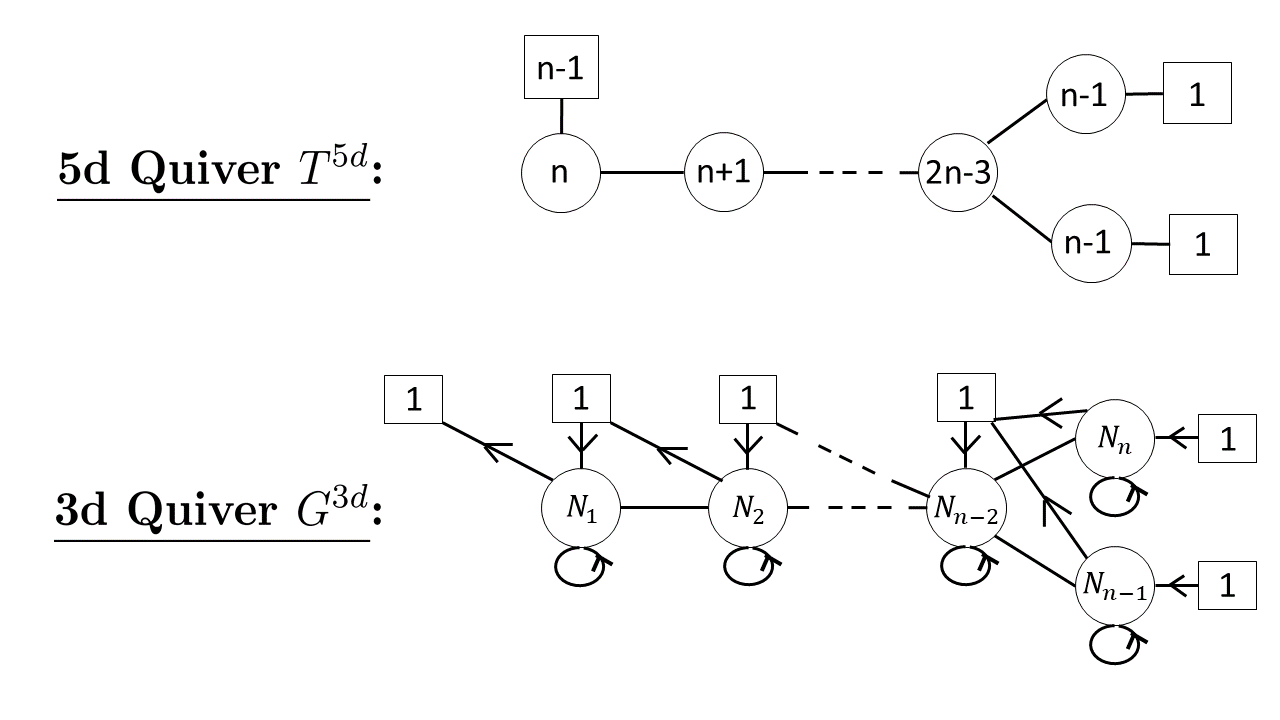}
        	\end{center}
        \caption{Sphere with 3 full $D_n$  punctures: 5d theory $T^{5d}$ and 3d theory $G^{3d}$ resulting from $\cW_{\cS}$.}
\label{DnFull}
\end{figure}

We will be more brief for the rest of the simply-laced cases. For ${\cal W}_{\cal S}$, we take the following collection of $n+1$ weights of $D_n$:
\beq
 \begin{aligned}\label{fundwd}
&\omega_1= - w_1^\vee + \alpha_1^\vee +\alpha_{2}^\vee + \ldots +\alpha_{n-2}^\vee + \alpha_{n-1}^\vee + \alpha_{n}^\vee\\
& \omega_i = \omega_{i-1} + \alpha_{n-i}^\vee\; , \qquad i=2, \ldots n-1\\
&  \omega_{n}= -{w}_{n-1}^\vee \\
 & \omega_{n+1} = -{w}_n^\vee 
\end{aligned}
\eeq
Writing each coweight above in terms of fundamental coweights, it is clear that $\cW_{\cS}$ has no common zeros, and acting on  $\cW_{\cS}$ with the Weyl group will not change that, so the set is distinguished and this is indeed a full puncture.\\

The complex dimension of the Coulomb branch of $T^{5d}$ (or the number of vacua of $G^{3d}$) is
$$\sum_{a=1}^n d_a=\frac{(n-1)(3n-2)}{2}=\sum\limits_{\langle e_\gamma,\omega_i\rangle<0} \left\vert\langle e_\gamma,\omega_i\rangle\right\vert \; .
$$
In the above sum, one counts all positive roots that have a negative inner product with at least one of the coweights. Here, some of the positive roots of $D_n$ satisfy this condition with multiplicity 1, while others satisfy it with multiplicity 2, so the answer is necessarily bigger than the number of positive roots of $D_n$. This is also the number of supersymmetric vacua (or equivalently, integration contours) of the 3d theory, and the number of parameters needed to  specify the 3-point of the deformed $\cW_{q,t}(D_n)$ algebra.\\

In the CFT limit, when $m_s\rightarrow\infty$, the counting is done without the multiplicity 2 for some of the positive roots; the Coulomb branch dimension of the D5 brane theory therefore decreases and becomes equal to the number of positive roots of $D_n$, which is $n^2-n$. The Coulomb branch of the resulting theory $T^{4d}$ is therefore the maximal nilpotent orbit of $D_n$. 
The 5d and 3d theories are shown in figure \ref{DnFull}.

\subsubsection{$E_n$ Full Puncture}

\begin{figure}[h!]
	\begin{center}
		\includegraphics[width=0.95\textwidth]{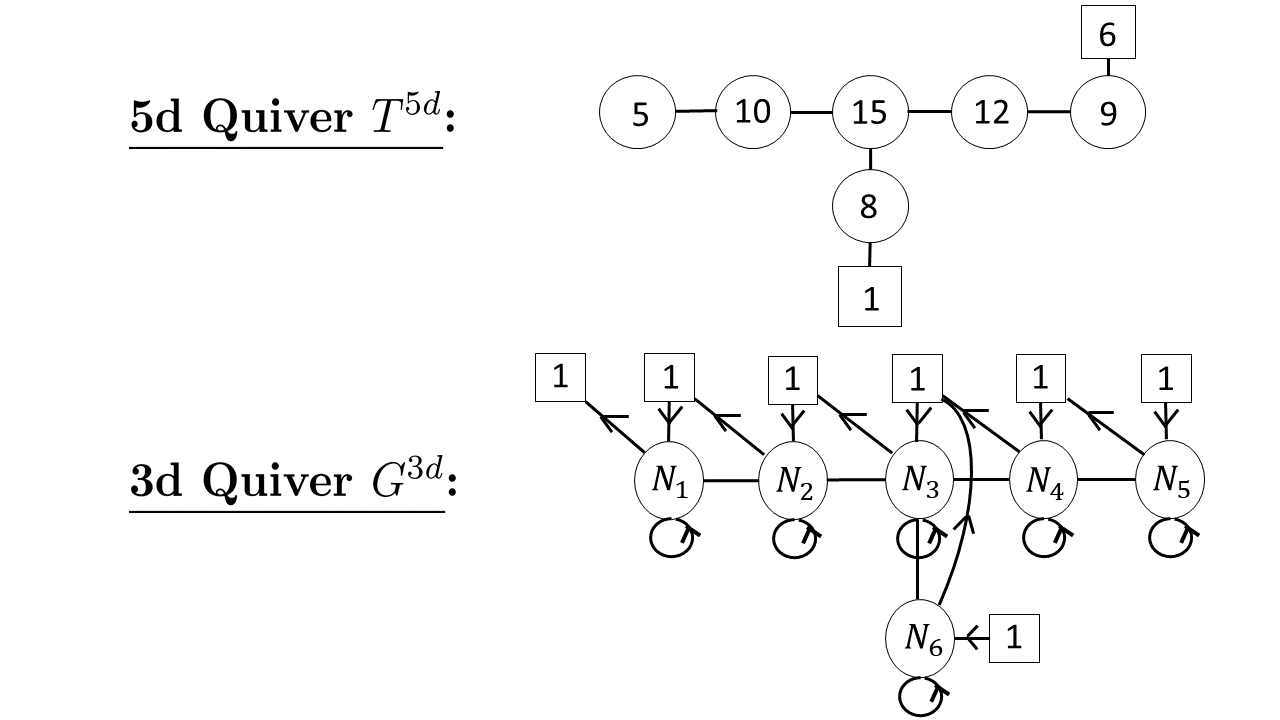}
        	\end{center}
        \caption{Sphere with 3 full $E_6$  punctures: 5d theory $T^{5d}$ and 3d theory $G^{3d}$ resulting from $\cW_{\cS}$.}
\label{E6Full}
\end{figure}

In the case of $E_6$, we take the set ${\cal W}_{\cal S}$ to be the following collection of 7 coweights:
\beq\label{funde6}
\begin{aligned}
 & \omega_1 = -{w}^\vee_5 \\
 & \omega_2= -{w}^\vee_5 + \alpha^\vee_5\\
& \omega_3= -{w}^\vee_5 + \alpha_1^\vee + 2\alpha^\vee_2 + 3\alpha^\vee_3 + 3\alpha^\vee_4  + 2\alpha^\vee_5 + 2\alpha^\vee_6\\
& \omega_4= -{w}^\vee_5 + \alpha^\vee_1 + 2\alpha^\vee_2 + 4\alpha^\vee_3 + 3\alpha^\vee_4  + 2\alpha^\vee_5 + 2\alpha^\vee_6 \\
& \omega_5= -{w}^\vee_5 + \alpha^\vee_1 + 3\alpha^\vee_2 + 4\alpha^\vee_3 + 3\alpha^\vee_4  + 2\alpha^\vee_5 + 2\alpha^\vee_6 \\
& \omega_6= -{w}^\vee_5 + 2\alpha^\vee_1 + 3\alpha^\vee_2 + 4\alpha^\vee_3 + 3\alpha^\vee_4  + 2\alpha^\vee_5 + 2\alpha^\vee_6 \\
& \omega_7= -{w}^\vee_6 
\end{aligned}
\eeq 

\begin{figure}[h!]
	\begin{center}
		\includegraphics[width=0.95\textwidth]{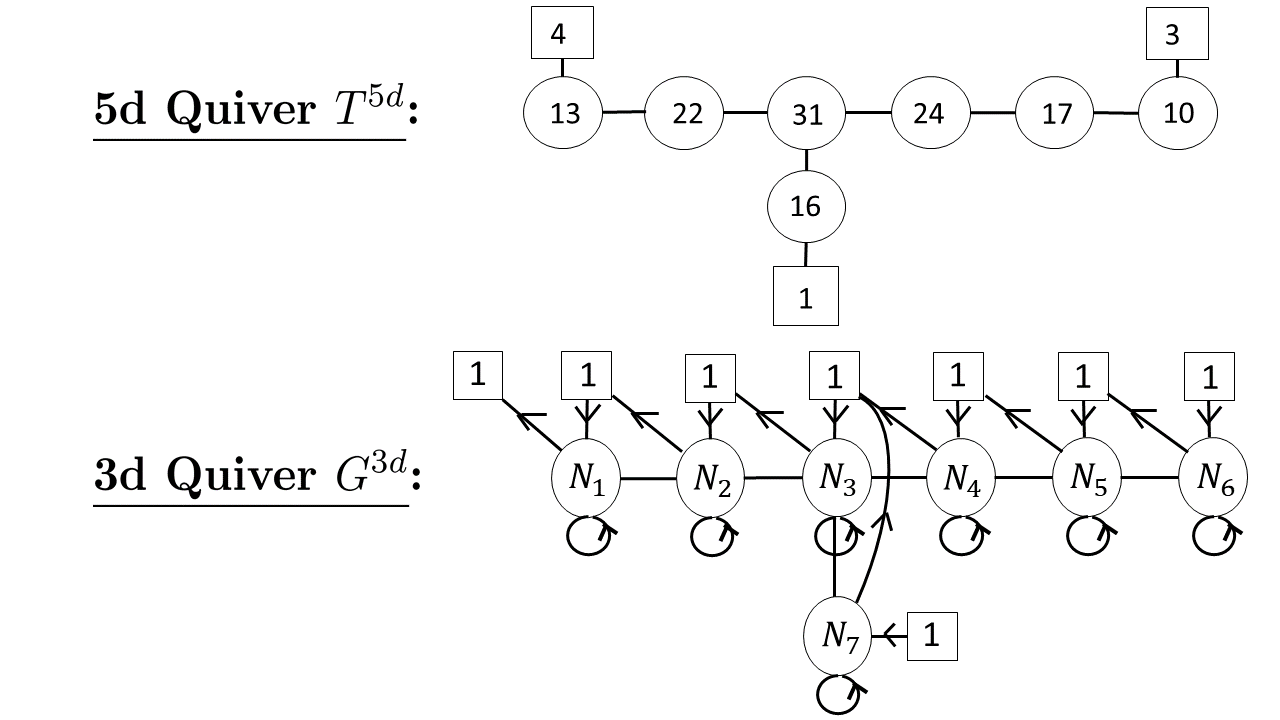}
        	\end{center}
        \caption{Sphere with 3 full $E_7$  punctures: 5d theory $T^{5d}$ and 3d theory $G^{3d}$ resulting from $\cW_{\cS}$.}
\label{E7Full}
\end{figure}

In the case of $E_7$, we take the set ${\cal W}_{\cal S}$ to be the following collection of 8 coweights:
\beq\label{funde7}
\begin{aligned}
&\omega_1= -{w}^\vee_1 + 3\alpha^\vee_1 + 5\alpha^\vee_2 + 7\alpha^\vee_3 + 6\alpha^\vee_4  + 4\alpha^\vee_5 + 2\alpha^\vee_6 + 4\alpha^\vee_7\\
&\omega_2 = -{w}^\vee_1 + 3\alpha^\vee_1 + 5\alpha^\vee_2 + 8\alpha^\vee_3 + 6\alpha^\vee_4  + 4\alpha^\vee_5 + 2\alpha^\vee_6 + 4\alpha^\vee_7 \\
 &\omega_3 = -{w}^\vee_1 + 3\alpha^\vee_1 + 6\alpha^\vee_2 + 8\alpha^\vee_3 + 6\alpha^\vee_4  + 4\alpha^\vee_5 + 2\alpha^\vee_6 + 4\alpha^\vee_7 \\
& \omega_4 = -{w}^\vee_1 + 4\alpha^\vee_1 + 6\alpha^\vee_2 + 8\alpha^\vee_3 + 6\alpha^\vee_4  + 4\alpha^\vee_5 + 2\alpha^\vee_6 + 4\alpha^\vee_7\\
&\omega_5 = -{w}^\vee_6 \\
&\omega_6 = -{w}^\vee_6 + \alpha^\vee_6 \\
& \omega_7 = -{w}^\vee_6 + \alpha^\vee_5 + \alpha^\vee_6\\
 &\omega_8 = -{w}^\vee_7
\end{aligned}
\end{equation} 

\begin{figure}[h!]
	\begin{center}
		\includegraphics[width=0.95\textwidth]{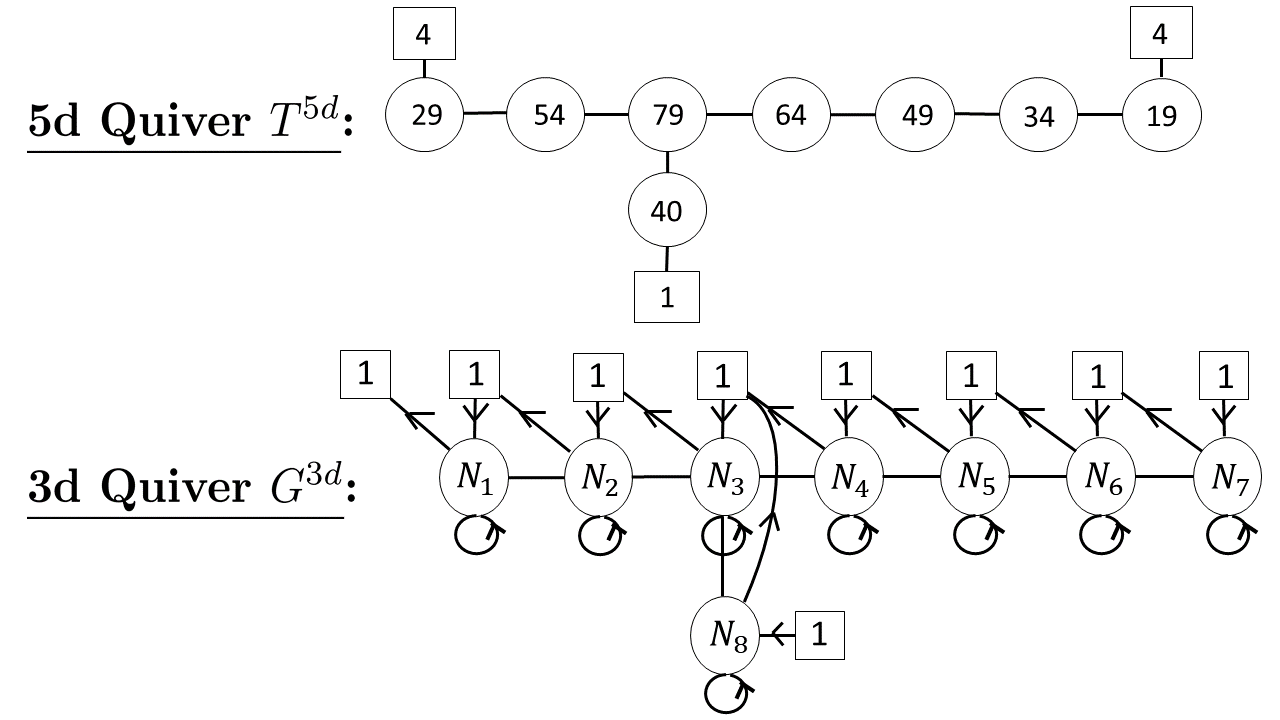}
        	\end{center}
        \caption{Sphere with 3 full $E_8$  punctures: 5d theory $T^{5d}$ and 3d theory $G^{3d}$ resulting from $\cW_{\cS}$.}
\label{E8Full}
\end{figure}

In the case of $E_8$, we take the set ${\cal W}_{\cal S}$ to be the following collection of 9 coweights:
\beq\label{funde8}
\begin{aligned}
& \omega_1 = -{w}^\vee_1 + 7\alpha^\vee_1 + 13\alpha^\vee_2 + 19\alpha^\vee_3 + 16\alpha^\vee_4  + 12\alpha^\vee_5 + 8\alpha^\vee_6 + 4\alpha^\vee_7 + 10\alpha^\vee_8\\
 & \omega_2 = -{w}^\vee_1 + 7\alpha^\vee_1 + 13\alpha^\vee_2 + 20\alpha^\vee_3 + 16\alpha^\vee_4  + 12\alpha^\vee_5 + 8\alpha^\vee_6 + 4\alpha^\vee_7 + 10\alpha^\vee_8\\
& \omega_3= -{w}^\vee_1 + 7\alpha^\vee_1 + 14\alpha^\vee_2 + 20\alpha^\vee_3 + 16\alpha^\vee_4  + 12\alpha^\vee_5 + 8\alpha^\vee_6 + 4\alpha^\vee_7 + 10\alpha^\vee_8\\
& \omega_4= -{w}^\vee_1 + 8\alpha^\vee_1 + 14\alpha^\vee_2 + 20\alpha^\vee_3 + 16\alpha^\vee_4  + 12\alpha^\vee_5 + 8\alpha^\vee_6 + 4\alpha^\vee_7 + 10\alpha^\vee_8\\
& \omega_5= -{w}^\vee_7 \\
& \omega_6 = -{w}^\vee_7 + \alpha^\vee_7 \\
 & \omega_7 = -{w}^\vee_7 + \alpha^\vee_6 + \alpha^\vee_7 \\
& \omega_8 = -{w}^\vee_7 + \alpha^\vee_5 + \alpha^\vee_6 + \alpha^\vee_7 \\
 & \omega_9= -{w}^\vee_8 
 \end{aligned}
\end{equation} 
Once again, one can check that these sets  are distinguished and have no common zeros in their Dynkin labels, so these are indeed full punctures of $E_n$.

The complex dimension of the Coulomb branch of $T^{5d}$ is
$$\sum_{a=1}^n d_a=\sum\limits_{\langle e_\gamma,\omega_i\rangle<0} \left\vert\langle e_\gamma,\omega_i\rangle\right\vert \; .
$$
For $E_6$, we therefore find (using either sum) that the  Coulomb branch dimension is 59. For $E_7$, we find that the Coulomb branch dimension is 63. For $E_8$, we find that the Coulomb branch dimension is 368.\\

In the CFT limit, when $m_s\rightarrow\infty$, the counting is done without multiplicity in the sum on the right-hand side; the Coulomb branch dimension therefore decreases and becomes equal to the number of positive roots of $E_n$; for $E_6$, this is 36. For $E_7$, this is 63. For $E_8$, this is 120. The Coulomb branch of the resulting theory $T^{4d}$ is the maximal nilpotent orbit of $E_n$. 
The 5d and 3d theories are shown in figures \ref{E6Full}, \ref{E7Full}, and \ref{E8Full}.

\subsubsection{$G_2$ Full Puncture}

Consider a $D_4$ theory with two D5 branes wrapping the non-compact 2-cycle on the central $D_4$ node, and a single D5 brane wrapping each non-compact 2-cycle on the external $D_4$ nodes. Add further the required number of compact D5 branes to ensure the net flux vanishes. Then, we impose the branes to be invariant under the $\mathbb{Z}_3$-outer automorphism action and fold, resulting in a $G_2$ quiver gauge theory. The theory can be described by the following set ${\cal W}_{\cal S}$ of 3 coweights:

\begin{figure}[h!]
	\begin{center}
		\includegraphics[width=0.95\textwidth]{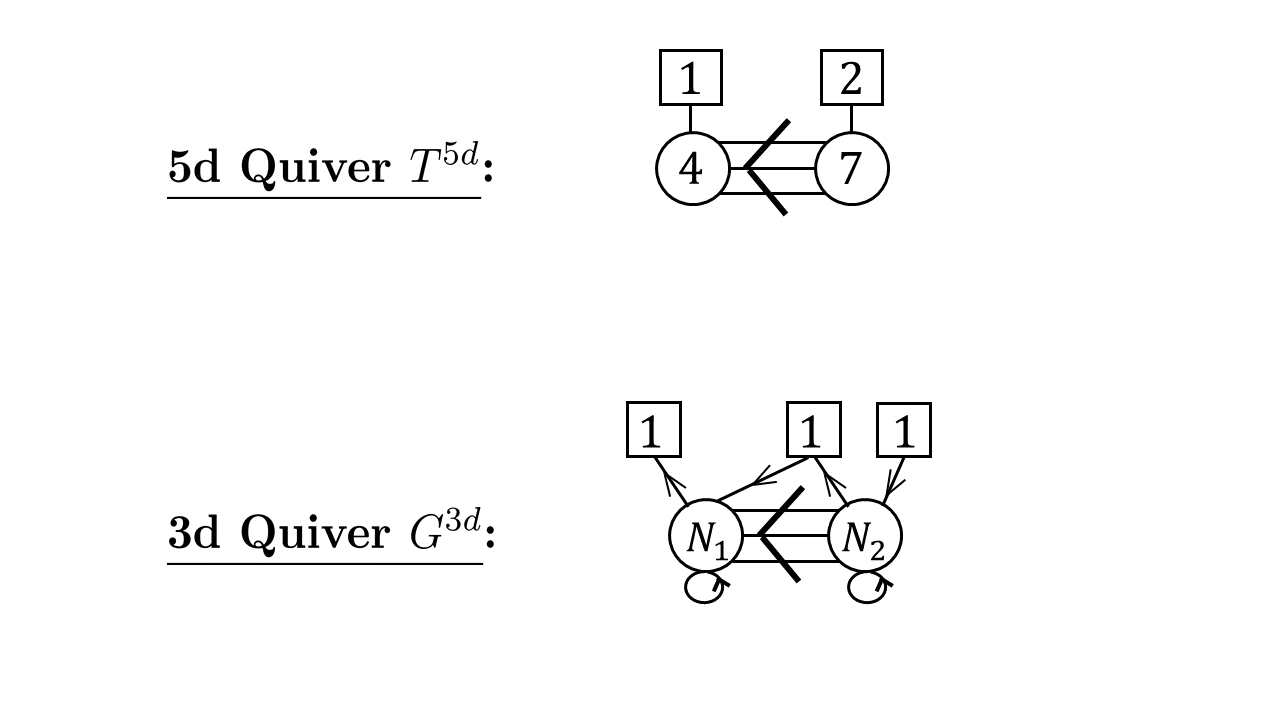}
        	\end{center}
        \caption{Sphere with 3 full $G_2$  punctures: 5d theory $T^{5d}$ and 3d theory $G^{3d}$ resulting from $\cW_{\cS}$.}
\label{G2Full}
\end{figure}

\beq\label{fundg2}
\begin{aligned}
 & \omega_1 = -{w}^\vee_1 + 4\alpha^\vee_1 + 6\alpha^\vee_2\\
 & \omega_2= -{w}^\vee_2 + \alpha^\vee_2\\
& \omega_3= -{w}^\vee_2 
\end{aligned}
\eeq 
After freezing the Coulomb moduli, the Dynkin labels of the coweights \eqref{fundg2} expanded in terms of fundamental coweights encode the precise matter content of the 3d theory:
\beq
 \begin{aligned}
\omega_1 &= [\phantom{-}1,\phantom{-}0]\\
\omega_2 &= [-1, \phantom{-}1]\\
\omega_{3} &= [\phantom{-}0,  -1]
\end{aligned}
\eeq
We obtain the 3d $\cN=2$ quiver gauge theory $G^{3d}$ shown in figure \ref{G2Full}. The deformed vertex operator that realizes the full puncture is the product $:\prod_{i=1}^{3} V^\vee_{\omega_i}(x_i):$, where
\beq
 \begin{aligned}\label{vertg2}
V^\vee_{\omega_1} (x) =&  : W_1^{-1}(x)  E_1(xv^{-1}) E_2(xv^{-2}q^{-1/2}) E_2(xv^{-2}q^{-3/2}) E_2(xv^{-2}q^{-5/2}) E_1(xv^{-3}q^{-1})\\
& \;\;\;E_1(xv^{-3}q^{-2}) E_2(xv^{-4}q^{-3/2})  E_2(xv^{-4}q^{-5/2})  E_2(xv^{-4}q^{-7/2}) E_1(xv^{-5}q^{-3}): \\
V^\vee_{\omega_2} (x)=&  : W_2^{-1}(x)  E_2(xq^{-1/2}): \\
V^\vee_{\omega_3} (x)=& : W_2^{-1}(x) :
\end{aligned}
\eeq
The above ``fundamental coweight" and ``simple coroot" vertex operators were defined in section \ref{ssec:Toda}; the expression is a refinement of the relation \eqref{fundg2}. The dependence on the $v$ and $q$ factors encodes the value of the Coulomb moduli at the triality point. Namely, let $v^{\#_{a,I}}\, q^{\#'_{a,I}}$ be the various $v$ and $q$ factors  appearing  in the $E_a$ operators of \eqref{vertg2}. Then, the Coulomb moduli of the 5d gauge theory that truncate the partition function to the $G_2$ deformed conformal block are given by:
$$e_{a,I}=f_{I}\, t^{N_{a,I}} \,v^{\#_{a,I}}\, q^{\#'_{a,I}} \, v^{2-a}\, q^{(a-1)/2}\, , \qquad \;a=1, 2 \; .$$
Recall that in our notation, $a=1$ is gauge node designating the short root, while $a=2$ designates the long root.\\

The Coulomb branch of the 5d theory has complex dimension:
$$\sum_{a=1}^2 d_a=11=\sum\limits_{\langle e_\gamma,\omega_i\rangle<0} \left\vert\langle e_\gamma,\omega_i\rangle\right\vert \; ,
$$ 
with $d_a$ the ranks of the 2 gauge groups. 
In the right-hand sum, one counts all positive roots that have a negative inner product with at least one of the coweights. This is also the number of supersymmetric vacua (or equivalently, integration contours) of the 3d theory, and the number of parameters needed to  specify the 3-point of the deformed $\cW_{q,t}(G_2)$ algebra.\\

In the CFT limit, when $m_s\rightarrow\infty$, the counting is done without multiplicity. The Coulomb branch dimension therefore decreases and becomes equal to the number of positive roots of $G_2$, which is 6. The Coulomb branch of the resulting theory $T^{4d}$ is the maximal nilpotent orbit of $G_2$, with Bala-Carter label $G_2$. Because the defect is polarized, its Coulomb branch must be in the image of the Spaltenstein map. In our case, the full puncture Coulomb branch is the image of the orbit denoted by $\varnothing$. This pre-image Bala--Carter label $\varnothing$ is identified at once by acting on  $\cW_{\cS}$ with the Weyl group and noticing the set never has any common zeros in the Dynkin labels of the different coweights.

\subsubsection{$F_4$ Full Puncture}

We start with an $E_6$ singluarity with a collection of non-compact D5 branes and fold to $F_4$ using the $\mathbb{Z}_2$ outer automorphism action. The resulting set ${\cal W}_{\cal S}$ we consider is a collection of 5 coweights:

\begin{figure}[h!]
	\begin{center}
		\includegraphics[width=0.95\textwidth]{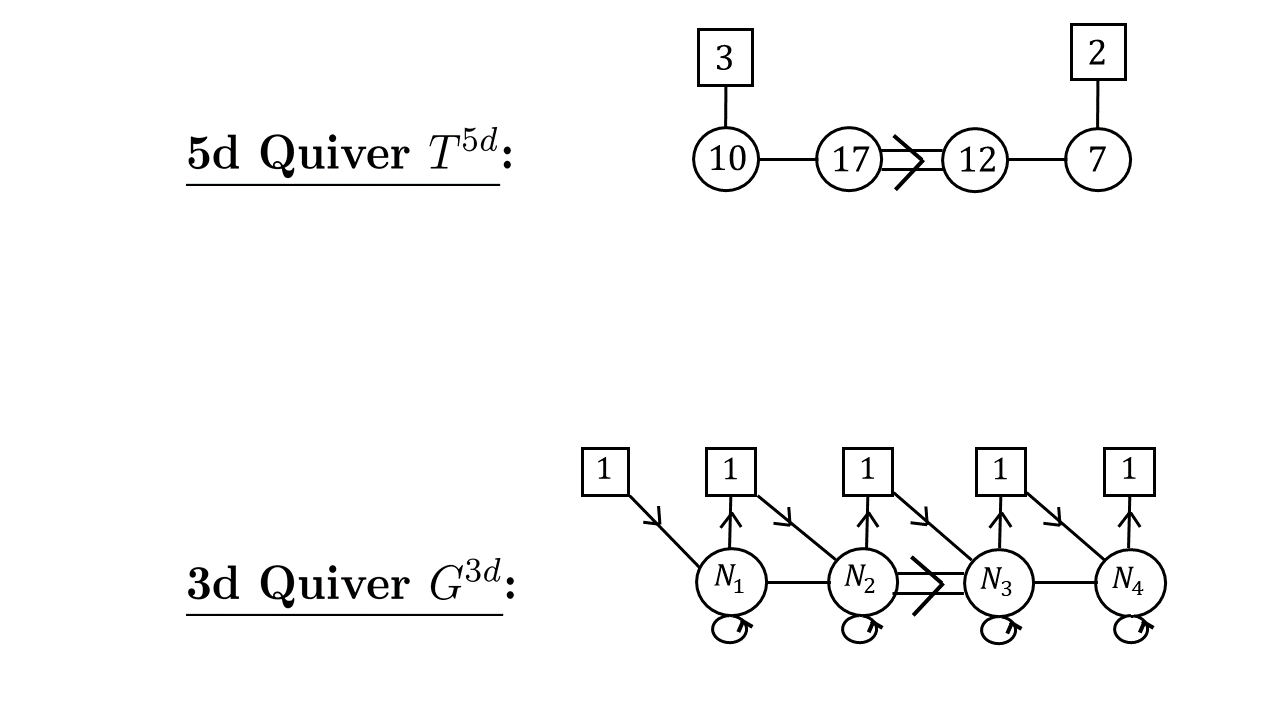}
        	\end{center}
        \caption{Sphere with 3 full $F_4$  punctures: 5d theory $T^{5d}$ and 3d theory $G^{3d}$ resulting from $\cW_{\cS}$.}
\label{F4Full}
\end{figure}

\beq\label{fundf4}
\begin{aligned}
 & \omega_1 = -{w}^\vee_4 + 4\alpha^\vee_1 + 8\alpha^\vee_2 + 6\alpha^\vee_3 + 4\alpha^\vee_4\\
 & \omega_2 = -{w}^\vee_4 + 4\alpha^\vee_1 + 8\alpha^\vee_2 + 6\alpha^\vee_3 + 3\alpha^\vee_4\\
 & \omega_3 = -{w}^\vee_1 + \alpha^\vee_1 + \alpha^\vee_2\\
 & \omega_4 = -{w}^\vee_1 + \alpha^\vee_1\\
 & \omega_5 = -{w}^\vee_1
\end{aligned}
\eeq 

After freezing the Coulomb moduli, the  Dynkin labels of the coweights \eqref{fundf4} expanded in terms of fundamental coweights encode the precise matter content of the 3d theory:
\beq
 \begin{aligned}
\omega_1 &= [\phantom{-}0,\phantom{-}0, \phantom{-}0, \phantom{-}1]\\
\omega_2 &= [\phantom{-}0, \phantom{-}0, \phantom{-}1, -1]\\
\omega_{3} &= [\phantom{-}0, \phantom{-}1, -1, \phantom{-}0]\\
\omega_{4} &= [\phantom{-}1, -1,\phantom{-}0, \phantom{-}0]\\
\omega_{5} &= [-1, \phantom{-}0, \phantom{-}0, \phantom{-}0]
\end{aligned}
\eeq
We obtain the  quiver gauge theory $G^{3d}$ shown in figure \ref{F4Full}.

The deformed vertex operator that realizes the full puncture is the product $:\prod_{i=1}^{5} V^\vee_{\omega_i}(x_i):$, where
\beq
 \begin{aligned}\label{vertf4}
V^\vee_{\omega_1} (x) =&  : W_4^{-1}(x)  E_4(xv^{-1}) E_3(xv^{-2}) E_2(xv^{-3}q^{-1/2}) E_2(xv^{-3}q^{-3/2}) E_1(xv^{-4}q^{-1})\\
& \;\;\;E_3(xv^{-4}q^{-1})  E_1(xv^{-4}q^{-2})  E_4(xv^{-5}q^{-1})  E_2(xv^{-5}q^{-3/2}) E_2(xv^{-5}q^{-5/2}) E_3(xv^{-6}q^{-1})\\
&\;\;\; E_3(xv^{-6}q^{-2}) E_2(xv^{-7}q^{-3/2}) E_4(xv^{-7}q^{-2}) E_2(xv^{-7}q^{-5/2}) E_1(xv^{-8}q^{-2}) E_1(xv^{-8}q^{-3})\\  & \;\;\; E_3(xv^{-8}q^{-3}) E_2(xv^{-9}q^{-5/2})  E_2(xv^{-9}q^{-7/2}) E_3(xv^{-10}q^{-3}) E_4(xv^{-11}q^{-3}): \\
V^\vee_{\omega_2} (x)=&  : W_4^{-1}(x)  E_4(xv^{-1}) E_3(xv^{-2}) E_2(xv^{-3}q^{-1/2}) E_2(xv^{-3}q^{-3/2}) E_1(xv^{-4}q^{-1})\\
&\;\;\; E_3(xv^{-4}q^{-1})  E_1(xv^{-4}q^{-2})  E_4(xv^{-5}q^{-1})  E_2(xv^{-5}q^{-3/2}) E_2(xv^{-5}q^{-5/2}) E_3(xv^{-6}q^{-1})\\
&\;\;\; E_3(xv^{-6}q^{-2}) E_2(xv^{-7}q^{-3/2}) E_4(xv^{-7}q^{-2}) E_2(xv^{-7}q^{-5/2}) E_1(xv^{-8}q^{-2}) E_1(xv^{-8}q^{-3})\\ &\;\;\; E_3(xv^{-8}q^{-3}) E_2(xv^{-9}q^{-5/2})  E_2(xv^{-9}q^{-7/2}) E_3(xv^{-10}q^{-3}): \\
V^\vee_{\omega_3} (x)=& : W_1^{-1}(x) E_1(xv^{-4}q^{-1}) E_2(xv^{-5}q^{-3/2}):\\
V^\vee_{\omega_4} (x)=& : W_1^{-1}(x)  E_1(xv^{-4}q^{-1}):\\
V^\vee_{\omega_{5}}(x)=& : W_1^{-1}(x):
\end{aligned}
\eeq
The above expression is a refinement of the relation \eqref{fundf4}. The dependence on the $v$ and $q$ factors encodes the value of the Coulomb moduli at the triality point. Namely, let $v^{\#_{a,I}}\, q^{\#'_{a,I}}$ be the various $v$ and $q$ factors  appearing  in the $E_a$ operators of \eqref{vertf4}. Then, the Coulomb moduli of the 5d gauge theory that truncate the partition function to the $F_4$ deformed conformal block are given by:
\begin{align}\nonumber
e_{a,I} &=f_{I}\, t^{N_{a,I}} \,v^{\#_{a,I}}\, q^{\#'_{a,I}} \, v^{5-a}\, q^{(3-a)/2} \, , \qquad \;a=1,\ldots, 4
\end{align}
In our notation, $a=1,2$ designate the long roots, while $a=3, 4$ designate the short roots.\\

The Coulomb branch of the 5d theory has complex dimension:
$$\sum_{a=1}^4 d_a=46=\sum\limits_{\langle e_\gamma,\omega_i\rangle<0} \left\vert\langle e_\gamma,\omega_i\rangle\right\vert \; ,
$$
with $d_a$ the ranks of the 4 gauge groups. 
In the right-hand sum, one counts all positive roots that have a negative inner product with at least one of the coweights. This is also the number of supersymmetric vacua (or equivalently, integration contours) of the 3d theory, and the number of parameters needed to  specify the 3-point of the deformed $\cW_{q,t}(F_4)$ algebra.\\

In the CFT limit, when $m_s\rightarrow\infty$, the counting is done without multiplicity. The Coulomb branch dimension therefore decreases and becomes equal to the number of positive roots of $F_4$, which is 24. The Coulomb branch of the resulting theory $T^{4d}$ is the maximal nilpotent orbit of $F_4$, with Bala--Carter label $F_4$. Because the defect is polarized, its Coulomb branch must be in the image of the Spaltenstein map. In our case, the full puncture Coulomb branch is the image of the orbit denoted by $\varnothing$. This pre-image Bala--Carter label $\varnothing$ is identified at once by acting on  $\cW_{\cS}$ with the Weyl group and noticing the set never has any common zeros in the Dynkin labels of the different coweights.

\subsubsection{$B_n$ Full Puncture}

For $B_n$, we take the set ${\cal W}_{\cal S}$ to be the following collection of $n+1$ coweights:

\begin{figure}[h!]
	\begin{center}
		\includegraphics[width=0.95\textwidth]{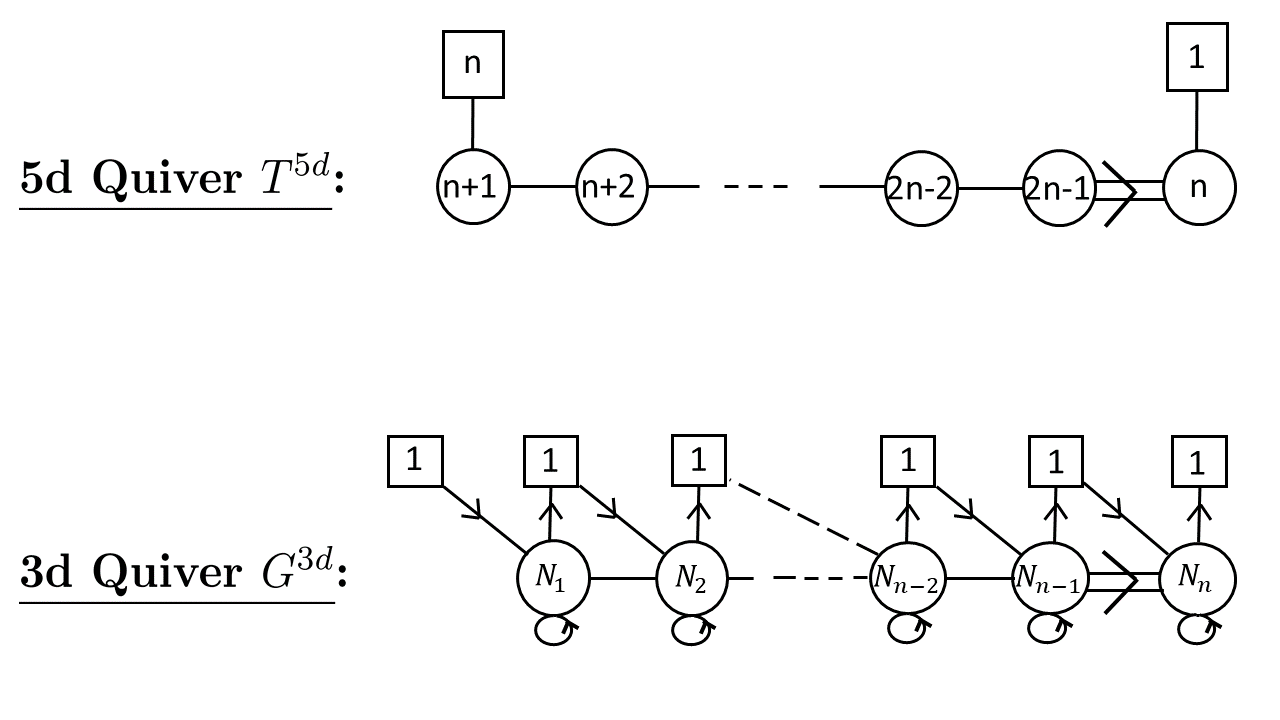}
        	\end{center}
        \caption{Sphere with 3 full $B_n$  punctures: 5d theory $T^{5d}$ and 3d theory $G^{3d}$ resulting from $\cW_{\cS}$.}
\label{BnFull}
\end{figure}

\beq\label{fundBn}
\begin{aligned}
&\omega_1= - w_1^\vee + \alpha_1^\vee +\alpha_{2}^\vee + \ldots  + \alpha_{n-1}^\vee + \alpha_{n}^\vee\\
& \omega_i = \omega_{i-1} + \alpha_{n-i+1}^\vee\; , \qquad i=2, \ldots n\\
 & \omega_{n+1} = -{w}_n^\vee 
\end{aligned}
\eeq 
After freezing the Coulomb moduli, the  Dynkin labels of the coweights \eqref{fundBn} expanded in terms of fundamental coweights encode the precise matter content of the 3d theory:

\beq
 \begin{aligned}
\omega_1 &= [\phantom{-}0,\phantom{-}0, \phantom{-}0, \ldots,  \phantom{-}0, -1,  \phantom{-}1]\\
\omega_2 &= [\phantom{-}0,\phantom{-}0, \phantom{-}0, \ldots,  -1, \phantom{-}1,  \phantom{-}0]\\
&  \vdots\\ 
\omega_{n-2} &= [\phantom{-}0, -1,\phantom{-}1, \ldots,  \phantom{-}0, \phantom{-}0,\phantom{-}0]\\
\omega_{n-2} &= [\phantom{-}0, -1,\phantom{-}1, \ldots,  \phantom{-}0, \phantom{-}0,\phantom{-}0]\\
\omega_{n-1} &= [-1,\phantom{-}1, \phantom{-}0, \ldots,  \phantom{-}0, \phantom{-}0,\phantom{-}0]\\
\omega_{n+1} &= [\phantom{-}0,\phantom{-}0, \phantom{-}0, \ldots,  \phantom{-}0, \phantom{-}0, -1]
\end{aligned}
\eeq
We obtain the  quiver gauge theory $G^{3d}$ shown in figure \ref{BnFull}.

The deformed vertex operator that realizes the full puncture is the product $:\prod_{i=1}^{n+1} V^\vee_{\omega_i}(x_i):$, where
\beq
 \begin{aligned}\label{vertbn}
V^\vee_{\omega_1} (x) &= : W_1^{-1}(x) E_1(xv^{-1}q^{-1/2}) E_2(x v^{-2}q^{-1}) \ldots  E_{n-1}(xv^{1-n}q^{(-n+1)/2}) E_{n}(xv^{-n}q^{(-n+1)/2}):\\
V^\vee_{\omega_i} (x) &= :V^\vee_{\omega_{i-1}}(x) E_{n-i+1}(xv^{-n-i+1}q^{(-n-i+3)/2}) : \qquad\; i=2, \ldots, n\\
V^\vee_{\omega_{n+1}} (x) &= : W_n^{-1}(x) :
\end{aligned}
\eeq
The above expression is a refinement of the relation \eqref{fundBn}. The dependence on the $v$ and $q$ factors encodes the value of the Coulomb moduli at the triality point. Namely, let $v^{\#_{a,I}}\, q^{\#'_{a,I}}$ be the various $v$ and $q$ factors  appearing  in the $E_a$ operators of \eqref{vertbn}. Then, the Coulomb moduli of the 5d gauge theory that truncate the partition function to the $B_n$ deformed conformal block are given by:
\begin{align}\nonumber
e_{a,I} &=f_{I}\, t^{N_{a,I}} \,v^{\#_{a,I}}\, q^{\#'_{a,I}} \, v^{2-a}  \, q^{(2-a)/2} \, , \qquad \; a=1, \ldots, n
\end{align}
In our notation, $a=1,\ldots, n-1$ designate the long roots, while $a=n$ designates the short root.\\

The Coulomb branch of the 5d theory has complex dimension:
$$\sum_{a=1}^n d_a=\frac{n(3n-1)}{2}=\sum\limits_{\langle e_\gamma,\omega_i\rangle<0} \left\vert\langle e_\gamma,\omega_i\rangle\right\vert \; ,
$$
with $d_a$ the ranks of the $n$ gauge groups. 
In the right-hand sum, one counts all positive roots that have a negative inner product with at least one of the coweights. This is also the number of supersymmetric vacua (or equivalently, integration contours) of the 3d theory, and the number of parameters needed to  specify the 3-point of the deformed $\cW_{q,t}(B_n)$ algebra.\\

In the CFT limit, when $m_s\rightarrow\infty$, the counting is done without multiplicity. The Coulomb branch dimension therefore decreases and becomes equal to the number of positive roots of $B_n$, which is $n^2$. The Coulomb branch of the resulting theory $T^{4d}$ is the maximal nilpotent orbit of $B_n$, with Bala-Carter label $B_n$. Because the defect is polarized, its Coulomb branch must be in the image of the Spaltenstein map. In our case, the full puncture Coulomb branch is the image of the orbit denoted by $\varnothing$. This pre-image Bala--Carter label $\varnothing$ is identified at once by acting on  $\cW_{\cS}$ with the Weyl group and noticing the set never has any common zeros in the Dynkin labels of the different coweights.

\subsubsection{$C_n$ Full Puncture}

For $C_n$, we take the set ${\cal W}_{\cal S}$ to be the following collection of $n+1$ coweights:

\begin{figure}[h!]
	\begin{center}
		\includegraphics[width=0.95\textwidth]{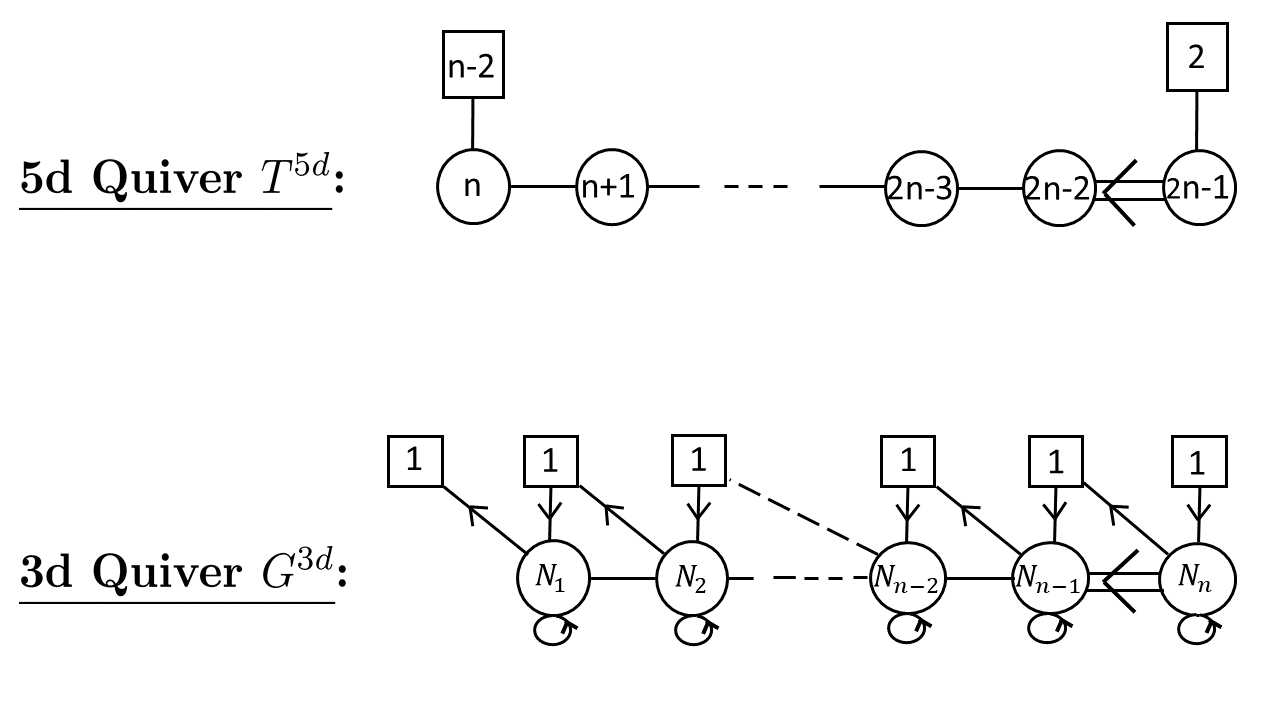}
        	\end{center}
        \caption{Sphere with 3 full $C_n$  punctures: 5d theory $T^{5d}$ and 3d theory $G^{3d}$ resulting from $\cW_{\cS}$.}
\label{CnFull}
\end{figure}

\beq\label{fundcn}
\begin{aligned}
&\omega_1= - w_1^\vee + \alpha_1^\vee +\alpha_{2}^\vee + \ldots  + \alpha_{n-2}^\vee + 2\alpha_{n-1}^\vee + 2\alpha_{n}^\vee\\
& \omega_i = \omega_{i-1} + \alpha_{n-i}^\vee\; , \qquad i=2, \ldots n-1\\
& \omega_{n} = -{w}_n^\vee + \alpha_{n}^\vee\\ 
 & \omega_{n+1} = -{w}_n^\vee 
\end{aligned}
\eeq 
After freezing the Coulomb moduli, the  Dynkin labels of the coweights \eqref{fundcn} expanded in terms of fundamental coweights encode the precise matter content of the 3d theory:

\beq
 \begin{aligned}
\omega_1 &= [\phantom{-}0,\phantom{-}0,  \ldots, \phantom{-}0, -1,  \phantom{-}1,  \phantom{-}0]\\
\omega_2 &= [\phantom{-}0, \phantom{-}0, \ldots, -1,  \phantom{-}1, \phantom{-}0,  \phantom{-}0]\\
&  \vdots\\ 
\omega_{n-2} &= [-1, \phantom{-}1, \ldots, \phantom{-}0,  \phantom{-}0, \phantom{-}0,  \phantom{-}0]\\
\omega_{n-1} &= [\phantom{-}1, \phantom{-}0, \ldots, \phantom{-}0,  \phantom{-}0, \phantom{-}0,  \phantom{-}0]\\
\omega_{n} &= [\phantom{-}0,\phantom{-}0,  \ldots, \phantom{-}0, \phantom{-}0, -1, \phantom{-}1]\\
\omega_{n+1} &= [\phantom{-}0,\phantom{-}0,  \ldots, \phantom{-}0, \phantom{-}0,  \phantom{-}0, -1]
\end{aligned}
\eeq
We obtain the quiver gauge theory $G^{3d}$ shown in figure \ref{CnFull}.

The deformed vertex operator that realizes the full puncture is the product $:\prod_{i=1}^{n+1} V^\vee_{\omega_i}(x_i):$, where
\beq
 \begin{aligned}\label{vertcn}
V^\vee_{\omega_1} (x) &= : W_1^{-1}(x) E_1(xv^{-1}) E_2(x v^{-2}) \ldots  E_{n-1}(xv^{-n}) E_{n}(xv^{-n})\\
&\;\;\;\; E_{n}(xv^{1-n}q^{-1}) E_{n-1}(xv^{-n-1}q^{-1}):\\
V^\vee_{\omega_i} (x) &= :V^\vee_{\omega_{i-1}}(x)E_{n-i}(xv^{-n-i}q^{-1}) : \qquad\; i=2, \ldots, n-1\\
V^\vee_{\omega_{n}} (x) &= : W_n^{-1}(x) E_n(xv^{2}):\\
V^\vee_{\omega_{n+1}} (x) &= : W_n^{-1}(x) :
\end{aligned}
\eeq
The above expression is a refinement of the relation \eqref{fundcn}. The dependence on the $v$ and $q$ factors encodes the value of the Coulomb moduli at the triality point. Namely, let $v^{\#_{a,I}}\, q^{\#'_{a,I}}$ be the various $v$ and $q$ factors  appearing  in the $E_a$ operators of \eqref{vertcn}. Then, the Coulomb moduli of the 5d gauge theory that truncate the partition function to the $C_n$ deformed conformal block are given by:
\begin{align}\nonumber
e_{a,I} &=f_{I}\, t^{N_{a,I}} \,v^{\#_{a,I}}\, q^{\#'_{a,I}} \, v^{2-a} \, , \qquad \;a=1, \ldots, n
\end{align}
In our notation, $a=1,\ldots, n-1$ designate the short roots, while $a=n$ designates the long root.\\

The Coulomb branch of the 5d theory has complex dimension:
$$\sum_{a=1}^n d_a=\frac{n(3n-1)}{2}=\sum\limits_{\langle e_\gamma,\omega_i\rangle<0} \left\vert\langle e_\gamma,\omega_i\rangle\right\vert \; ,
$$
with $d_a$ the ranks of the $n$ gauge groups. Note it is the same as for the $B_n$ full puncture.
In the right-hand sum, one counts all positive roots that have a negative inner product with at least one of the coweights. This is also the number of supersymmetric vacua (or equivalently, integration contours) of the 3d theory, and the number of parameters needed to  specify the 3-point of the deformed $\cW_{q,t}(C_n)$ algebra.\\

In the CFT limit, when $m_s\rightarrow\infty$, the counting is done without multiplicity. The Coulomb branch dimension therefore decreases and becomes equal to the number of positive roots of $C_n$, which is $n^2$. The Coulomb branch of the resulting theory $T^{4d}$ is the maximal nilpotent orbit of $C_n$, with Bala-Carter label $C_n$. Because the defect is polarized, its Coulomb branch must be in the image of the Spaltenstein map. In our case, the full puncture Coulomb branch is the image of the orbit denoted by $\varnothing$. This pre-image Bala--Carter label $\varnothing$ is identified at once by acting on  $\cW_{\cS}$ with the Weyl group and noticing the set never has any common zeros in the Dynkin labels of the different coweights.

\subsection{All Punctures of the $G_2$ Little String and CFT Limit}
\label{ssec:g2list}

In this section, we present a classification of defects of the $\fg=G_2$ little string theory, according to their conjectured Bala--Carter label in the CFT limit. The defects are generated by D5 branes wrapping non-compact 2-cycles of a resolved $D_4$ singularity; these D5 branes must remain invariant under the $\mathbb{Z}_3$ outer automorphism group action. 

The resulting defects after folding are labeled by coweights of $G_2$, which are  weights of $^L G_2=G_2$. We find that there are at least two ``distinct" unpolarized defects, both generated by the null coweight $[\phantom{-}0,\phantom{-}0]$, considered in each of the two fundamental representations. The Coulomb branch of each featured quiver gauge theory $T_{5d}$ flows to a nilpotent orbit of $G_2$ in the CFT limit.\\

\begin{figure}[h]
        \centering
        \begin{subfigure}[b]{1.02\textwidth}
                \includegraphics[width=1.00\textwidth]{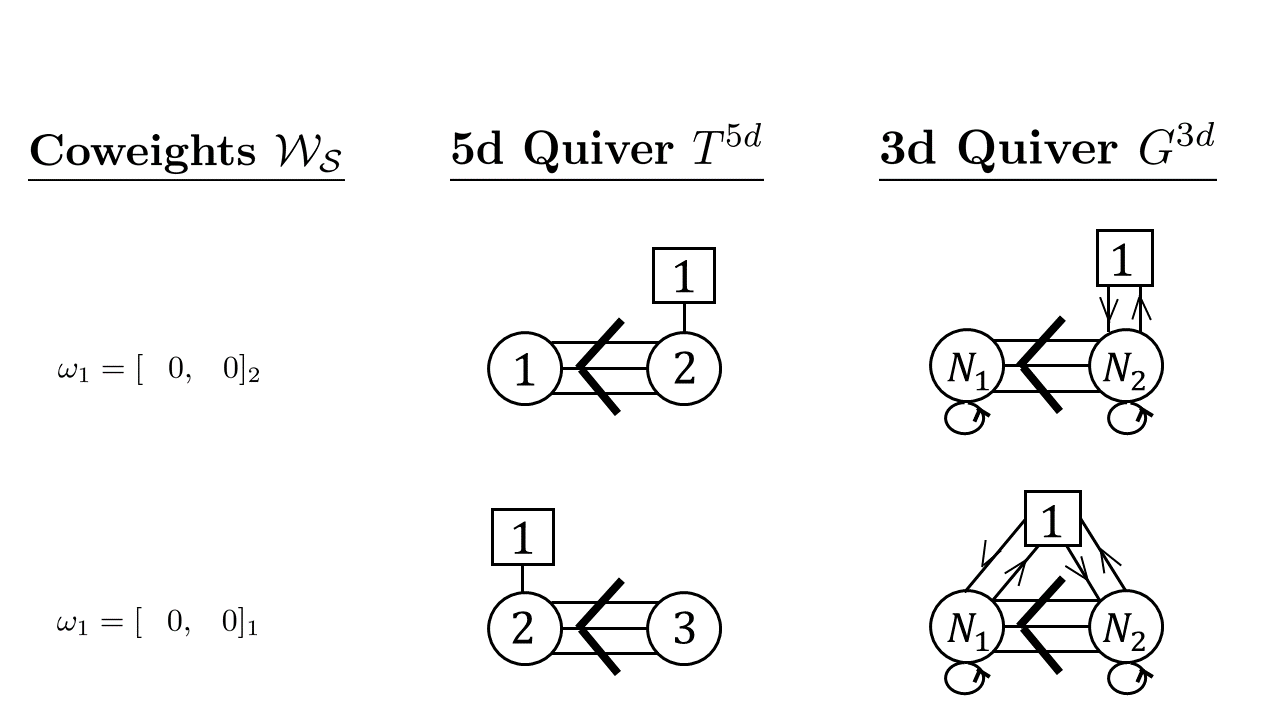}
        \end{subfigure}%
          
        \begin{subfigure}[b]{1.02\textwidth}
                \includegraphics[width=1.00\textwidth]{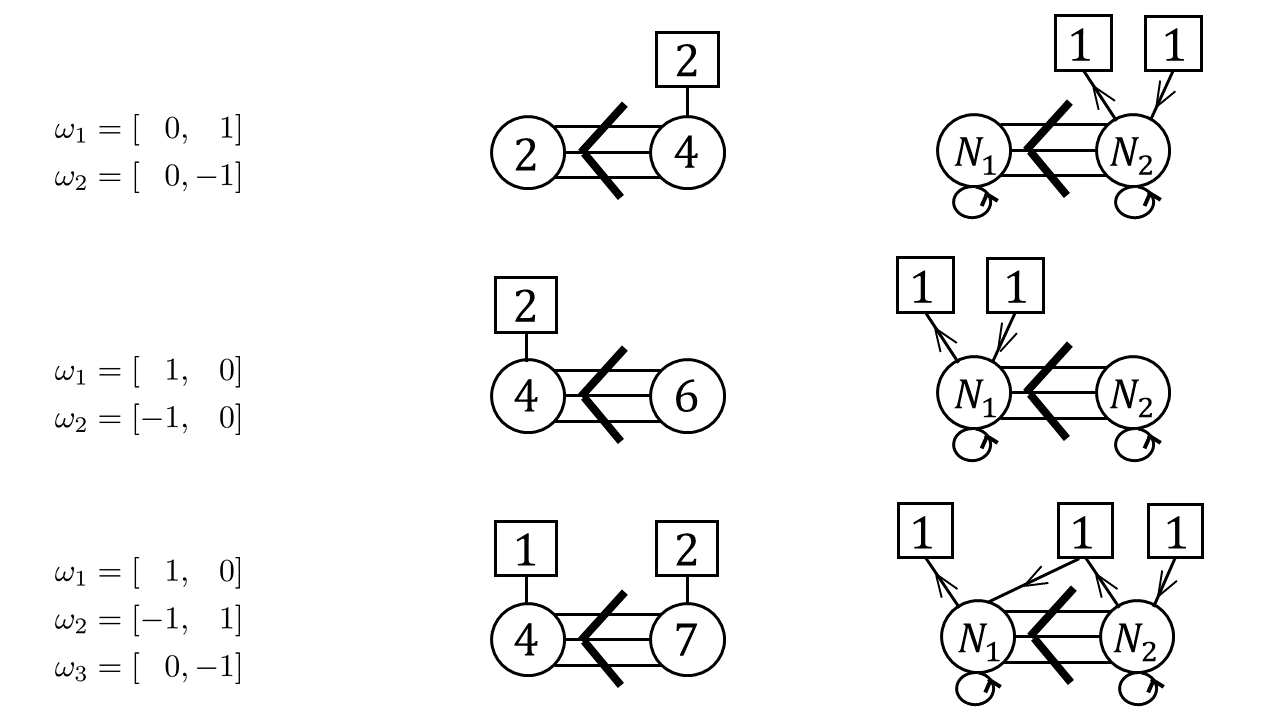}
        \end{subfigure}
       
        \caption{Distinct defects of the $G_2$ little string.}\label{G2plots}
\end{figure}

Furthermore, we illustrate $G_2$ triality, by considering the little string on a sphere with two full punctures and one of the punctures of figure \ref{G2plots}.

\subsubsection{The $\varnothing$ Orbit}

The first puncture we study was introduced in the previous section, so we will be brief: it is realized with three coweights, in the following set $\cW_{\cS}$:
\begin{align*}
& \omega_1 = -{w}^\vee_1 + 4\alpha^\vee_1 + 6\alpha^\vee_2= [\phantom{-}1,\phantom{-}0]\\
& \omega_2= -{w}^\vee_2 + \alpha^\vee_2= [-1, \phantom{-}1]\\
& \omega_3= -{w}^\vee_2 = [\phantom{-}0,  -1]
\end{align*}
All the elements of $\cW_{\cS}$ are in the Weyl group orbit of the fundamental representation they belong in, so the defect is polarized. 
Moreover, the set is distinguished, as can be checked by acting on all elements of $\cW_{\cS}$ simultaneously with the Weyl group of $G_2$. The fundamental matter content of $T^{5d}$ is:
\begin{align*}
\prod_{1\leq I\leq d_1} N_{\emptyset \mu^{1}_{I}}( v^2\, f_{1}/e_{1, I};q)\prod_{1\leq I\leq d_2} N_{\emptyset \mu^{2}_{I}}( v_3^2\, f_{2}/e_{2, I};q^3) \prod_{1\leq I\leq d_2} N_{\emptyset \mu^{2}_{I}}( v_3^2\, f_{3}/e_{2, I};q^3).
\end{align*}
The truncation of  $T^{5d}$'s  partition function to a 3d theory's partition function is achieved by setting: 
\begin{align*}
e_{1,1} = q^{-0} v^{-0} t^{N_{1,1}}\, f_1\qquad &e_{2,1} = q^{-0} v^{-2} t^{N_{2,1}}\, f_1\\
e_{1,2} = q^{-1} v^{-2} t^{N_{1,2}}\, f_1\qquad &e_{2,2} = q^{-1} v^{-2} t^{N_{2,2}}\, f_1\\
e_{1,3} = q^{-2} v^{-2} t^{N_{1,3}}\, f_1\qquad &e_{2,3} = q^{-2} v^{-2} t^{N_{2,3}}\, f_1\\
e_{1,4} = q^{-3} v^{-4} t^{N_{1,4}}\, f_1\qquad &e_{2,4} = q^{-1} v^{-4} t^{N_{2,4}}\, f_1\\
                                          &e_{2,5} = q^{-2} v^{-4} t^{N_{2,5}}\, f_1\\
                                          &e_{2,6} = q^{-3} v^{-4} t^{N_{2,6}}\, f_1\\
                                          &e_{2,7} = q^{-0} v^{-0} t^{N_{2,7}}\, f_2
\end{align*}
The resulting 3d partition function has fundamental matter content given exactly by \eqref{hyper}. 

We now turn to the CFT limit: there are no common zeros in the distinguished set $\cW_{\cS}$, so the  Bala--Carter label associated to this polarized defect is $\varnothing$. 
The Coulomb branch is then given by the Spaltenstein dual of this orbit, which is $G_2$, of complex dimension 6. 
This is confirmed by the fact that all 6 positive roots have a negative inner product with at least one of the weights. The momentum of the associated vertex operator in $G_2$-Toda theory is $\beta=\sum_{i=1}^{3}\beta_i \, \omega_i$.

\subsubsection{The $A_1$ Orbit}

Consider the two coweights in the following set $\cW_{\cS}$:
\begin{align*}
& \omega_1 = -{w}^\vee_2 + 2\alpha^\vee_1 + 4\alpha^\vee_2= [\phantom{-}0,\phantom{-}1]\\
& \omega_2= -{w}^\vee_2 = [\phantom{-}0,  -1]
\end{align*} 
All the elements of $\cW_{\cS}$ are in the Weyl group orbit of the fundamental representation they belong in, so the defect is polarized. 
Moreover, the set is distinguished, as can be checked easily. The fundamental matter content of $T^{5d}$ is:
\begin{align*}
\prod_{1\leq I\leq d_2} N_{\emptyset \mu^{2}_{I}}( v_3^2\, f_{1}/e_{2, I};q^3) \prod_{1\leq I\leq d_2} N_{\emptyset \mu^{2}_{I}}( v_3^2\, f_{2}/e_{2, I};q^3).
\end{align*}
The truncation of  $T^{5d}$'s  partition function to a 3d theory's partition function is achieved by setting:
\begin{align*}
e_{1,1} = q^{-0} v^{-0} t^{N_{1,1}}\, f_1\qquad &e_{2,1} = q^{-0} v^{-0} t^{N_{2,1}}\, f_1\\
e_{1,2} = q^{-1} v^{-2} t^{N_{1,2}}\, f_1\qquad &e_{2,2} = q^{-0} v^{-2} t^{N_{2,2}}\, f_1\\                      &e_{2,3} = q^{-1} v^{-2} t^{N_{2,3}}\, f_1\\
&e_{2,4} = q^{-1} v^{-4} t^{N_{2,4}}\, f_1
\end{align*}
The resulting 3d partition function has fundamental matter content given exactly by \eqref{hyper}. 

We now turn to the CFT limit: there is a common zero in the first Dynkin label of the distinguished set $\cW_{\cS}$, so the  Bala--Carter label associated to this polarized defect is $A_1$. 
The Coulomb branch  is then given by the Spaltenstein dual of this orbit, which is $G_2(a_1)$, of complex dimension 5. 
This is confirmed by the fact that all but the positive (simple) root $\alpha_1$ have a negative inner product with $\omega_2$, and $\omega_1$  does not have a negative inner product with $\alpha_1$ either, so one of the six positive roots is not counted. The momentum of the associated vertex operator in $G_2$-Toda theory is $\beta=\sum_{i=1}^{2}\beta_i \, \omega_i$. Note that this defect characterizes a level 1 null state of $G_2$-Toda:
\[
\langle\beta , \alpha_1\rangle =0
\]

\subsubsection{The $A_{1_s}$ Orbit}

Consider the two coweights in the following set $\cW_{\cS}$:
\begin{align*}
& \omega_1 = -{w}^\vee_1 + 4\alpha^\vee_1 + 6\alpha^\vee_2= [\phantom{-}1,\phantom{-}0]\\
& \omega_2= -{w}^\vee_1 = [-1,\phantom{-}0]
\end{align*} 
All the elements of $\cW_{\cS}$ are in the Weyl group orbit of the fundamental representation they belong in, so the defect is polarized. 
Moreover, the set is distinguished, as can be checked easily. The fundamental matter content of $T^{5d}$ is:
\begin{align*}
\prod_{1\leq I\leq d_1} N_{\emptyset \mu^{1}_{I}}( v^2\, f_{1}/e_{1, I};q) \prod_{1\leq I\leq d_1} N_{\emptyset \mu^{1}_{I}}( v^2\, f_{2}/e_{1, I};q).
\end{align*}
The truncation of  $T^{5d}$'s  partition function to a 3d theory's partition function is achieved by setting:
\begin{align*}
e_{1,1} = q^{-0} v^{-0} t^{N_{1,1}}\, f_1\qquad &e_{2,1} = q^{-0} v^{-2} t^{N_{2,1}}\, f_1\\
e_{1,2} = q^{-1} v^{-2} t^{N_{1,2}}\, f_1\qquad &e_{2,2} = q^{-1} v^{-2} t^{N_{2,2}}\, f_1\\
e_{1,3} = q^{-2} v^{-2} t^{N_{1,3}}\, f_1\qquad &e_{2,3} = q^{-2} v^{-2} t^{N_{2,3}}\, f_1\\
e_{1,4} = q^{-3} v^{-4} t^{N_{1,4}}\, f_1\qquad &e_{2,4} = q^{-1} v^{-4} t^{N_{2,4}}\, f_1\\ &e_{2,5} = q^{-2} v^{-4} t^{N_{2,5}}\, f_1\\
&e_{2,6} = q^{-3} v^{-4} t^{N_{2,6}}\, f_1
\end{align*}
The resulting 3d partition function has fundamental matter content given exactly by \eqref{hyper}. 

We now turn to the CFT limit: there is a common zero in the second Dynkin label of the distinguished set $\cW_{\cS}$, so the  Bala--Carter label associated to this polarized defect is $A_{1,s}$. It is a distinct label from  $A_{1}$ in the previous example, since we must distinguish between the short and the long root.
The Coulomb branch is then given by the Spaltenstein dual of this orbit, which is $G_2(a_1)$, of complex dimension 5; this is the same as in the $A_{1}$ case.
This is confirmed by the fact that all but the positive (simple) root $\alpha_2$ have a negative inner product with $\omega_2$, and $\omega_1$  does not have a negative inner product with $\alpha_2$ either, so one of the six positive roots is not counted. The momentum of the associated vertex operator in $G_2$-Toda theory is $\beta=\sum_{i=1}^{2}\beta_i \, \omega_i$. Note that this defect characterizes a level 1 null state of $G_2$-Toda:
\[
\langle\beta , \alpha_2\rangle =0
\]

\subsubsection{The $G_2(a1)$ Orbit}

Consider the single coweight in the following set $\cW_{\cS}$:
\begin{align*}
& \omega_1 = -{w}^\vee_2 + 1\alpha^\vee_1 + 2\alpha^\vee_2= [\phantom{-}0,\phantom{-}0]_2
\end{align*} 
The null coweight is in its own Weyl group orbit, so the defect is unpolarized. The fundamental matter content of $T^{5d}$ is:
\begin{align*}
\prod_{1\leq I\leq d_2} N_{\emptyset \mu^{2}_{I}}( v_3^2\, f_{1}/e_{2, I};q^3).
\end{align*}
The truncation of  $T^{5d}$'s  partition function to a 3d theory's partition function is achieved by setting:
\begin{align*}
e_{1,1} = q^{-0} v^{-0} t^{N_{1,1}}\, f_1\qquad &e_{2,1} = q^{-0} v^{-0} t^{N_{2,1}}\, f_1\\
&e_{2,2} = q^{-0} v^{-2} t^{N_{2,2}}\, f_1\\
\end{align*}
A 3d chiral matter contribution survives this truncation, even with $\cW_{\cS}$ containing only the zero coweight. This is because the defect is unpolarized, so refinement due to $q$ and $v$ factors crucially enter the computation. The resulting matter content is  pictured in the ``teeth" of the 3d quiver \ref{G2plots}; see also Section \ref{sec:unpolarizedG2} below.\\

We now turn to the CFT limit: the coweight $[\phantom{-}0,\phantom{-}0]_2$ has only zeros as Dynkin labels, so part of the Bala--Carter label is $G_2$. The extra label ``2" on the coweight, denoting the fundamental representation of $^L G_2$ the coweight is taken in, is in one-to-one correspondence with an extra simple root label in the Bala--Carter classification. All in all, the label is $G_2(a_1)$. Because this is an unpolarized defect, there is no reason to expect that the Coulomb branch of $T^{4d}$ should be  given by the Spaltenstein dual of this orbit (namely, $G_2(a_1)$ itself), and this is indeed not the case. The Coulomb branch of $T^{4d}$ is in fact the orbit $A_1$, of complex dimension 3. This can be argued from the $T^{5d}$ quiver, which already has complex Coulomb dimension 3. Since the dimension of the Coulomb branch can only decrease in the $m_s\rightarrow\infty$ limit, this is the right orbit. Note  $A_1$ is not in the image of the Spaltenstein map. 

\subsubsection{Bala--Carter Labels and Unpolarized Classification}

From the above discussion, it may seem like one of the nilpotent orbits of $G_2$ is not realized as the Coulomb branch of some defect theory $T^{4d}$; namely, the orbit $A_{1_s}$ has complex dimension 4, is not in the image of the Spaltenstein map, and did not appear so far as a Coulomb branch. However, we conjecture that this orbit can be realized as follows:

Consider one coweight labeled by the  set $\cW_{\cS}$:
\begin{align*}
& \omega_1 = -{w}^\vee_1 + 2\alpha^\vee_1 + 3\alpha^\vee_2= [\phantom{-}0,\phantom{-}0]_1
\end{align*} 
This is a different way to produce the null coweight, taken this time in the representation labeled by the first fundamental coweight $[\phantom{-}1,\phantom{-}0]$. In our terminology, this  defect must be unpolarized.  The fundamental matter content of $T^{5d}$ is:
\begin{align*}
\prod_{1\leq I\leq d_1} N_{\emptyset \mu^{1}_{I}}( v^2\, f_{1}/e_{1, I};q).
\end{align*}
The truncation of  $T^{5d}$'s  partition function to a deformed conformal block with the appropriate vertex operator is achieved by setting:
\begin{align*}
e_{1,1} = q^{-0} v^{-0} t^{N_{1,1}}\, f_1\qquad &e_{2,1} = q^{-0} v^{-2} t^{N_{2,1}}\, f_1\\
e_{1,2} = q^{-1} v^{-2} t^{N_{1,2}}\, f_1\qquad &e_{2,2} = q^{-1} v^{-2} t^{N_{2,2}}\, f_1\\
&e_{2,3} = q^{-1} v^{-4} t^{N_{2,3}}\, f_1
\end{align*}
A potential survives this truncation, even with $\cW_{\cS}$ containing only the zero coweight. This is because the defect is unpolarized, so refinement due to $q$ and $v$ factors crucially enter the computation. As a result, a potential survives, as pictured in the teeth of the quiver \ref{G2plots}; see also section \ref{sec:unpolarizedG2} below.\\

We now turn to the CFT limit: we claim that the $[\phantom{-}0,\phantom{-}0]_1$ defect is distinct from the previous unpolarized one, which was engineered by $[\phantom{-}0,\phantom{-}0]_2$. We conjecture in this case that the Coulomb branch of $T^{4d}$ should be the orbit $A_{1_s}$, of complex dimension 4.  This would be consistent with dimension counting as we have described it in this paper. Note this orbit is  \textit{not} in the image of the Spaltenstein map.\\

We end this section with an important remark: we presented two truncation schemes for the null coweight of $G_2$, one for $[\phantom{-}0,\phantom{-}0]_1$ and one for $[\phantom{-}0,\phantom{-}0]_2$, but there exist more. In particular, it would be important to understand what defects are engineered by the remaining truncations. A natural guess is that they should correspond to the remaining null coweights of the fundamental representations of the quantum affine algebra $U_{\hbar}(\widehat{G_2})$.

\subsection{Unpolarized Defect of $G_2$: Determining the 3d Theory}
\label{sec:unpolarizedG2}

In this section, we show how to determine a 3d theory after truncation of a 5d theory when the defect under consideration is unpolarized. The vector multiplet and bifundamental multiplet contributions are universal for all defects, so we only need to determine the chiral matter content of the theory, the ``teeth" in the 3d quiver. 

For concreteness, let  $\cW_{\cS}$ be the set made up of the $G_2$ coweight $\omega=[\phantom{-}0, \phantom{-}0]_2$ we studied above, with $T^{5d}$ the 5d quiver engineered at the top of figure \ref{G2plots}.
Because the only coweight of $\cW_{\cS}$ is the null coweight, one would naively think that there should be no chiral matter surviving the truncation of the 5d partition function to 3d,; put differently, the D5 brane would not be felt by the compact D3 branes. This is however not the case: in the little string theory, a refinement due to $q$ and $v$ factors results in chiral and anti-chiral matter in $G^{3d}$, and one ends up with a ``quantization" of the coweight $[\phantom{-} 0,\phantom{-} 0]_2$. Namely, we perform the truncation of the 5d partition function by setting:
\begin{align*}
e_{1,1} = q^{-0} v^{-0} t^{N_{1,1}}\, f_1\qquad &e_{2,1} = q^{-0} v^{-0} t^{N_{2,1}}\, f_1\\
&e_{2,2} = q^{-0} v^{-2} t^{N_{2,2}}\, f_1 \; .
\end{align*}
A short computation shows that after some cancellations of Nekrasov functions, we obtain the partition function of a 3d theory $G^{3d}$ with  matter content $z^{3d}_{H_{2}}(x_{\mu^{2}})/z^{3d}_{H_{2}}(x_{\varnothing})$, where 
\beq\label{null2}
z^{3d}_{H_{2}}(x_{2})= \prod_{1\leq I \leq N_2} \frac{(v^2 \, q  \, e^{x^{(2)}_I}/f_{1};q^{3})_{\infty}}{( q^2 \, e^{x^{(2)}_I}/f_{1};q^{3})_{\infty}} \; .
\eeq
This is the coweight $[\phantom{-} 0,\phantom{-} 0]_2$ as it would appear in a fundamental representation of the quantum affine algebra $U_{\hbar}(\widehat{G_2})$. This refinement of the zero coweight can be recovered from other methods, such as the computation of the $G_2$ $qq$-characters, see \cite{Kimura:2017hez, Frenkel:1998} for details. Note that in the  unrefined limit $q v^{-2}=1$, the matter content becomes trivial and the D5 brane contribution decouples.

\subsection{Non Simply-Laced Triality from Folding}

In this section, we illustrate that the folding of a simply-laced defect to a non simply-laced one commutes with triality. Namely, the truncation of Young diagrams in the 5d theory is preserved by folding, so triality to a 3d non simply-laced gauge theory can be explicitly described as a folding operation. Let us consider here the example of a $\mathbb{Z}_2$-folding from a $E_6$ defect to a $F_4$ defect.\\

Our starting theory will be the following polarized defect  $\cW_{\cS}$  of $E_6$, engineered by two D5 branes:

\begin{align*}
& \omega_1 = -{w}^\vee_6 + 2\alpha^\vee_1 + 4\alpha^\vee_2 + 6\alpha^\vee_3 + 4\alpha^\vee_4 + 2\alpha^\vee_5 + 4\alpha^\vee_6\\
& \omega_2= -{w}^\vee_6
\end{align*} 

\begin{figure}[h!]
	\begin{center}
		\includegraphics[width=0.95\textwidth]{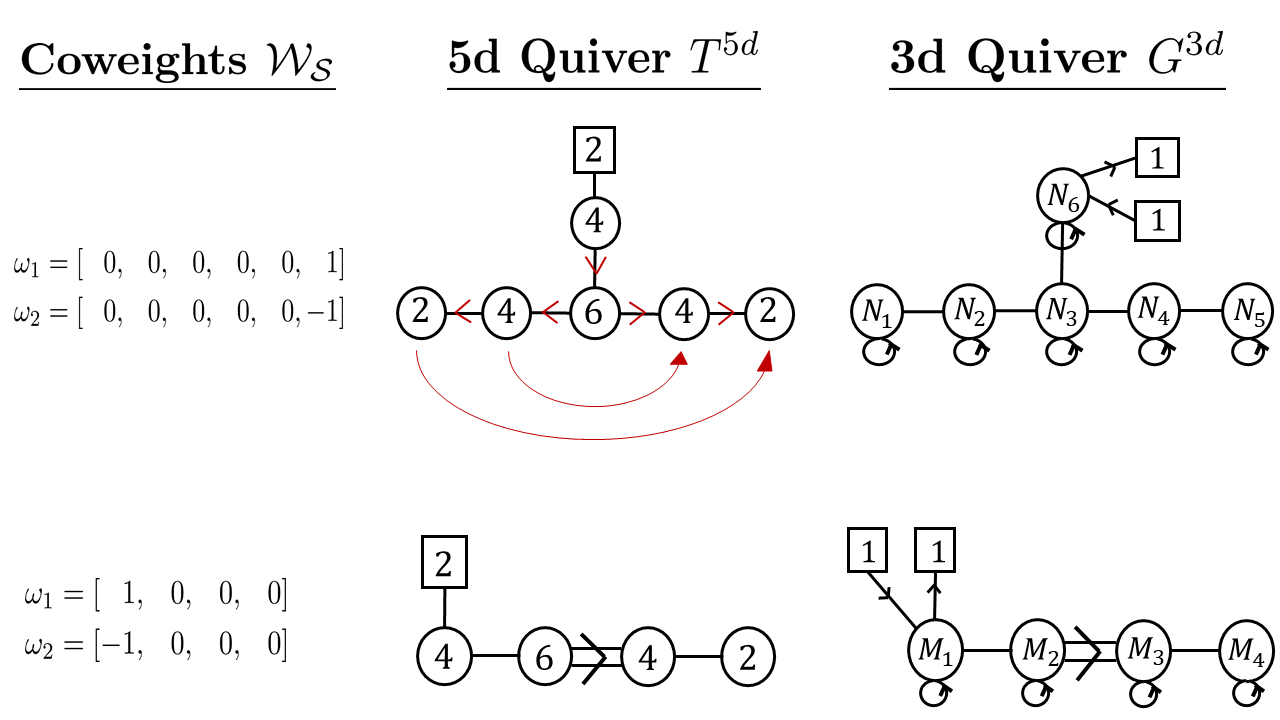}
	\end{center}
    \caption{Folding of a $E_6$ little string defect and the resulting $F_4$ defect. The 3d theory at the triality locus is shown on the right.}
	\label{F4folding}
\end{figure}

Equivalently, decomposed in terms of fundamental coweights, these read:
\begin{align*}
& \omega_1 = [\phantom{-}0,\phantom{-}0,\phantom{-}0,\phantom{-}0,\phantom{-}0,\phantom{-}1]\\
& \omega_2= [\phantom{-}0,\phantom{-}0,\phantom{-}0,\phantom{-}0,\phantom{-}0,-1]
\end{align*} 
The resulting 5d quiver $T^{5d}_{E_6}$ is shown in figure \ref{F4folding}. After $\mathbb{Z}_2$-folding, the nodes 3 and 6 now designate long roots (being invariant under the outer automorphism action) and we obtain an $F_4$ defect theory $T^{5d}_{F_4}$, with coweights:
\begin{align*}
& \omega_1' = -{w}^\vee_1 + 4\alpha^\vee_1 + 6\alpha^\vee_2 + 4\alpha^\vee_3 + 2\alpha^\vee_4\\
& \omega_2'= -{w}^\vee_1
\end{align*} 

Equivalently, decomposed in terms of fundamental coweights, these read:
\begin{align*}
& \omega_1 = [\phantom{-}1,\phantom{-}0,\phantom{-}0,\phantom{-}0]\\
& \omega_2= [-1,\phantom{-}0,\phantom{-}0,\phantom{-}0]
\end{align*}

The fundamental matter content of $T^{5d}_{E_6}$ is
\begin{align*}
\prod_{1\leq I\leq d_6} N_{\emptyset \mu^{6}_{I}}( v^2\, f_{1}/e_{6, I};q).
\end{align*}

We now show that the truncation scheme of the $T^{5d}_{E_6}$ theory at the triality point are preserved by the folding operation. Namely, we set the Coulomb parameters of  $T^{5d}_{E_6}$ to:
\begin{align*}
e_{1,1} &=  v^{-6} t^{N_{1,1}}\, f_1 &  e_{2,1} &=  v^{-4} t^{N_{2,1}}\, f_1 &  e_{3,1} &=  v^{-2} t^{N_{3,1}}\, f_1\\
e_{1,2} &=  v^{-10} t^{N_{1,2}}\, f_1 &  e_{2,2} &=  v^{-6} t^{N_{2,2}}\, f_1 &  e_{3,2} &=  v^{-4} t^{N_{3,2}}\, f_1\\
&\left. \right.  &  e_{2,3} &=  v^{-8} t^{N_{2,3}}\, f_1 &  e_{3,3} &=  v^{-6} t^{N_{3,3}}\, f_1\\
&\left. \right.  &  e_{2,4} &=  v^{-10} t^{N_{2,4}}\, f_1 &  e_{3,4} &=  v^{-6} t^{N_{3,4}}\, f_1\\
&\left. \right.  &  \left. \right. & \left. \right. &  e_{3,5} &=  v^{-8} t^{N_{3,5}}\, f_1\\
&\left. \right.  &  \left. \right. & \left. \right. &  e_{3,6} &=  v^{-10} t^{N_{3,6}}\, f_1\\
&\left. \right.  &  \left. \right. & \left. \right. &  \left. \right. & \left. \right.\\
e_{4,1} &=  v^{-4} t^{N_{4,1}}\, f_1 &  e_{5,1} &=  v^{-6} t^{N_{5,1}}\, f_1 &  e_{6,1} &=  v^{-0} t^{N_{6,1}}\, f_1\\
e_{4,2} &=  v^{-6} t^{N_{4,2}}\, f_1 &  e_{5,2} &=  v^{-10} t^{N_{5,2}}\, f_1 &  e_{6,2} &=  v^{-4} t^{N_{6,2}}\, f_1\\
e_{4,3} &=  v^{-8} t^{N_{4,3}}\, f_1 &  \left. \right. & \left. \right. &  e_{6,3} &=  v^{-6} t^{N_{6,3}}\, f_1\\
e_{4,4} &=  v^{-10} t^{N_{4,4}}\, f_1 & \left. \right. & \left. \right. &  e_{6,4} &=  v^{-10} t^{N_{6,4}}\, f_1
\end{align*}
The resulting 3d theory is shown in the right column of figure \ref{F4folding}. Now, notice that the above $v$ factors used in the truncation are identical on nodes 1 and 5, and identical and nodes 2 and 4. These become the $v$ factors of $T^{5d}_{F_4}$ for the two short roots. The $v$ factors on nodes 3 and 6 become the $v$ factors of $T^{5d}_{F_4}$ for the two long roots.

There is one slight caveat: a bifundamental hypermultiplet truncating a partition starting from  the long root 1 to the long root 2  is now accompanied by a $q^{-1}$ factor in the argument of the Nekrasov factors in the 5d partition function\footnote{This could be undone by introducing explicit masses for the bifundamental hypermultiplets, which we chose not to do in this work.}. With this adjustment, we obtain the following truncation scheme for  $T^{5d}_{F_4}$:
\begin{align*}
e_{1,1} &= q^{-0} v^{-0} t^{M_{1,1}}\, f_1 &  e_{2,1} &=  q^{-1} v^{-2} t^{M_{2,1}}\, f_1\\
e_{1,2} &=  q^{-1} v^{-4} t^{M_{1,2}}\, f_1 &  e_{2,2} &= q^{-1} v^{-4} t^{M_{2,2}}\, f_1\\
e_{1,3} &= q^{-1} v^{-6} t^{M_{1,3}}\, f_1 &  e_{2,3} &= q^{-1} v^{-6} t^{M_{2,3}}\, f_1\\
e_{1,4} &= q^{-2} v^{-10} t^{M_{1,4}}\, f_1 &  e_{2,4} &= q^{-2} v^{-6} t^{M_{2,4}}\, f_1\\
&\left. \right.  &  e_{2,5} &= q^{-2} v^{-8} t^{M_{2,5}}\, f_1\\
&\left. \right.  &  e_{2,6} &= q^{-2} v^{-10} t^{M_{2,6}}\, f_1\\
& \left. \right. &  \left. \right. & \left. \right.\\
e_{3,1} &= q^{-1} v^{-4} t^{M_{3,1}}\, f_1 &  e_{4,1} &= q^{-1} v^{-6} t^{M_{4,1}}\, f_1\\
e_{3,2} &=   q^{-1} v^{-6} t^{M_{3,2}}\, f_1 &  e_{4,2} &= q^{-2} v^{-10} t^{M_{4,2}}\, f_1\\
e_{3,3} &=   q^{-2} v^{-8} t^{M_{3,3}}\, f_1 & \left. \right. &\left. \right.\\
e_{3,4} &=  q^{-2} v^{-10} t^{M_{3,4}}\, f_1 &  \left. \right. &  \left. \right.
\end{align*}
The resulting 3d theory $G^{3d}$ is shown in the right column of figure \ref{F4folding}.

\section*{Acknowledgments}
We thank Taro Kimura and Vasily Pestun for the enlightening correspondence about their work. We thank Mina Aganagic, Amihay Hanany, Amer Iqbal, Peter Koroteev, Christian Schmid and Shing-Tung Yau   for discussions and comments at various stages of this project. We would like to thank the Center for Mathematical Sciences and Applications (CMSA) at Harvard University where the project was initiated. The work of C.K. was supported by CMSA until his current affiliation started.
The research of N. H. is supported in part by the Berkeley Center for Theoretical
Physics,  by  the  National  Science  Foundation  (award PHY-1521446) and by the
US Department of Energy under Contract DE-AC02-05CH11231.

\newpage
\bibliography{summaryABCDEFG}

\providecommand{\href}[2]{#2}\begingroup\raggedright\begin{thebibliography}{10}

\bibitem{Alday:2009aq}
L.~F. Alday, D.~Gaiotto and Y.~Tachikawa, \emph{{Liouville Correlation
  Functions from Four-dimensional Gauge Theories}},
  \href{http://dx.doi.org/10.1007/s11005-010-0369-5}{\emph{Lett. Math. Phys.}
  {\bf 91} (2010) 167--197}, [\href{https://arxiv.org/abs/0906.3219}{{\tt
  0906.3219}}].

\bibitem{Aganagic:2015cta}
M.~Aganagic and N.~Haouzi, \emph{{ADE Little String Theory on a Riemann Surface
  (and Triality)}},  \href{https://arxiv.org/abs/1506.04183}{{\tt 1506.04183}}.

\bibitem{Shiraishi:1995rp}
J.~Shiraishi, H.~Kubo, H.~Awata and S.~Odake, \emph{{A Quantum deformation of
  the Virasoro algebra and the Macdonald symmetric functions}},
  \href{http://dx.doi.org/10.1007/BF00398297}{\emph{Lett. Math. Phys.} {\bf 38}
  (1996) 33--51}, [\href{https://arxiv.org/abs/q-alg/9507034}{{\tt
  q-alg/9507034}}].

\bibitem{Awata:1995zk}
H.~Awata, H.~Kubo, S.~Odake and J.~Shiraishi, \emph{{Quantum W(N) algebras and
  Macdonald polynomials}},
  \href{http://dx.doi.org/10.1007/BF02102595}{\emph{Commun. Math. Phys.} {\bf
  179} (1996) 401--416}, [\href{https://arxiv.org/abs/q-alg/9508011}{{\tt
  q-alg/9508011}}].

\bibitem{Frenkel:1998}
E.~Frenkel and N.~Reshetikhin, \emph{Deformations of {$\mathcal W$}-algebras
  associated to simple {L}ie algebras}, {\emph{Comm. Math. Phys.} {\bf 197}
  (1998) 1--32}, [\href{https://arxiv.org/abs/q-alg/9708006}{{\tt
  q-alg/9708006}}].

\bibitem{Kimura:2017hez}
T.~Kimura and V.~Pestun, \emph{{Fractional quiver W-algebras}},
  \href{http://dx.doi.org/10.1007/s11005-018-1087-7}{\emph{Lett. Math. Phys.}
  {\bf 108} (2018) 2425--2451}, [\href{https://arxiv.org/abs/1705.04410}{{\tt
  1705.04410}}].

\bibitem{Chacaltana:2012zy}
O.~Chacaltana, J.~Distler and Y.~Tachikawa, \emph{{Nilpotent orbits and
  codimension-two defects of 6d {$\mathcal{N}=(2,0)$} theories}},
  \href{http://dx.doi.org/10.1142/S0217751X1340006X}{\emph{Int. J. Mod. Phys.}
  {\bf A28} (2013) 1340006}, [\href{https://arxiv.org/abs/1203.2930}{{\tt
  1203.2930}}].

\bibitem{Chacaltana:2010ks}
O.~Chacaltana and J.~Distler, \emph{{Tinkertoys for Gaiotto Duality}},
  \href{http://dx.doi.org/10.1007/JHEP11(2010)099}{\emph{JHEP} {\bf 11} (2010)
  099}, [\href{https://arxiv.org/abs/1008.5203}{{\tt 1008.5203}}].

\bibitem{Chacaltana:2011ze}
O.~Chacaltana and J.~Distler, \emph{{Tinkertoys for the $D_N$ series}},
  \href{http://dx.doi.org/10.1007/JHEP02(2013)110}{\emph{JHEP} {\bf 02} (2013)
  110}, [\href{https://arxiv.org/abs/1106.5410}{{\tt 1106.5410}}].

\bibitem{Chacaltana:2013oka}
O.~Chacaltana, J.~Distler and A.~Trimm, \emph{{Tinkertoys for the Twisted
  D-Series}},  \href{https://arxiv.org/abs/1309.2299}{{\tt 1309.2299}}.

\bibitem{Chacaltana:2014jba}
O.~Chacaltana, J.~Distler and A.~Trimm, \emph{{Tinkertoys for the E$_{6}$
  theory}}, \href{http://dx.doi.org/10.1007/JHEP09(2015)007}{\emph{JHEP} {\bf
  09} (2015) 007}, [\href{https://arxiv.org/abs/1403.4604}{{\tt 1403.4604}}].

\bibitem{Chacaltana:2015bna}
O.~Chacaltana, J.~Distler and A.~Trimm, \emph{{Tinkertoys for the Twisted $E_6$
  Theory}}, \href{http://dx.doi.org/10.1007/JHEP04(2015)173}{\emph{JHEP} {\bf
  04} (2015) 173}, [\href{https://arxiv.org/abs/1501.00357}{{\tt 1501.00357}}].

\bibitem{Chacaltana:2016shw}
O.~Chacaltana, J.~Distler and A.~Trimm, \emph{{Tinkertoys for the Z3-twisted D4
  Theory}},  \href{https://arxiv.org/abs/1601.02077}{{\tt 1601.02077}}.

\bibitem{Chacaltana:2012ch}
O.~Chacaltana, J.~Distler and Y.~Tachikawa, \emph{{Gaiotto duality for the
  twisted A$_{2N−1}$ series}},
  \href{http://dx.doi.org/10.1007/JHEP05(2015)075}{\emph{JHEP} {\bf 05} (2015)
  075}, [\href{https://arxiv.org/abs/1212.3952}{{\tt 1212.3952}}].

\bibitem{Aganagic:2017smx}
M.~Aganagic, E.~Frenkel and A.~Okounkov, \emph{{Quantum q-Langlands
  Correspondence}},  \href{https://arxiv.org/abs/1701.03146}{{\tt 1701.03146}}.

\bibitem{Seiberg:1997zk}
N.~Seiberg, \emph{{New theories in six-dimensions and Matrix description of M
  theory on $T^5$ and $T^5 / \mathbb{Z}_2$}},
  \href{http://dx.doi.org/10.1016/S0370-2693(97)00805-8}{\emph{Phys. Lett.}
  {\bf B408} (1997) 98--104}, [\href{https://arxiv.org/abs/hep-th/9705221}{{\tt
  hep-th/9705221}}].

\bibitem{Losev:1997hx}
A.~Losev, G.~W. Moore and S.~L. Shatashvili, \emph{{M \& m's}},
  \href{http://dx.doi.org/10.1016/S0550-3213(98)00262-4}{\emph{Nucl. Phys.}
  {\bf B522} (1998) 105--124},
  [\href{https://arxiv.org/abs/hep-th/9707250}{{\tt hep-th/9707250}}].

\bibitem{Aharony:1999ks}
O.~Aharony, \emph{{A Brief review of 'little string theories'}},
  \href{http://dx.doi.org/10.1088/0264-9381/17/5/302}{\emph{Class. Quant.
  Grav.} {\bf 17} (2000) 929--938},
  [\href{https://arxiv.org/abs/hep-th/9911147}{{\tt hep-th/9911147}}].

\bibitem{Kim:2015gha}
J.~Kim, S.~Kim and K.~Lee, \emph{{Little strings and T-duality}},
  \href{http://dx.doi.org/10.1007/JHEP02(2016)170}{\emph{JHEP} {\bf 02} (2016)
  170}, [\href{https://arxiv.org/abs/1503.07277}{{\tt 1503.07277}}].

\bibitem{Bhardwaj:2015oru}
L.~Bhardwaj, M.~Del~Zotto, J.~J. Heckman, D.~R. Morrison, T.~Rudelius and
  C.~Vafa, \emph{{F-theory and the Classification of Little Strings}},
  \href{http://dx.doi.org/10.1103/PhysRevD.93.086002}{\emph{Phys. Rev.} {\bf
  D93} (2016) 086002}, [\href{https://arxiv.org/abs/1511.05565}{{\tt
  1511.05565}}].

\bibitem{DelZotto:2015rca}
M.~Del~Zotto, C.~Vafa and D.~Xie, \emph{{Geometric engineering, mirror symmetry
  and $ 6{\mathrm{d}}_{\left(1,0\right)}\to
  4{\mathrm{d}}_{\left(\mathcal{N}=2\right)} $}},
  \href{http://dx.doi.org/10.1007/JHEP11(2015)123}{\emph{JHEP} {\bf 11} (2015)
  123}, [\href{https://arxiv.org/abs/1504.08348}{{\tt 1504.08348}}].

\bibitem{Lin:2015zea}
Y.-H. Lin, S.-H. Shao, Y.~Wang and X.~Yin, \emph{{Interpolating the Coulomb
  Phase of Little String Theory}},
  \href{http://dx.doi.org/10.1007/JHEP12(2015)022}{\emph{JHEP} {\bf 12} (2015)
  022}, [\href{https://arxiv.org/abs/1502.01751}{{\tt 1502.01751}}].

\bibitem{Hohenegger:2015btj}
S.~Hohenegger, A.~Iqbal and S.-J. Rey, \emph{{Instanton-monopole correspondence
  from M-branes on $\mathbb S^1$ and little string theory}},
  \href{http://dx.doi.org/10.1103/PhysRevD.93.066016}{\emph{Phys. Rev.} {\bf
  D93} (2016) 066016}, [\href{https://arxiv.org/abs/1511.02787}{{\tt
  1511.02787}}].

\bibitem{Hohenegger:2016eqy}
S.~Hohenegger, A.~Iqbal and S.-J. Rey, \emph{{Self-Duality and Self-Similarity
  of Little String Orbifolds}},
  \href{http://dx.doi.org/10.1103/PhysRevD.94.046006}{\emph{Phys. Rev.} {\bf
  D94} (2016) 046006}, [\href{https://arxiv.org/abs/1605.02591}{{\tt
  1605.02591}}].

\bibitem{Hohenegger:2016yuv}
S.~Hohenegger, A.~Iqbal and S.-J. Rey, \emph{{Dual Little Strings from F-Theory
  and Flop Transitions}},
  \href{http://dx.doi.org/10.1007/JHEP07(2017)112}{\emph{JHEP} {\bf 07} (2017)
  112}, [\href{https://arxiv.org/abs/1610.07916}{{\tt 1610.07916}}].

\bibitem{Aganagic:2016jmx}
M.~Aganagic and A.~Okounkov, \emph{{Elliptic stable envelope}},
  \href{https://arxiv.org/abs/1604.00423}{{\tt 1604.00423}}.

\bibitem{Bastian:2017ing}
B.~Bastian, S.~Hohenegger, A.~Iqbal and S.-J. Rey, \emph{{Dual Little Strings
  and their Partition Functions}},
  \href{https://arxiv.org/abs/1710.02455}{{\tt 1710.02455}}.

\bibitem{Bastian:2017ary}
B.~Bastian, S.~Hohenegger, A.~Iqbal and S.-J. Rey, \emph{{Triality in Little
  String Theories}},  \href{https://arxiv.org/abs/1711.07921}{{\tt
  1711.07921}}.

\bibitem{Douglas:1996sw}
M.~R. Douglas and G.~W. Moore, \emph{{D-branes, quivers, and ALE instantons}},
  \href{https://arxiv.org/abs/hep-th/9603167}{{\tt hep-th/9603167}}.

\bibitem{Aspinwall:1996nk}
P.~S. Aspinwall and M.~Gross, \emph{{The SO(32) heterotic string on a K3
  surface}}, \href{http://dx.doi.org/10.1016/0370-2693(96)01095-7}{\emph{Phys.
  Lett.} {\bf B387} (1996) 735--742},
  [\href{https://arxiv.org/abs/hep-th/9605131}{{\tt hep-th/9605131}}].

\bibitem{Haouzi:2016ohr}
N.~Haouzi and C.~Schmid, \emph{{Little String Origin of Surface Defects}},
  \href{http://dx.doi.org/10.1007/JHEP05(2017)082}{\emph{JHEP} {\bf 05} (2017)
  082}, [\href{https://arxiv.org/abs/1608.07279}{{\tt 1608.07279}}].

\bibitem{Haouzi:2016yyg}
N.~Haouzi and C.~Schmid, \emph{{Little String Defects and Bala-Carter Theory}},
   \href{https://arxiv.org/abs/1612.02008}{{\tt 1612.02008}}.

\bibitem{Aganagic:2013tta}
M.~Aganagic, N.~Haouzi, C.~Kozcaz and S.~Shakirov, \emph{{Gauge/Liouville
  Triality}},  \href{https://arxiv.org/abs/1309.1687}{{\tt 1309.1687}}.

\bibitem{Aganagic:2014oia}
M.~Aganagic, N.~Haouzi and S.~Shakirov, \emph{{$A_n$-Triality}},
  \href{https://arxiv.org/abs/1403.3657}{{\tt 1403.3657}}.

\bibitem{Moore:1997dj}
G.~W. Moore, N.~Nekrasov and S.~Shatashvili, \emph{{Integrating over Higgs
  branches}}, \href{http://dx.doi.org/10.1007/PL00005525}{\emph{Commun. Math.
  Phys.} {\bf 209} (2000) 97--121},
  [\href{https://arxiv.org/abs/hep-th/9712241}{{\tt hep-th/9712241}}].

\bibitem{Nekrasov:2002qd}
N.~A. Nekrasov, \emph{{Seiberg-Witten prepotential from instanton counting}},
  \href{http://dx.doi.org/10.4310/ATMP.2003.v7.n5.a4}{\emph{Adv. Theor. Math.
  Phys.} {\bf 7} (2003) 831--864},
  [\href{https://arxiv.org/abs/hep-th/0206161}{{\tt hep-th/0206161}}].

\bibitem{Nekrasov:2012xe}
N.~Nekrasov and V.~Pestun, \emph{{Seiberg-Witten geometry of four dimensional
  $\mathcal{N}=2$ quiver gauge theories}},
  \href{https://arxiv.org/abs/1211.2240}{{\tt 1211.2240}}.

\bibitem{Shadchin:2006yz}
S.~Shadchin, \emph{{On F-term contribution to effective action}},
  \href{http://dx.doi.org/10.1088/1126-6708/2007/08/052}{\emph{JHEP} {\bf 08}
  (2007) 052}, [\href{https://arxiv.org/abs/hep-th/0611278}{{\tt
  hep-th/0611278}}].

\bibitem{Hama:2011ea}
N.~Hama, K.~Hosomichi and S.~Lee, \emph{{SUSY Gauge Theories on Squashed
  Three-Spheres}}, \href{http://dx.doi.org/10.1007/JHEP05(2011)014}{\emph{JHEP}
  {\bf 05} (2011) 014}, [\href{https://arxiv.org/abs/1102.4716}{{\tt
  1102.4716}}].

\bibitem{Kapustin:2011jm}
A.~Kapustin and B.~Willett, \emph{{Generalized Superconformal Index for Three
  Dimensional Field Theories}},  \href{https://arxiv.org/abs/1106.2484}{{\tt
  1106.2484}}.

\bibitem{Beem:2012mb}
C.~Beem, T.~Dimofte and S.~Pasquetti, \emph{{Holomorphic Blocks in Three
  Dimensions}}, \href{http://dx.doi.org/10.1007/JHEP12(2014)177}{\emph{JHEP}
  {\bf 12} (2014) 177}, [\href{https://arxiv.org/abs/1211.1986}{{\tt
  1211.1986}}].

\bibitem{Bouwknegt:1992wg}
P.~Bouwknegt and K.~Schoutens, \emph{{W symmetry in conformal field theory}},
  \href{http://dx.doi.org/10.1016/0370-1573(93)90111-P}{\emph{Phys. Rept.} {\bf
  223} (1993) 183--276}, [\href{https://arxiv.org/abs/hep-th/9210010}{{\tt
  hep-th/9210010}}].

\bibitem{Dotsenko:1984nm}
V.~S. Dotsenko and V.~A. Fateev, \emph{{Conformal Algebra and Multipoint
  Correlation Functions in Two-Dimensional Statistical Models}},
  \href{http://dx.doi.org/10.1016/0550-3213(84)90269-4}{\emph{Nucl. Phys.} {\bf
  B240} (1984) 312}.

\bibitem{Dijkgraaf:2009pc}
R.~Dijkgraaf and C.~Vafa, \emph{{Toda Theories, Matrix Models, Topological
  Strings, and N=2 Gauge Systems}},
  \href{https://arxiv.org/abs/0909.2453}{{\tt 0909.2453}}.

\bibitem{Itoyama:2009sc}
H.~Itoyama, K.~Maruyoshi and T.~Oota, \emph{{The Quiver Matrix Model and 2d-4d
  Conformal Connection}},
  \href{http://dx.doi.org/10.1143/PTP.123.957}{\emph{Prog. Theor. Phys.} {\bf
  123} (2010) 957--987}, [\href{https://arxiv.org/abs/0911.4244}{{\tt
  0911.4244}}].

\bibitem{Mironov:2010zs}
A.~Mironov, A.~Morozov and S.~Shakirov, \emph{{Conformal blocks as
  Dotsenko-Fateev Integral Discriminants}},
  \href{http://dx.doi.org/10.1142/S0217751X10049141}{\emph{Int. J. Mod. Phys.}
  {\bf A25} (2010) 3173--3207}, [\href{https://arxiv.org/abs/1001.0563}{{\tt
  1001.0563}}].

\bibitem{Morozov:2010cq}
A.~Morozov and S.~Shakirov, \emph{{The matrix model version of AGT conjecture
  and CIV-DV prepotential}},
  \href{http://dx.doi.org/10.1007/JHEP08(2010)066}{\emph{JHEP} {\bf 08} (2010)
  066}, [\href{https://arxiv.org/abs/1004.2917}{{\tt 1004.2917}}].

\bibitem{Maruyoshi:2014eja}
K.~Maruyoshi, \emph{{$\beta$-Deformed Matrix Models and 2d/4d Correspondence}},
   \href{https://arxiv.org/abs/1412.7124}{{\tt 1412.7124}}.

\bibitem{Fateev:2007ab}
V.~A. Fateev and A.~V. Litvinov, \emph{{Correlation functions in conformal Toda
  field theory. I.}},
  \href{http://dx.doi.org/10.1088/1126-6708/2007/11/002}{\emph{JHEP} {\bf 11}
  (2007) 002}, [\href{https://arxiv.org/abs/0709.3806}{{\tt 0709.3806}}].

\bibitem{Nekrasov:2010ka}
N.~Nekrasov and E.~Witten, \emph{{The Omega Deformation, Branes, Integrability,
  and Liouville Theory}},
  \href{http://dx.doi.org/10.1007/JHEP09(2010)092}{\emph{JHEP} {\bf 09} (2010)
  092}, [\href{https://arxiv.org/abs/1002.0888}{{\tt 1002.0888}}].

\bibitem{Frenkel:qch}
E.~Frenkel and N.~Reshetikhin, \emph{The q-characters of representations of
  quantum affine algebras and deformations of w-algebras}, {\emph{in
  Contemporary Math} {\bf 248} (2000) },
  [\href{https://arxiv.org/abs/math/9810055}{{\tt math/9810055}}].

\bibitem{Nekrasov:2015wsu}
N.~Nekrasov, \emph{{BPS/CFT correspondence: non-perturbative Dyson-Schwinger
  equations and qq-characters}},
  \href{http://dx.doi.org/10.1007/JHEP03(2016)181}{\emph{JHEP} {\bf 03} (2016)
  181}, [\href{https://arxiv.org/abs/1512.05388}{{\tt 1512.05388}}].

\bibitem{Kimura:2015rgi}
T.~Kimura and V.~Pestun, \emph{{Quiver W-algebras}},
  \href{https://arxiv.org/abs/1512.08533}{{\tt 1512.08533}}.

\bibitem{Gukov:2008sn}
S.~Gukov and E.~Witten, \emph{{Rigid Surface Operators}},
  \href{http://dx.doi.org/10.4310/ATMP.2010.v14.n1.a3}{\emph{Adv. Theor. Math.
  Phys.} {\bf 14} (2010) 87--178}, [\href{https://arxiv.org/abs/0804.1561}{{\tt
  0804.1561}}].

\bibitem{Collingwood:1993}
D.~H. Collingwood and W.~M. McGovern, \emph{Nilpotent orbits in semisimple
  {L}ie algebras}.
\newblock Van Nostrand Reinhold Mathematics Series. Van Nostrand Reinhold Co.,
  New York, 1993.

\bibitem{Spaltenstein:1982}
N.~Spaltenstein, \emph{Classes unipotentes et sous-groupes de {B}orel},
  vol.~946 of \emph{Lecture Notes in Mathematics}.
\newblock Springer-Verlag, Berlin-New York, 1982.

\bibitem{10.2307/1971193}
D.~Barbasch and D.~A. Vogan, \emph{Unipotent representations of complex
  semisimple groups}, {\emph{Annals of Mathematics} {\bf 121} (1985) 41--110}.

\bibitem{bala:1976msaa}
P.~Bala and R.~W. Carter, \emph{Classes of unipotent elements in simple
  algebraic groups. {I}},
  \href{http://dx.doi.org/10.1017/S0305004100052403}{\emph{Math. Proc.
  Cambridge Philos. Soc.} {\bf 79} (1976) 401--425}.

\bibitem{bala:1976msab}
P.~Bala and R.~W. Carter, \emph{Classes of unipotent elements in simple
  algebraic groups. {II}},
  \href{http://dx.doi.org/10.1017/S0305004100052610}{\emph{Math. Proc.
  Cambridge Philos. Soc.} {\bf 80} (1976) 1--17}.

\bibitem{Kanno:2009ga}
S.~Kanno, Y.~Matsuo, S.~Shiba and Y.~Tachikawa, \emph{{$\mathcal{N}=2$ gauge
  theories and degenerate fields of Toda theory}},
  \href{http://dx.doi.org/10.1103/PhysRevD.81.046004}{\emph{Phys. Rev.} {\bf
  D81} (2010) 046004}, [\href{https://arxiv.org/abs/0911.4787}{{\tt
  0911.4787}}].

\bibitem{Gaiotto:2009we}
D.~Gaiotto, \emph{{$\mathcal{N}=2$ dualities}},
  \href{http://dx.doi.org/10.1007/JHEP08(2012)034}{\emph{JHEP} {\bf 08} (2012)
  034}, [\href{https://arxiv.org/abs/0904.2715}{{\tt 0904.2715}}].

\end{thebibliography}\endgroup
\bibliographystyle{JHEP}

\end{document}